\documentclass[fleqn,usenatbib]{mnras}

\usepackage{newtxtext,newtxmath}

\usepackage[T1]{fontenc}
\usepackage{threeparttablex}
\usepackage{subcaption}
\usepackage{caption}

\DeclareRobustCommand{\VAN}[3]{#2}
\let\VANthebibliography\thebibliography
\def\thebibliography{\DeclareRobustCommand{\VAN}[3]{##3}\VANthebibliography}

\usepackage{graphicx}	
\usepackage{amsmath}	
\usepackage{tablefootnote}
\usepackage{multirow}
\usepackage{adjustbox}

\newcommand{\um}{\mu\mathrm{m}}
\newcommand{\oiii}{[O\,\textsc{iii}]}
\newcommand{\oiiihb}{[O\,\textsc{iii}]$+\rm{H}\beta$}

\defcitealias{smit_high-precision_2015}{S15}
\defcitealias{roberts-borsani_z_2016}{R16}

\newcommand{\orcidsymb}[2]{#1\href{http://orcid.org/#2}{\adjustbox{trim={-.15\width} {0\height} {-.15\width} {0\height},clip}{\includegraphics[height=10pt]{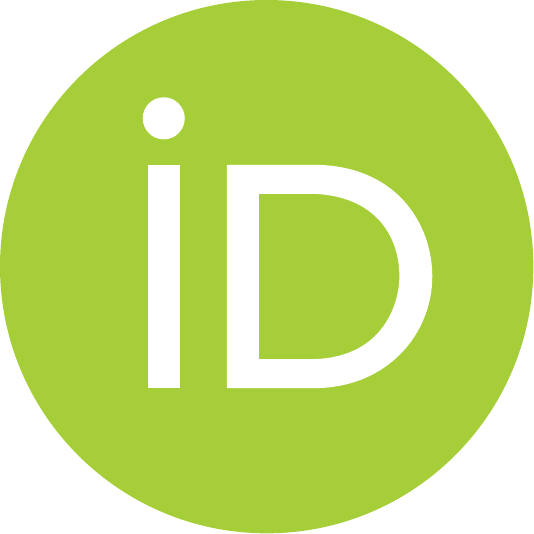}}}}

\title[IRAC-Excess Galaxies with JWST/NIRSpec]{IRAC-Excess Galaxies in the NIRSpec Wide Survey: physical properties of bright EELGs at $z\sim7$ and implications for Reionization}

\author[K. Ormerod et al.]{\orcidsymb{Katherine Ormerod}{0000-0003-2000-3420}$^{1}$\thanks{E-mail: \href{mailto:arikorme@ljmu.ac.uk}{arikorme@ljmu.ac.uk}},
\orcidsymb{Renske Smit}{0000-0001-8034-7802}$^{1}$,
\orcidsymb{Joris Witstok}{0000-0002-7595-121X}$^{2,3}$,
\orcidsymb{Anna de Graaff}{0000-0002-2380-9801}$^{4, 5}$\thanks{Clay Fellow}, 
\orcidsymb{Michael V. Maseda}{0000-0003-0695-4414}${^6}$, \newauthor
\orcidsymb{Andrew J.\ Bunker }{0000-0002-8651-9879}${^7}$,
\orcidsymb{Chris Willott}{0000-0002-4201-7367}$^{8}$
\\
$^{1}$Astrophysics Research Institute, Liverpool John Moores University, 146 Brownlow Hill, Liverpool, L3 5RF, UK \\
$^{2}$Cosmic Dawn Center (DAWN), Copenhagen, Denmark \\
${^3}$Niels Bohr Institute, University of Copenhagen, Jagtvej 128, DK-2200, Copenhagen, Denmark\\
${^4}$Max-Planck-Institut f\"ur Astronomie, K\"onigstuhl 17, D-69117 Heidelberg, Germany \\
$^5$Center for Astrophysics $|$ Harvard \& Smithsonian, 60 Garden St., Cambridge MA 02138 USA \\
$^6$Department of Astronomy, University of Wisconsin-Madison, 475 N. Charter St., Madison, WI 53706, USA\\
$^7$Department of Physics, University of Oxford, Denys Wilkinson Building, Keble Road, Oxford, OX1 3RH, UK \\
${^8}$NRC Herzberg, 5071 West Saanich Rd, Victoria, BC V9E 2E7, Canada \\
}

\date{Accepted XXX. Received YYY; in original form ZZZ}

\pubyear{2026}

\begin{document}
\label{firstpage}
\pagerange{\pageref{firstpage}--\pageref{lastpage}}
\maketitle

\begin{abstract}
We present an analysis of 23 \emph{Spitzer}/IRAC excess galaxies at $5.66 < z < 7.64$, observed as part of the NIRSpec Wide GTO survey. With NIRSpec PRISM spectroscopy, we measure a median rest-frame {\oiiihb} EW $=1583$~{\AA}, consistent with predictions from the pre-\emph{JWST} selection and we identify $17/23$ $(\sim74\%)$ of our sample as extreme emission line galaxies (EELGs), based on a selection of {\oiiihb} EW $\geq1000$~{\AA}. 
The typical EELG is powered by star-formation rather than active galactic nuclei (AGN) based on their line widths, line ratios and absence of high-ionization lines, though we cannot exclude an AGN contribution in a few individual sources.
We find that our luminous ($\langle M_\mathrm{UV}\rangle=-20.7$) EELG sample is characterised by sub-solar metallicities $(7.38 < \rm 12 + \log(O/H) < 8.12)$,  young stellar populations ($t_\star \sim 20$ Myr), and enhanced ionizing photon production efficiencies, with median $\log_{10}(\xi_\mathrm{ion}/\mathrm{erg}^{-1}~\mathrm{Hz}) = 25.62$. 
We find that more UV luminous ($\sim L_\mathrm{UV, z=7}^\ast$) EELGs exhibit systematically lower O32 ratios than fainter ($\sim 0.1\times L_\mathrm{UV}^\ast$) EELGs at fixed {\oiiihb} EW, indicating less extreme radiation fields and/or lower gas densities  than their fainter counterparts. Furthermore, bright EELGs exhibit low neutral hydrogen column densities, consistent with a scenario in which they have largely ionized their circumgalactic medium (CGM). Despite this, we find very low Lyman-$\alpha$ escape fractions ($<2$\%), suggesting these systems reside in relatively small ($\lesssim 0.25$~pMpc) ionized regions of the intergalactic medium (IGM),
possibly suggesting that bright galaxies have a minor contribution to reionization.

\end{abstract}




\section{Introduction}
\label{sec:intro}

Galaxies that exhibit strong rest-frame optical emission lines relative to their stellar continuum, with rest-frame equivalent widths (EW) in the hundreds or thousands of Angstroms (\AA) are known as extreme emission line galaxies \citep[EELGs;][]{van_der_wel_extreme_2011}. 
Such extreme emission lines are most commonly seen in the rest-optical {\oiii}$\lambda5007$ and H$\alpha$ emission lines, and are driven by ionizing photons emitted from massive O and B type stars, or active galactic nuclei (AGN). Conversely, the surrounding rest-optical continuum includes contribution from older, less massive stars \citep{eldridge_new_2022, boyett_extreme_2024, llerena_physical_2024}. Therefore, high equivalent widths are seen in galaxies with high specific star formation rates (sSFRs), and may be used to identify galaxies undergoing an intense starburst or recent upturn in star formation history \citep{van_der_wel_extreme_2011, endsley_star-forming_2024}. 
EELGs have been studied extensively across cosmic time, from the local Universe \citep[e.g.,][]{cardamone_galaxy_2009, amorin_oxygen_2010, izotov_green_2011, liu_strong_2022}, at intermediate redshifts \citep[e.g.,][]{van_der_wel_extreme_2011, maseda_nature_2014, amorin_extreme_2015, maseda_number_2018, tang_mmtmmirs_2019, tran_mosel_2020, boyett_o_2022}, and within the Epoch of Reionization (EoR) and beyond \citep[e.g.,][]{smit_evidence_2014, smit_high-precision_2015, roberts-borsani_z_2016, de_barros_greats_2019,endsley_o_2021, chen_jwst_2024, boyett_extreme_2024, llerena_physical_2024, davis_census_2024}.

Studies of EELGs across cosmic time have revealed that these systems become increasingly common at higher redshift \citep[e.g.,][]{smit_evidence_2014, boyett_o_2022, withers_spectroscopy_2023, boyett_extreme_2024, Begley_2025}. At fixed redshift, the prevalence of EELGs is also found to scale with the UV luminosity ($L_\mathrm{UV}$). \citet{endsley_star-forming_2024} demonstrated that {\oiiihb} EWs of $z\sim6-9$ galaxies show a significant decline ($\gtrsim5\sigma$) towards lower UV luminosities, which may be explained by a combination of decreasing metallicity with UV luminosity and a larger fraction of systems undergoing a burst of star-formation at higher luminosities. 

The Epoch of Reionization is the last major phase transition of the Universe, where hydrogen in the intergalactic medium (IGM) transitioned from completely neutral to completely ionized \citep[e.g.,][]{Robertston_2022, ellis_galaxies_2025, Jones_reionization_2025,stark_observations_2026}. Current observational constraints suggest that reionization was approximately halfway complete by $z=7.7\pm0.7$ \citep{planck_collaboration_planck_2020}, and was complete by $z\sim5.3-6$ \citep[e.g.,][]{bosman_hydrogen_2022, zhu_long_2022, zhu_damping_2024, giovinazzo_reionization_2026}.
It is thought that star forming galaxies were responsible for reionizing the majority of hydrogen in the Universe, with AGN playing a subdominant role \citep[e.g.,][]{Robertson_2015, atek_most_2024, dayal_uncovering_2025}.  
Whether reionization is driven primarily by faint \citep{atek_most_2024} or bright \citep{naidu_rapid_2020} galaxies remains an active area of debate. Recent work has shown that the relative contribution from the faint or bright galaxy populations varies with redshift, with brighter galaxies contributing more to the cosmic ionizing emissivity ($\dot{n}_\mathrm{ion}$) at lower redshift, and the $M_\mathrm{UV} < -20$ population providing the largest contribution by the end of reionization \citep[][]{summerfield_thesan-zoom_2026, giovinazzo_reionization_2026}. 

Prior to the launch of the \emph{James Webb Space Telescope} \citep[\emph{JWST;}][]{Gardner_JWST}, studies of the rest-frame optical properties of galaxies within the EoR relied on \emph{Spitzer's} Infrared Array Camera (IRAC).
Through IRAC observations, a subset of galaxies exhibiting extremely blue or red $[3.6]-[4.5]$ micron ($\mu$m) colours was discovered, and are often referred to as \emph{Spitzer}/IRAC excess sources \citep[e.g.,][]{smit_evidence_2014, roberts-borsani_z_2016}. 
Due to the limited sensitivity of \emph{Spitzer}/IRAC, these galaxies are typically UV bright sources ($\sim L_\mathrm{UV}^{*}$).
The IRAC colour excess has been shown to be redshift dependent \citep[e.g.][]{labbe_spectral_2013, wilkins_theoretical_2013, laporte_first_2014, smit_evidence_2014}, and is often interpreted as arising from strong rest-frame optical nebular emission lines such as [O\,\textsc{iii}]$\lambda\lambda4959,5007$\AA\ ([O\,\textsc{iii}] hereafter) and H$\alpha$ moving through the IRAC filters. This results in moderately blue colours at $z\sim6$ where both the $[3.6]$ and $[4.5]$ bands are contaminated by emission lines, very blue colours at $z\sim 6.8$ where only the $[3.6]$ band is contaminated, and very red colours at $z > 7$ where the $[4.5]$ band is contaminated \citep[e.g.][]{schaerer_impact_2009, labbe_spectral_2013, smit_evidence_2014}. At $z\sim6$, very blue IRAC colours can also be caused by blue continua in young, dust free galaxies, or high [O\,\textsc{iii}]/H$\alpha$ ratios. 
These IRAC excess colours have been used to infer average rest-frame [O\,\textsc{iii}]$+\rm{H}\beta$ EWs of $>630${\AA} at $z\sim6-8$ \citep{labbe_spectral_2013, smit_evidence_2014, smit_high-precision_2015, de_barros_greats_2019}, with individual galaxies showing extreme EWs of $>1000${\AA} \citep{smit_evidence_2014, smit_high-precision_2015, roberts-borsani_z_2016}, although whether these extreme emission lines were powered by AGN or star formation remained debated.
Before the launch of \emph{JWST}, spectroscopic observations of luminous galaxies with red IRAC colours revealed high fractions of Lyman-$\alpha$ (Ly$\alpha$) detections \citep[][and references therein]{Roberts-Borsani_2023_lya}. Notably, \citet{roberts-borsani_z_2016} found a 100 per cent detection rate, providing strong evidence for luminous IRAC-excess sources carving out ionized bubbles. Additionally, \citet{Endsley_2021_MMT} find increased Ly$\alpha$ EWs in strong {\oiiihb} emitters, attributed to a larger ionizing photon production efficiency caused by extremely young stars. Together, these findings indicate that IRAC-excess sources may be efficient contributors to the ionizing photon budget.

However, the origin of the IRAC excess itself is debated. Rather than strong line emission, it has been suggested that the red colour of $4.5\mu\mathrm{m}$ excess sources at $7 < z < 9$ could be caused by the presence of older stellar populations resulting in a Balmer break \citep[e.g.][]{roberts-borsani_interpreting_2020}. The distinction between these two interpretations is of vital importance to understanding galaxy formation and evolution, as attributing the IRAC excess of galaxies at $z>7$ to the presence of a Balmer break, rather than extreme rest-optical emission lines, suggests that galaxies may be more massive and host older stellar populations than previously thought, requiring an earlier onset of star formation. 

With the launch of \emph{JWST}, spectroscopic observations of the {\oiiihb} emission lines are now possible out to $z\sim9.5$ with the Near-infrared Spectrograph \citep[NIRSpec;][]{jakobsen_near-infrared_2022, ferruit_near-infrared_2022,boker_-orbit_2023}. This enables the spectroscopic identification of EELGs in the early Universe, revealing the prevalence of these systems at early times.
Spectroscopic observations from \emph{JWST} focusing on faint EELGs ($L\sim 0.03-0.3 L^\ast_{z=7}$) have shown that EELGs are typically characterised by a high ionizing photon production efficiency, low dust attenuation, subsolar oxygen abundances, and a recent starburst, with massive stars favoured as the dominant ionizing source \citep[e.g.,][]{llerena_physical_2024, llerena_ionising_2025, Llerena_2026, boyett_extreme_2024}. 

In this work, we present an analysis of \emph{Spitzer}/IRAC excess sources observed as part of the NIRSpec Wide Guaranteed Time Observations (GTO) Survey \citep[Wide;][]{maseda_nirspec_2024}, covering all five Cosmic Assembly Near-infrared Deep Extragalactic Legacy Survey (CANDELS) fields \citep{grogin_candels_2011, koekemoer_candels_2011}. Due to their inherent brightness $(L\sim L^{*}_{z=7})$ compared to previous studies, they provide insights into some of the most extreme, bursty systems in the EoR. 
This paper is organized as follows: Section \ref{sec:data} discusses the observations and sample selection, Section \ref{sec:analysis} describes the analysis of our sample and results, and in Section \ref{sec:discussion} we discuss our results in the context of the Epoch of Reionization. Throughout this paper, we use a standard cosmology of $H_0 = 70~{\rm km~s}^{-1}{\rm Mpc}^{-1}$, $\Omega_m = 0.3$, and $\Omega_\Lambda = 0.7$. All magnitudes are quoted in the AB magnitude system \citep{oke_secondary_1983}. We assume a solar metallicity of $12+\log(\mathrm{O/H})=8.69$ \citep{asplund_chemical_2021}. All equivalent widths are quoted in the rest frame.

\section{Data and Sample Selection}
\label{sec:data}

\subsection{Sample Selection}
\label{sec:selection}

Our initial sample is comprised of 24 IRAC-excess sources, which were originally identified in \citet{smit_high-precision_2015} and \citet{roberts-borsani_z_2016} (hereafter \citetalias{smit_high-precision_2015} and \citetalias{roberts-borsani_z_2016}), 
and followed up with \emph{JWST} as part of the NIRSpec Wide GTO Survey. 
IRAC-excess sources were selected as one of the Wide top priority `P1' targets, as these bright galaxies are sufficiently rare that Wide is the only GTO tier that can probe the full population \citep[][]{maseda_nirspec_2024}.

Sources from \citetalias{smit_high-precision_2015} exhibit blue IRAC colours, and were selected using the criterion $P([3.6]-[4.5] < -0.5) > 84\%$, with a  signal-to-noise cut on the \emph{Spitzer}/IRAC data, resulting in sources with typical $m_{\rm UV}\lesssim26$. We use both the fiducial $z\sim 6.8$ sample and the additional $z\sim6$ sample. Sources from \citetalias{roberts-borsani_z_2016} display red IRAC colours meeting the criterion $[3.6]-[4.5] > 0.5$. The primary sample was selected using a magnitude cut of $m_{\rm UV} <25.5$, however, we also include the supporting sample, which may be as faint as $m_{\rm UV} \lesssim 26.5$ \citepalias[Tables 3 and 5 of][]{roberts-borsani_z_2016}. An additional IRAC-excess candidate is included, with an IRAC colour of $[3.6]-[4.5] = 0.22 \pm 0.06$ measured by \citetalias{roberts-borsani_z_2016}, but $[3.6]-[4.5] = 0.53 \pm 0.26$ measured by \citet{Skelton_2014}. The sample is summarised in Table \ref{tab:general}, and the redshift distribution is shown in Figure \ref{fig:general_hist}.

We note that three galaxies were observed twice. However, we include both observations separately within our analysis as the NIRSpec slit placements sample different regions of the galaxy. The NIRSpec slit locations for the duplicate observations are shown in Figure \ref{fig:slit locs}, while the slit locations for the remaining galaxies with public NIRCam coverage are shown in Figure \ref{fig:individual slit locs}. Finally, we exclude Wide-000005 from our analysis due to the lack of detected emission lines, caused 
by the source falling in the corner of the NIRSpec shutter (see Figure \ref{fig:individual slit locs}). As a result of this, Wide-000005 has substantial slit losses and a low signal-to-noise ratio (SNR).  This results in a final sample of 26 spectra (23 galaxies).. We show a number of properties as a function of redshift coloured by IRAC colour in Figure \ref{fig:irac_colours} to investigate potential selection biases between the samples selected from \citetalias{smit_high-precision_2015} and \citetalias{roberts-borsani_z_2016}. The median UV magnitudes, UV continuum slopes, and Balmer break strengths are very similar within both samples. Despite the considerable scatter seen in the NIRSpec derived {\oiiihb} equivalent width distributions of each sample, the \citetalias{smit_high-precision_2015} sample exhibits a higher median value ($1687_{-843}^{+472}\,\text{\AA}$ for \citealp{smit_high-precision_2015} compared to $1124_{-405}^{+1192}\,\text{\AA}$ for \citealp{roberts-borsani_z_2016}). This difference may be explained by \citealp{smit_high-precision_2015} including the $1\sigma$ uncertainty on the $[3.6]-[4.5]$ colour in their selection criterion, consistent with their sample exhibiting a more extreme median IRAC colour ($[3.6]-[4.5]=-1.00^{+0.10}_{-0.38}$) compared to the median IRAC colour ($[3.6]-[4.5]=0.70^{+0.32}_{-0.20}$) of the sample selected from \citealp{roberts-borsani_z_2016}.

\renewcommand{\arraystretch}{1.5}
\begin{table*}
    \centering
    \caption{Overview of the Wide IRAC-excess sample. Columns show the galaxy ID referred to throughout this work, the literature ID of each galaxy, $M_\mathrm{UV}$ from \emph{HST} F160W photometry (with the exception of Wide-001576, where we quote the \emph{JWST} F150W value from \citet{ormerod_detection_2025}), right ascension (RA) and declination (Dec) of each galaxy, $z_\mathrm{prism}$ obtained from \textsc{ppxf} fitting in Section \ref{sec:ppxf}, and [3.6]-[4.5] IRAC colours. The final column indicates the study in which each galaxy was identified as an IRAC-excess source. }
    \begin{ThreePartTable}
    \begin{tabular}{l|c|c|c|c|c|c|c} \hline
        ID & Literature ID & $M_\mathrm{UV}$ & RA (J2000) & Dec (J2000)  & $z_\mathrm{prism}$ & $[3.6]-[4.5]$ (mag)& Reference \\ \hline
        Wide-000003 & UDSY-4133353345 & -21.18 & 02:17:41.33 & -05:15:33.41 & $7.284\pm0.001$ & $0.5^{+ 0.2}_{- 0.2}$ & \citetalias{roberts-borsani_z_2016} \\
        Wide-000001 & UDSZ-4199355469 & -20.46 & 02:17:41.99 & -05:15:54.61 & $7.283\pm0.001$ & $1.8^{+ 2.6}_{- 0.6}$ & \citetalias{roberts-borsani_z_2016} \\
        Wide-000005$^a$ & UDSY-4308785165 & -20.75 & 02:17:43.08 & -05:08:51.61 & $7.619\pm0.004$ & $0.7^{+0.8}_{-0.5}$ &  \citetalias{roberts-borsani_z_2016} \\
        Wide-000036 & UDSI-5125113373 & -21.13 & 02:17:51.25 & -05:11:33.68 & $6.096\pm0.003$ & $-0.9^{+ 0.3}_{- 0.3}$ & \citetalias{smit_high-precision_2015} \\
        Wide-001018 & UDS-5417745460 & -21.25 & 02:17:54.18 & -05:14:54.60 & $6.555\pm0.001$ & $-0.7^{+ 0.2}_{- 0.2}$ & \citetalias{smit_high-precision_2015} \\
        Wide-002631 & ERS-2115344329 & -19.73 & 03:32:11.55 & -27:44:33.31 & $6.275\pm0.002$ & $-1.0^{+ 0.4}_{- 0.4}$ & \citetalias{smit_high-precision_2015} \\
        Wide-000440 & GSD-2234402156 & -19.81 & 03:32:23.46 & -27:50:21.84 & $5.774\pm0.003$ & $-1.0^{+ 0.3}_{- 0.3}$ & \citetalias{smit_high-precision_2015} \\
        Wide-008401 & GSD-2237749136 & -20.32 & 03:32:23.79 & -27:49:13.86 & $6.895\pm0.003$ & $-1.3^{+ 0.4}_{- 0.4}$ & \citetalias{smit_high-precision_2015} \\
        Wide-007082 & GSDZ-2254450533 & -20.66 & 03:32:25.45 & -27:50:53.58 & $6.910\pm0.003$ & $-1.3^{+ 0.4}_{- 0.4}$ & \citetalias{smit_high-precision_2015} \\
        Wide-007437 & COSZ-0301815598 & -22.06 & 10:00:30.18 & +02:15:59.80 & $6.853\pm0.001$ & $-1.2^{+ 0.1}_{- 0.1}$ & \citetalias{smit_high-precision_2015} \\
        Wide-006482 & COSY-0439027359 & -20.40 & 10:00:43.90 & +02:27:35.89 & $7.046\pm0.004$ & $0.7^{+ 0.2}_{- 0.2}$ & \citetalias{roberts-borsani_z_2016} \\
        Wide-000733$^*$ & GNDZ-6372717115 & -20.57 & 12:36:37.27 & +62:17:11.28 & $6.813\pm0.001$ & $-0.9^{+ 0.3}_{- 0.3}$ & \citetalias{smit_high-precision_2015} \\
        Wide-000749$^*$ & GNDZ-6372717115 & -20.57 & 12:36:37.27 & +62:17:11.28 & $6.808\pm0.001$ & $-0.9^{+ 0.3}_{- 0.3}$ & \citetalias{smit_high-precision_2015} \\
        Wide-002815 & GNDY-6379018085 & -21.57 & 12:36:37.90 & +62:18:08.55 & $7.506\pm0.002$ & $0.88^{+ 0.11}_{- 0.11}$ & \citetalias{roberts-borsani_z_2016} \\
        Wide-000715 & GNDZ-6427518385 & -19.89 & 12:36:42.76 & +62:18:38.30 & $6.571\pm0.001$ & $-1.5^{+ 0.4}_{- 0.4}$ & \citetalias{smit_high-precision_2015} \\
        Wide-000770 & GNDY-6487514332 & -20.70 & 12:36:48.74 & +62:14:32.94 & $7.644\pm0.007$ & $0.6^{+ 0.7}_{- 0.6}$ & \citetalias{roberts-borsani_z_2016} \\
        Wide-000705$^{\dagger}$ & GNDZ-7035815571 & -20.66 & 12:37:03.59 & +62:15:56.83 & $6.934\pm0.002$ & $-1.0^{+ 0.3}_{- 0.3}$ & \citetalias{smit_high-precision_2015} \\
        Wide-001408$^{\dagger}$ & GNDZ-7035815571 & -20.66 & 12:37:03.59 & +62:15:56.83 & $6.937\pm0.002$ & $-1.0^{+ 0.3}_{- 0.3}$ & \citetalias{smit_high-precision_2015} \\
        Wide-000767 & GNDY-7048017191 & -20.88 & 12:37:04.81 & +62:17:18.80 & $7.633\pm0.001$ & $1.1^{+ 0.3}_{- 0.2}$ & \citetalias{roberts-borsani_z_2016} \\
        Wide-001576 & GNWY-7379420231 & -20.29$\pm$0.05 & 12:37:37.95 & +62:20:22.78 & $7.113\pm0.001$ & $0.5^{+ 1.5}_{- 0.5}$ & \citetalias{roberts-borsani_z_2016} \\
        Wide-001559$^{\ddagger}$ & GNWZ-7455218088 & -20.58 & 12:37:45.53 & +62:18:08.55 & $7.579\pm0.003$ & $0.7^{+ 0.2}_{- 0.2}$ & \citetalias{roberts-borsani_z_2016} \\
        Wide-001648$^{\ddagger}$ & GNWZ-7455218088 & -20.58 & 12:37:45.53 & +62:18:08.55 & $7.572\pm0.001$ & $0.7^{+ 0.2}_{- 0.2}$ & \citetalias{roberts-borsani_z_2016} \\
        Wide-006214 & EGS-1350184593 & -20.33 & 14:19:13.50 & +52:48:45.90 & $6.740\pm0.001$ & $-1.6^{+ 0.6}_{- 0.6}$ & \citetalias{smit_high-precision_2015} \\
        Wide-004221 & EGS-4434164969 & -20.34 & 14:19:44.33 & +52:56:49.54 & $6.654\pm0.001$ & $-1.3^{+ 0.5}_{- 0.5}$ & \citetalias{smit_high-precision_2015} \\
        Wide-006308 & EGSY-9597563148 & -22.00 & 14:19:59.74 & +52:56:31.40 & $7.549\pm0.001$ & $0.22^{+ 0.06}_{- 0.06}$ & \citetalias{roberts-borsani_z_2016} \\
        Wide-004267 & EGS-zs8-2 & -21.89 & 14:20:12.08 & +53:00:26.89 & $7.478\pm0.003$ & $0.96^{+ 0.17}_{- 0.17}$ & \citetalias{roberts-borsani_z_2016} \\
        Wide-007695 & EGS-2802701763 & -20.09 & 14:20:28.02 & +53:00:17.60 & $5.656\pm0.001$ & $-1.4^{+ 0.5}_{- 0.5}$ & \citetalias{smit_high-precision_2015} \\
        \hline
    \end{tabular}
    \label{tab:general}

        \begin{tablenotes}[]
        \footnotesize
        \item \textit{Notes:} \\ 
        \item $^{*}$,  $^{\dagger}$ and $^{\ddagger}$ denote duplicate observations of the same galaxy.
        \item $^{a}$ Excluded from the analysis sample.
    \end{tablenotes}
    \end{ThreePartTable}

\end{table*}

\subsection{NIRSpec Observations}
\label{sec:survey}
The observations presented in this work were obtained as part of the NIRSpec Wide GTO Program \citep[][]{maseda_nirspec_2024}. The Wide survey covers the CANDELS fields \citep[][]{grogin_candels_2011, koekemoer_candels_2011} and consists of 31 pointings, covering $\approx 320~$ arcmin$^2$ in $105$ hours.  We refer the reader to \citet{maseda_nirspec_2024} for a full description of the Wide survey observing strategy, but summarise below. 

Low-resolution spectra were obtained using PRISM/CLEAR (PRISM hereafter), spanning a wavelength range of $0.6\mu\mathrm{m}-5.3\mu\mathrm{m}$ with a spectral resolution of $R~\approx~100$ (ranging from $30-300$ for a uniformly illuminated shutter; \citealt{jakobsen_near-infrared_2022}). High-resolution spectra were obtained using the G235H and G395H gratings with the associated F170LP and F290LP filters, covering wavelength ranges of $1.66\mu\mathrm{m}-3.05\um$ and $2.78\um-5.14\um$, respectively, at a spectral resolution of $R~\approx~2700$.

For the PRISM observations, a single exposure with 55 groups was taken, using the NRSIRS2RAPID read-out mode \citep{rauscher_improved_2017}. A 3-point nodding pattern was used to cycle through the 3 shutter-slitlet, resulting in an effective exposure time of $2451\mathrm{\,s}$. High resolution gratings were nodded between the two outer shutter positions. For the G235H grating, one exposure with 55 groups was taken (1634s total) and for the G395H grating one exposure with 60 groups was taken (1780s total).

\subsection{Data Reduction}
\label{sec:reduction}
We adopt the same core reduction process as other NIRSpec Multi-Object Spectroscopy (MOS) GTO surveys \citep[e.g.,][]{curtis-lake_spectroscopic_2023, cameron_jades_2023, carniani_jades_2024, bunker_jades_2024, saxena_jades_2024} developed by the ESA NIRSpec Science Operations Team (SOT) and GTO teams, as described in \citet{carniani_jades_2024}. The majority of the reduction follows the official Space Telescope Science Institute (STScI) pipeline \citep{alves_de_oliveira_preparing_2018, ferruit_near-infrared_2022}, with small survey-specific modifications \citep{maseda_nirspec_2024}. These modifications include the use of a finer wavelength grid with regular steps in the 2D rectification process and path loss estimations for each source that account for the relative intra-shutter position and assume a point-like morphology, as in \citet{bunker_jades_2024, curti_jades_2024}. The sole difference between the Wide reduction and other GTO reductions lies in the sigma-clipping algorithm used to exclude outliers when creating the 1D spectrum. 
As some sources appear extended (see Figures \ref{fig:individual slit locs} and \ref{fig:slit locs} of Appendix \ref{sec:duplicate obs}), we apply an aperture correction to the stellar masses and star formation rates derived from SED fitting, detailed in Section \ref{sec:sed fitting}.
Unlike other GTO reduction pipelines, we omit the sigma clipping step when creating the 1D spectrum, as it does not work well with a low number of exposures, and does not account for Poisson noise from bright pixels, both of which are typical of Wide sources. 

\section{Analysis and Results}
\label{sec:analysis}

\subsection{UV Magnitudes and Slopes}
\label{sec:UV mag slopes}
As not all sources are covered by NIRCam imaging, we convert the \emph{HST} F160W apparent magnitudes ($m_{\rm UV}$) to UV absolute magnitudes ($M_{\rm UV}$), with the exception of Wide-001567, where we report the $M_\mathrm{UV}$ from \citet{ormerod_detection_2025, Ormerod_2026}. Additionally, we measure UV absolute magnitudes from the PRISM spectra ($M_{\rm UV, prism}$), in a $100$\AA\ window centred on rest-frame $1500$\AA. 

We measure the UV continuum slopes $(\beta)$ by fitting a power law $F_\lambda \propto \lambda^{\beta}$ to the rest-UV section of the SED, following the methodology commonly adopted in the literature \citep[e.g.,][]{austin_epochs_2024, Saxena_2024_slopes, Topping_2024_slopes}. 
The fitting is performed within the windows defined in \citet{calzetti_dust_1994} which are defined to exclude UV absorption and emission features, such as the C\textsc{iii}]$\lambda\lambda1907,1909$ emission line. For the two sources that exhibit a UV bump feature, Wide-001567 and Wide-006214, we report the values obtained in \citet{ormerod_detection_2025}, which employs slightly modified fitting windows to account for the presence of the UV bump feature.

We adopt a Bayesian fitting procedure to fit the rest-frame UV continuum using a \textsc{python} implementation of the \textsc{multinest} nested sampling algorithm \citep{feroz_multinest_2009}, \textsc{pymultinest} \citep{buchner_x-ray_2014}. We use a Gaussian prior distribution for the power-law index, centred on $\mu_\beta = -2$, with a standard deviation of $\sigma_\beta = 0.5$, and a flat prior on the normalization at $\lambda_{\rm rest} = 1500$\AA\ (between 0 and twice the maximum flux value of the spectrum in the fitting regions). 
The choice of $\mu_B=-2$ is motivated by studies such as \citet{austin_epochs_2024} and \citet{Saxena_2024_slopes}, who find an average UV continuum slope at $z\sim7$ of $\beta\sim-2$.
The best-fit values of $\beta$ are given by the 50th percentile (median) of the posterior distribution, with the 16th and 84th percentiles given as a $\pm 1 \sigma$ confidence range. 

The resulting $M_\mathrm{UV}$ and $\beta$ values are shown in Figure \ref{fig:general_hist}. These IRAC-excess systems are brighter on average than those examined in previous studies ($\langle M_\mathrm{UV} \rangle \sim -20.58$ for the full sample, $\langle M_\mathrm{UV} \rangle \sim -20.66$ for EELGs), compared to $\langle M_\mathrm{UV} \rangle \sim -19.17$ from \citet{Llerena_2026} and $\langle M_\mathrm{UV} \rangle \sim -18.70$ from \citet{boyett_extreme_2024}. As expected from $\beta$-$M_\mathrm{UV}$ relations \citep[][]{bouwens_2014_uv_slopes, Topping_2024_slopes}, these systems also show redder UV continuum slopes than the fainter EELGs from \citet{boyett_extreme_2024}. However, the UV continuum slopes seen in these IRAC-excess systems are similar to those seen in EELGs from \citet{Llerena_2026}.

\subsection{Emission Line Properties}
In this section, we describe our analysis of emission line properties for our sample, including emission line fitting, equivalent width measurements, and line ratio calculations. 

\subsubsection{\textsc{ppxf} Spectral Fitting}
\label{sec:ppxf}

We use the spectral fitting software \textsc{ppxf} \citep{cappellari_improving_2017, cappellari_full_2023} to fit the PRISM spectra as a linear combination of spectral templates, as in \citet{cameron_jades_2023, deugenio_jades_2024}. We use simple stellar population (SSP) spectra with a resolution of $R=10000$ from \textsc{fsps} \citep{conroy_propagation_2009}, calculated using \textsc{mist} isochrones \citep{choi_mesa_2016} and C3K model atmospheres \citep{conroy_resolving_2019}. We consider a subset of the templates grid which logarithmically spans ages of $0.03-20$ Gyr and metallicities of $\mathrm{[Z/H]}=-2.5 ~\mathrm{to} ~0$. 
Prior to running \textsc{ppxf}, input templates are smoothed to twice the spectral resolution of the data, and the templates are truncated to match the approximate rest-frame wavelength of the data.
For each galaxy, the age grid is adjusted such that the oldest available SSP is consistent with the age of the Universe at the given redshift and the flux bluewards of the Lyman break is set to zero. The continuum is modelled as a linear superposition of SSP spectra using non-negative weights. Finally, we use a fifth-order multiplicative Legendre polynomial to account for systematic differences between the SSPs and the data such as dust, flux calibration effects, and the variation between SSPs and high redshift stellar populations \citep[e.g.,][]{cameron_jades_2023}.

Nebular emission lines are modelled using three types of Gaussian templates: single Gaussian templates that represent individual emission lines that are spectrally isolated, single Gaussian templates that represent spectrally blended lines (e.g., H$\alpha + \mathrm{[N\textsc{ii}]}\lambda\lambda 6548, 6583$), and doublet Gaussian templates representing emission line doublets with fixed ratios, such as {\oiii}$\lambda\lambda4959, 5007$.
\textsc{ppxf} is run twice for each galaxy: the first run ties the emission line templates in kinematic subsets which are constrained to have the same redshift and velocity dispersion. There are four kinematic groups: the Balmer lines and stellar templates, rest-frame UV emission lines, rest-frame optical emission lines, and rest-frame near-infrared emission lines. After this first run, all emission lines detected at $3\sigma$ significance are kept, while others are discarded. The second run fixes the kinematics of stellar continuum absorption, and uses only previously detected emission line templates. Additionally, all kinematic groups are removed for the second run leaving the kinematics of each line free, with the exception of the blend group H$\alpha + \mathrm{[N\,\textsc{ii}]}\lambda\lambda 6548, 6583$, the group formed by {\oiii}$\lambda\lambda4959, 5007 + \mathrm{H}\beta$, and the group formed by $\mathrm{H}\gamma + [\mathrm{O\,\textsc{iii}}]\lambda4363$. The kinematics of emission lines within these groups remain tied together.
Additionally, a step function is included to capture strong Balmer jumps \citep[e.g.,][]{cameron_nebular_2024}. 

Due to the spectral resolution of the NIRSpec PRISM, the H\,$\alpha$ emission line is blended with the $\mathrm{[N\,\textsc{ii}]}\lambda\lambda 6548, 6583$ doublet. To estimate the contribution of the [N\,\textsc{ii}] doublet to the blended H$\alpha + \mathrm{[N\,\textsc{ii}]}$ flux, we stack the spectra of all galaxies in which this blend is detected. We fit a triple Gaussian model to the blended lines, tying the line widths, fixing the line wavelengths, and fixing the ratio of the [N\,\textsc{ii}] lines to 3.049 \citep{Storey_2000} to reduce the number of free parameters. From this, we find the [N\,\textsc{ii}] contribution to the total flux to be 12\%, and therefore multiply the measured H$\alpha + \mathrm{[N\,\textsc{ii}]}$ flux of each galaxy by a correction factor of $0.88$ to obtain the H\,$\alpha$ flux.

\begin{figure*}
    \centering
    \includegraphics[width=0.95\textwidth]{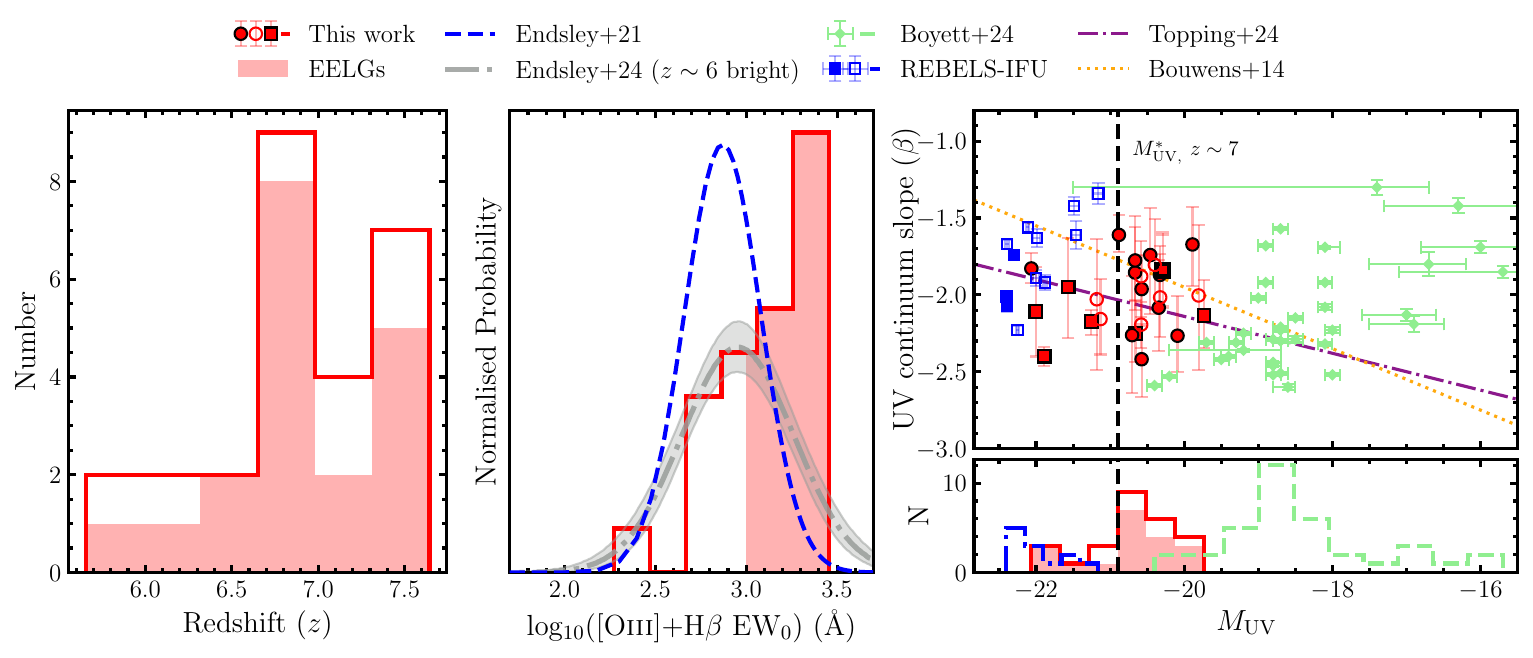}
    \caption{General properties of our sample. \textit{Left:} The redshift distribution of our sample, where the red shaded regions represent EELGs.
     \textit{Centre:} The distribution of rest-frame \oiiihb\ equivalent widths for our IRAC-excess selected sample shown in red, where the red shaded region represents the EELG selection of \oiiihb\ EW$_0 > 1000${\AA}. The blue dashed line shows a comparison to the EW distribution from \citet{endsley_o_2021}. The grey dot-dashed line represents the bright ($\langle M_\mathrm{UV} \rangle = -20.0$) $z\sim6$ sample from \citet{endsley_star-forming_2024}. \textit{Right:} The top panel shows UV continuum slopes ($\beta$) as a function of $M_\mathrm{UV}$, with EELGs shown as red circles, non-EELGs as open circles, and LAEs shown as square markers. The UV bump galaxy from \citet{ormerod_detection_2025} is shown with a thicker black border. We compare to the $3 < z < 9.5$ JADES EELGs \citep[green triangles;][]{boyett_extreme_2024} and $z\sim7$ REBELS-IFU galaxies \citep[blue squares, where solid markers represent EELGs and open markers represent non-EELGs;][]{fisher_rebels-ifu_2025, Rowland_Metallicity_2025}. We show the $z\sim7$ $\beta-M_\mathrm{UV}$ relations from \citet{Topping_2024_slopes} and \citet{bouwens_2014_uv_slopes} with the purple dot-dashed and orange dotted lines, respectively. The bottom panel shows the $M_\mathrm{UV}$ distribution of our sample in red, with red shading representing EELGs. We show a comparison to \citet{boyett_extreme_2024}, shown by the green dashed line, and the REBELS-IFU galaxies in blue. In both panels, the vertical black dashed line shows the characteristic magnitude ($M_\mathrm{UV}^*=-20.89$) of the $z\sim7$ UV luminosity function (UVLF)  from \citet{Harikane_2025}.}
    \label{fig:general_hist}
\end{figure*}

\begin{figure}
    \centering
    \includegraphics[width=1.0\linewidth]{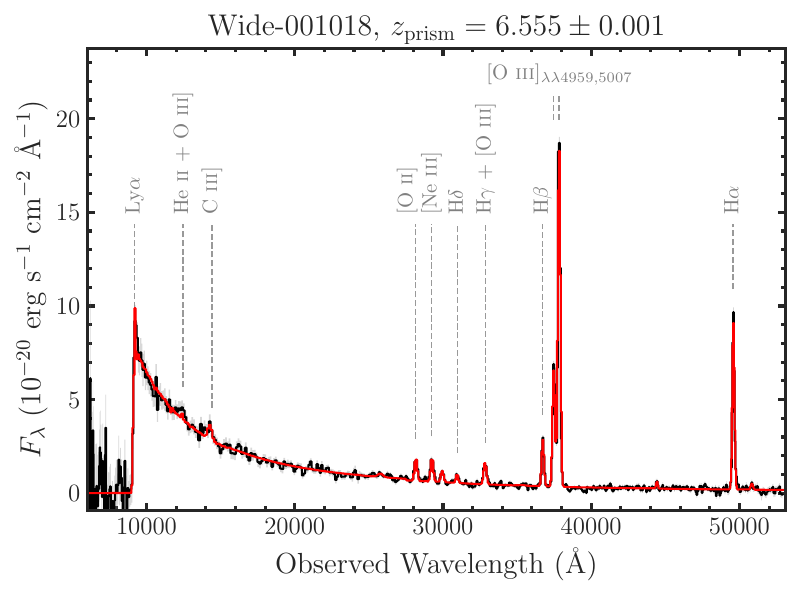}
    \caption{The PRISM spectrum of Wide-001018 is shown as a black line, with $1\sigma$ errors indicated by grey shading. The \textsc{ppxf} best fit is shown by the red line. The locations of key emission lines are indicated by grey dashed lines.}
    \label{fig:ppxf_fit}
\end{figure}

The resulting fits provide the PRISM redshifts ($z_\mathrm{prism}$) and emission line fluxes used in this work, with the exception of Wide-001576, which are presented in \citet{Ormerod_2026}. The PRISM redshifts are presented in Table \ref{tab:general} and Figure \ref{fig:general_hist}. 
An example fit is shown in Figure \ref{fig:ppxf_fit}. This fit is typical of sources within our sample, characterised by an absence of prominent rest-UV emission lines, with C\,\textsc{iii}]$\lambda \lambda1907,1909$ emission detected in only $4/26$ galaxies.

As H$\alpha$ is redshifted out of NIRSpec PRISM spectral coverage for galaxies at $z\gtrsim7$, we estimate the stellar dust attenuation from the value of $A_V$ obtained from \textsc{bagpipes} fitting (Section \ref{sec:sed fitting}). We convert to a nebular $A_V$ following \citet{reddy_mosdef_2020}, and correct for dust attenuation assuming the \citet{cardelli_relationship_1989} attenuation curve for nebular emission. This approach allows us to adopt a consistent dust correction methodology across the sample. 

\subsubsection{[O\,\textsc{iii}]+H$\beta$ Equivalent Widths}
\label{sec: EW measurement}
The IRAC excess colours observed within the EoR are often attributed to extreme \oiiihb\ EWs \citep[e.g.,][]{smit_evidence_2014, roberts-borsani_z_2016}, which are seen in galaxies undergoing intense, recent star formation \citep[][]{boyett_o_2022, boyett_extreme_2024, Endsley_23_faint_uv, Llerena_2026}.
We measure the [O\,\textsc{iii}]+H$\beta$ equivalent widths for our sample by integrating the flux density over the wavelength range covering the [O\,\textsc{iii}]+H$\beta$  emission lines, and normalise to the \textsc{ppxf} continuum fit. EWs are quoted as the rest-frame values throughout.

The equivalent width of the \oiiihb\ emission lines, or the [O\,\textsc{iii}]$\lambda5007$ emission line alone, are commonly used to determine whether a galaxy is an EELG, with varying EW thresholds used in the literature \citep[e.g.,][]{boyett_o_2022,boyett_extreme_2024, Laseter_2025}. We adopt a criterion of $\mathrm{EW}_0([\mathrm{O\,\textsc{iii}}+\mathrm{H}\beta])\geq1000$ {\AA}, similar to the threshold of $\mathrm{EW}_0([\mathrm{O\,\textsc{iii}}])> 750$ {\AA} utilised by \citet{boyett_extreme_2024}. From this, we identify 17/23 galaxies ($\sim74\%$) in our sample as EELGs, hereafter referred to as the EELG sample which is made up of 19 spectra, as two galaxies are observed twice. 
The remaining 6/23  ($\sim26\%$) galaxies are not classified as EELGs and are referred to as the non-EELG sample, made up of 7 spectra, as one non-EELG is observed twice.  
The distribution of EWs is shown in Figure \ref{fig:general_hist}, with individual values presented in Table \ref{tab:emission_lines}. Even for a bright galaxy selection our sample exhibits remarkably extreme {\oiiihb} emission, with a median EW of {\oiiihb}~$=1583$~{\AA} for the full sample, and {\oiiihb}~$=1834$~{\AA} for the EELG sample. This exceeds expectations from \citet{smit_high-precision_2015} who estimate 1375~{\AA} in the average IRAC-excess source (1085~{\AA} after correcting for the the bias due to the noisy IRAC colour selection).

We note that, despite a selection based on extreme IRAC-excesses, 7/26 spectra (six galaxies; Wide-001559 and Wide-001648 are observations of the same galaxy) are classified within our `non-EELG' sample. Of these non-EELGs, 5/6 galaxies show equivalent widths strong enough ($\mathrm{EW}_0([\mathrm{O\,\textsc{iii}}+\mathrm{H}\beta])>700$ {\AA}) to account for an IRAC excess - given the large uncertainty in the IRAC colours. 
Furthermore, due to the large PSF of \textit{Spitzer}, blending of sources can be a significant issue with IRAC colour measurements. Even after deblending the flux, nearby sources on the sky can significantly affect the measured photometry \citep[e.g.][]{labbe_spectral_2013}. Given a bright neighbour $\sim 2$" to the North-East of Wide-008401/GSD-2237749136, we assume significant systematics in the IRAC photometry must be at the origin of the low $\mathrm{EW}_0([\mathrm{O\,\textsc{iii}}+\mathrm{H}\beta])$ in this particular source. 

\subsubsection{Emission Line Ratios}
\label{sec:line ratios}
From the dust corrected emission line fluxes, we calculate the following emission line ratios: 
\begin{equation}
\label{eqn: o32}
\mathrm{O} 32=\log_{10} \left( \frac{[\mathrm{O} \textsc{iii}] \lambda 5007}{[\mathrm{O\,\textsc{ii}}] \lambda \lambda 3727,3729} \right)
\end{equation}

\begin{equation}
\label{eqn: r23}
\mathrm{R} 23= \log_{10} \left(\frac{[\mathrm{O\,\textsc{ii}}] \lambda \lambda 3727,3729+[\mathrm{O\,\textsc{iii}}] \lambda \lambda 4959,5007}{\mathrm{H} \beta}\right)
\end{equation}

\begin{equation}
\label{eqn: r3}
\mathrm{R}3 =\log_{10} \left(\frac{[\mathrm{O\,\textsc{iii}}] \lambda 5007}{\mathrm{H} \beta} \right)
\end{equation}

\begin{equation}
    \label{eqn: s2}
    \mathrm{S}2 = \log_{10} \left( \frac{[\mathrm{S}\,\textsc{ii}]\lambda\lambda6718,32}{\mathrm{H}\alpha}\right)
\end{equation}

\begin{figure}
    \centering
    \includegraphics[width=0.95\columnwidth]{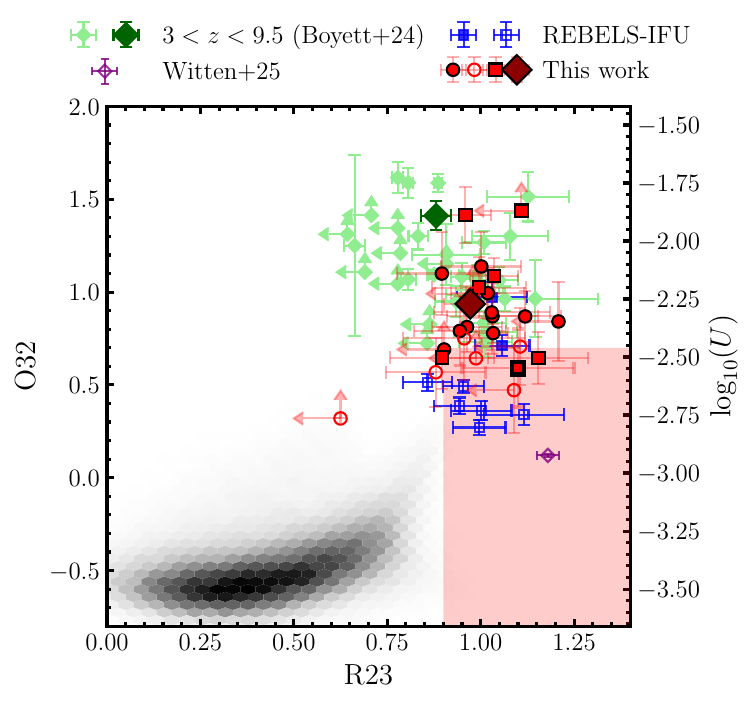}
    \caption{Dust corrected O32-R23 diagram, showing the Wide IRAC-excess sample in red. EELGs are shown by solid symbols, and non-EELGs by open symbols. LAEs are shown as square markers, while non-LAEs are shown as circular markers. Wide-001576, the UV bump source from \citet{ormerod_detection_2025}, is shown with a thicker black border.
    The dark red diamond represents the stack of the EELG sample.
    We compare to NIRSpec observations \citep[][]{boyett_extreme_2024,witten_rising_2025} and the Sloan Digital Sky Survey \citep[SDSS;][]{York_2000} Data Release 7 \citep[][]{SDSS_DR7}, shown in black. The light green diamonds represent individual galaxies from \citet{boyett_extreme_2024}, while the large dark green diamond represents their high-z EELG stack. Solid blue squares show EELGs from the REBELS-IFU survey, while non-EELGs are shown as open blue squares. We only show the REBELS-IFU galaxies where H$\alpha$ lies within the NIRSpec coverage, allowing a dust correction to be applied to emission line fluxes \citep{Rowland_Metallicity_2025}. 
    The red shading shows the region $\mathrm{O32}< 0.7$ and $\mathrm{R23}>0.9$, which may contain galaxies hosting an older stellar population \citep{witten_rising_2025}. 
    The right hand $y$-axis shows the corresponding ionization parameter, $\log_{10}(U)$, derived from the  $\log_{10}(U)-\mathrm{O32}$ relation in \citet{Witstok_2021_reionization}.
     }
    \label{fig:o32_r23}
\end{figure}

\begin{figure}
    \centering
    \includegraphics[width=0.95\columnwidth]{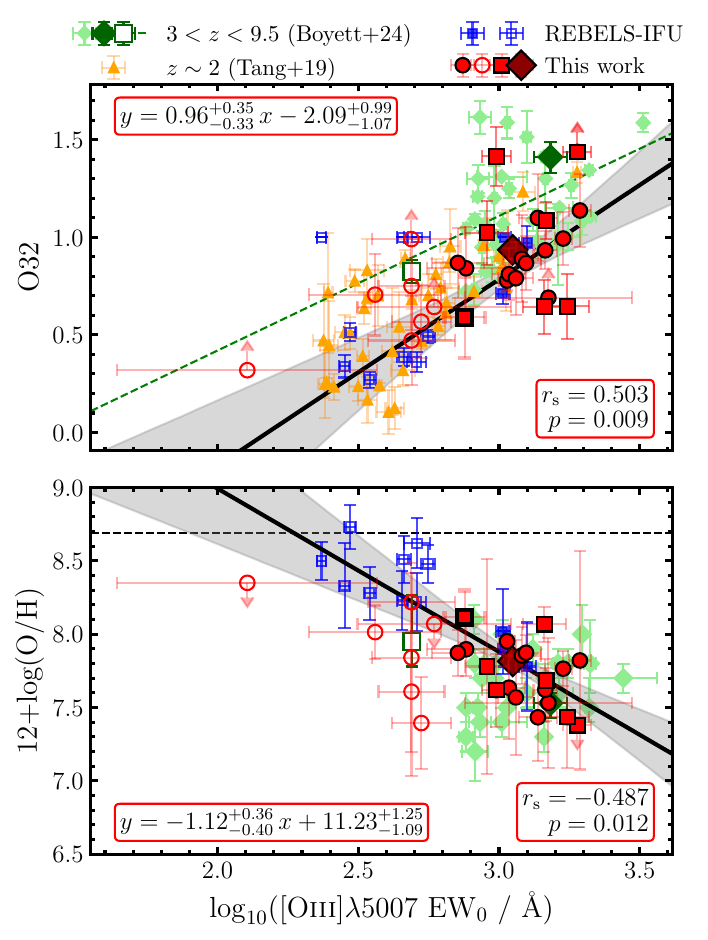}
    \caption{\textit{Top:} O32 ratio as a function of {\oiii}$\lambda5007$ EW. \textit{Bottom:} Oxygen abundance as a function of {\oiii}$\lambda5007$ EW.
    We show EELGs as solid red markers, non-EELGs as open markers, and the UV bump source from \citet{ormerod_detection_2025} with a thicker black border. LAEs are shown as square markers, with non-LAEs shown as circular markers. The dark red diamond represents the stack of the EELG sample.
    The JADES EELG sample from \citet{boyett_extreme_2024} is shown as light green diamonds, with the high-z EELG stack shown a large dark green diamond, and the high-z non-EELG stack shown as a large dark green open square.  We show the $z=1.3-2.4$ extreme \oiii\ emitters from \citet{tang_mmtmmirs_2019} as orange triangles. 
    Markers for the REBELS-IFU sample are the same as in previous figures, with EELGs shown as solid blue squares and non-EELGS shown as open blue squares. O32 upper limits from REBELS-IFU are not corrected for dust attenuation \citep{Rowland_Metallicity_2025}.
    The solid black line represents the best fit relations, fitting to both the EELGs and non-EELGs. The grey shading represents the 16th-84th percentiles. The Spearman correlation coefficients $(r_s)$ and $p$-values are shown in the bottom right corners, with the best fit relations indicated in the top and bottom left corners.
    The green dashed line in the top panel shows the relation between O32 and EW from \citet{boyett_extreme_2024}. 
    The black dashed line in the bottom panel indicates the solar metallicity of $12+\log \rm (O/H) = 8.69$ \citep[][]{asplund_chemical_2021}.}
    \label{fig:o32_ew}
\end{figure}

The O32 and R23 ratios, which trace ionization state and metallicity respectively, are shown in Figure \ref{fig:o32_r23}, while the relationship between the O32 line ratio and [O\,\textsc{iii}] EW is shown in Figure \ref{fig:o32_ew}. Additionally, we employ the R3 and S2 ratios as AGN diagnostics, shown in Figure \ref{fig:MEx} and Figure \ref{fig:s2-vo87} of the Appendix. We also use the O32 ratio to determine the ionization parameter ($\log_{10}U$) following the relation in \citet{Witstok_2021_reionization}.
Emission line ratios and ionization parameters are presented in Table \ref{tab:emission_lines}. 

Figures \ref{fig:o32_r23} and \ref{fig:o32_ew} show that the Wide IRAC-excess sample is systematically lower in O32 - even at fixed $\mathrm{EW}_0([\mathrm{O\,\textsc{iii}}+\mathrm{H}\beta])$ - than faint EELGs from JADES \citep[][]{boyett_extreme_2024}, but consistent with brighter EELGs from REBELS-IFU \citep{Rowland_Metallicity_2025} and lower redshift samples \citep{tang_mmtmmirs_2019}, which could suggest these are more evolved systems than their UV-faint counterparts. 
We use the \textsc{python} implementation of \textsc{linmix} \citep{Kelly2007_linmix} to perform a linear fit between O32 and [O\,\textsc{iii}]$\lambda5007$ EW\footnote{https://github.com/jmeyers314/linmix}. \textsc{linmix} accounts for uncertainties in both the $x$ and $y$ variables, and also for the inclusion of limits within the data. We fit to both the EELG and non-EELG sample.
To determine the significance of the relations examined in this work, we perform a Spearman rank correlation test \citep[e.g.,][]{Castellano_2023, llerena_physical_2024}. We consider a relation to be statistically significant if the correlation coefficient indicates at least a moderate correlation ($|r_\mathrm{s}|\geq 0.4$) and satisfies $p\leq0.05$ \citep[e.g.,][]{Komarova_REBELS_xi_ion}.
We find a strong correlation between O32 and [O\,\textsc{iii}]$\lambda5007$ EW, with the best fit relation given by: 
\begin{equation}
    \mathrm{O}32 = 0.96^{+0.35}_{-0.33}\times \log_{10}[\mathrm{EW_0 ([O\,\textsc{iii}]}\lambda5007/\text{\AA})] - 2.09^{+0.99}_{-1.07}.
    \label{eqn: O32_EW}
\end{equation}

\begin{table*}
    \caption{Emission line properties of our sample. Columns are: Wide ID, dust corrected O32 ratio, dust corrected R23 ratio, dust corrected R3 ratio, rest-frame \oiiihb\ EW, ionizing photon production efficiency, oxygen abundance, and ionization parameter. }
    \label{tab:emission_lines}
    \begin{tabular}{lccccccc}
    \hline
    Wide ID & O32 & R23  & R3 & [O\,\textsc{iii}]+H$\beta$ EW$_0$ ({\AA}) & $\log_{10}(\xi_\mathrm{ion}/\rm erg^{-1}~Hz)$ & $12+\log(\mathrm{O/H})$ & $\log_{10}U$ \\
    \hline
    Wide-000003 & $>0.64$ & $<0.99$  &$>0.79$& $865\pm542$ & $<25.88$ & $<8.07$ & $>-2.51$\\
    Wide-000001 & $0.84\pm0.21$ & $<1.21$  &$>1.04$& $1124\pm358$ & $<25.38$ & $7.90 \pm 0.18$ & $-2.35\pm0.17$\\
    Wide-000036 & $0.57\pm0.19$ & $0.88\pm0.13$  &$0.67\pm0.15$& $779\pm189$ & $25.51\pm0.08$  & $7.39 \pm 0.31$ & $-2.57\pm0.15$ \\
    Wide-001018 & $1.09\pm0.10$ & $1.04\pm0.06$  &$0.88\pm0.07$& $2163\pm96$ & $25.67\pm0.04$ & $7.68 \pm 0.08$ & $-2.15\pm0.08$ \\
    Wide-002631 & $1.02\pm0.16$ & $1.0\pm0.13$  &$0.84\pm0.13$& $1332\pm234$ & $25.47\pm0.06$ & $7.78 \pm 0.73$ & $-2.20\pm0.13$\\
    Wide-000440 & $0.47\pm0.23$ & $<1.09$  &$>0.86$& $719\pm255$ & $26.02\pm0.27$ & $8.22 \pm 0.20$ & $-2.64\pm0.18$ \\
    Wide-008401 & $>0.32$ & $<0.62$  &$>0.37$& $188\pm200$ & $<24.90$ & $<8.35$ & $>-2.76$ \\
    Wide-007082 & $>1.44$ & $<1.11$  &$0.97\pm0.11$& $2792\pm320$ & $25.62\pm0.07$ & $<7.38$ & $>-1.87$\\
    Wide-007437 & $0.78\pm0.05$ & $1.03\pm0.04$  &$0.86\pm0.04$& $1571\pm105$ & $25.67\pm0.03$& $7.95 \pm 0.04$ & $-2.40\pm0.04$\\
    Wide-006482 & $0.75\pm0.23$ & $0.96\pm0.16$  &$0.78\pm0.17$& $719\pm194$ & $25.40\pm0.14$ & $7.61 \pm 0.57$ & $-2.42\pm0.18$ \\
    Wide-000733 & $0.81\pm0.13$ & $0.96\pm0.09$  &$0.79\pm0.10$& $1594\pm215$ & $25.52\pm0.06$ & $7.64 \pm 0.35$ & $-2.37\pm0.11$\\
    Wide-000749 & $0.87\pm0.18$ & $1.03\pm0.12$  &$0.86\pm0.13$& $1050\pm224$ & $25.36\pm0.08$ & $7.87 \pm 0.16$ & $-2.32\pm0.14$\\
    Wide-002815 & $0.65\pm0.17$ & $0.9\pm0.14$  &$0.70\pm0.15$& $2569\pm453$ & $26.20\pm0.14$ & $7.44 \pm 0.35$ & $-2.50\pm0.13$\\
    Wide-000715 & $1.14\pm0.18$ & $1.0\pm0.11$  &$0.85\pm0.12$& $2850\pm257$ & $25.96\pm0.12$ & $7.82 \pm 0.74$ & $-2.11\pm0.15$\\
    Wide-000770 & $>0.69$ & $<0.9$  &$0.71\pm0.22$& $2203\pm1502$ & $25.44\pm0.22$ & $7.53 \pm 0.44$ & $>-2.47$\\
    Wide-000705 & $0.79\pm0.19$ & $0.94\pm0.12$  &$0.77\pm0.13$& $1687\pm242$ & $26.35\pm0.09$ & $7.57 \pm 0.39$ & $-2.39\pm0.15$\\
    Wide-001408 & $0.89\pm0.22$ & $1.03\pm0.16$  &$0.86\pm0.17$& $1772\pm272$ & $25.78\pm0.11$& $7.85 \pm 0.19$ & $-2.31\pm0.18$\\
    Wide-000767 & $0.99\pm0.14$ & $1.02\pm0.10$  &$0.86\pm0.11$& $2485\pm170$ & $26.26\pm0.08$ & $7.76 \pm 0.12$ & $-2.22\pm0.11$\\
    Wide-001576 & $0.59\pm0.21$& $1.10\pm0.18$&$0.90\pm0.16$& $1110\pm183$& $25.42\pm0.17$ & $8.12 \pm 0.18$ & $-2.55\pm0.17$\\
    Wide-001559 & $0.71\pm0.21$ & $<1.11$  &$>0.92$& $534\pm289$ & $<25.03$ & $8.01 \pm 0.18$ & $-2.45\pm0.17$\\
    Wide-001648 & $>0.99$ & $<0.98$  &$0.82\pm0.14$& $718\pm230$ & $25.21\pm0.13$ & $7.84 \pm 0.65$ & $>-2.23$\\
    Wide-006214 & $0.93\pm0.1$ & $0.96\pm0.06$  &$0.80\pm0.06$& $2146\pm207$ & $25.54\pm0.03$ & $7.62 \pm 0.21$  & $-2.27\pm0.08$\\
    Wide-004221 & $1.10\pm0.22$ & $0.9\pm0.12$  &$0.74\pm0.13$& $2020\pm312$ & $25.90\pm0.10$ & $7.43 \pm 0.30$  & $-2.14\pm0.18$\\
    Wide-006308 & $0.64\pm0.14$ & $1.15\pm0.13$  &$0.96\pm0.14$& $2130\pm374$ & $25.83\pm0.14$ & $8.07 \pm 0.12$ & $-2.50\pm0.11$\\
    Wide-004267 & $1.42\pm0.15$ & $0.96\pm0.07$  &$0.82\pm0.08$& $1439\pm168$ & $25.52\pm0.06$ & $7.62 \pm 0.25$ & $-1.89\pm0.12$\\
    Wide-007695 & $0.87\pm0.14$ & $1.12\pm0.1$  &$0.95\pm0.10$& $1834\pm230$ & $25.46\pm0.07$ & $7.87 \pm 0.12$ & $-2.33\pm0.11$\\
    \hline
    \end{tabular}
\end{table*}

\subsubsection{Stacking analysis}
\label{sec: stacks}
To create a stack, we first shift all spectra to their rest-frame wavelengths using $z_\mathrm{prism}$ and then interpolate to a common wavelength grid with 10{\AA} intervals \citep[e.g.,][]{roberts_borsani_between_the_extremes_2024, boyett_extreme_2024}. The individual spectra are normalised according to their UV continuum slopes by dividing the observed flux by the best-fitting power-law continuum model to give $F_\lambda/F_\mathrm{cont}$. The final stack is obtained through the median of fluxes, with the errors determined through bootstrapping. We do not apply weighting to the stack in order to prevent biasing the resulting spectrum towards brighter objects \citep[e.g.,][]{roberts_borsani_between_the_extremes_2024}.

\begin{figure}
    \centering
    \includegraphics[width=0.9\columnwidth]{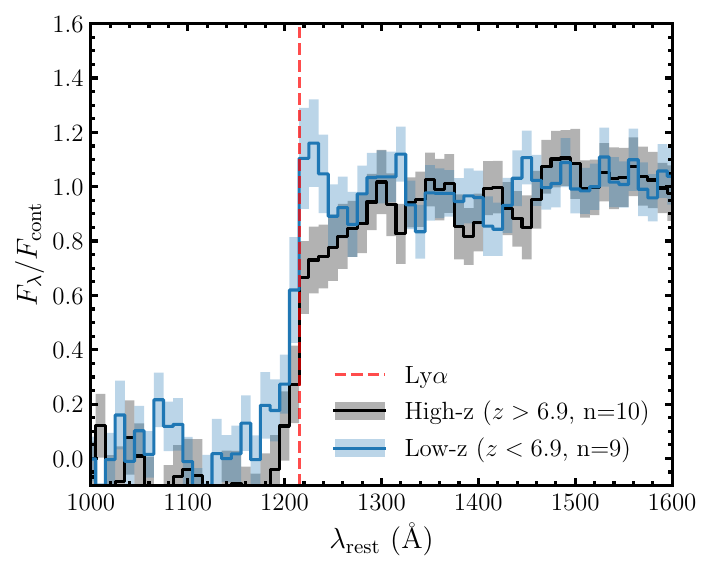}
    \caption{Zoom in on the region $\lambda_\mathrm{rest}=1000-1600$~{\AA} for the high-z stack ($z\geq6.91$; black line) and low-z stack ($z<6.91$; blue line) of EELGs. The dashed red line shows the systemic Ly$\alpha$ wavelength. Prior to stacking, galaxies are normalised according to their UV continuum slopes.}
    \label{fig:redshift_stack}
\end{figure}

We create stacked galaxy spectra by stacking the PRISM spectra of the EELG sample based on $z_\mathrm{prism}$. We create a `low-z' and `high-z' stack, separated based on the median EELG redshift of $z=6.91$, shown in Figure \ref{fig:redshift_stack}. We additionally create a stacked spectrum of the full sample, including both the EELG and non-EELG subsamples, shown in Figure \ref{fig:full stack}, alongisde a stack of the EELG sample. The stacked spectrum of the full sample does not show any evidence for strong rest-UV or high-ionization emission lines. The EELG stack shows a $\sim4\sigma$ detection of the blended C\,\textsc{iii}]$\lambda\lambda1907,1909$ emission line.

\subsubsection{Lyman-$\alpha$ emitters}
\label{sec:LAEs}
Ly$\alpha$ is one of the most useful emission lines for studying the EoR, with Ly$\alpha$ emitters (LAEs) often used to trace the reionization history of the Universe due to the sensitivity of Ly$\alpha$ as a tracer of the ionization state of gas surrounding galaxies \citep[e.g.,][]{Dijkstra_2014,Ouchi_2020, Jones_2024, Jones_reionization_2025, Witstok_2025_Lya, Witstok_2025_primeval_LAE}. 

To characterise Ly$\alpha$ emission within our sample, we carry out Bayesian emission line fitting similar to that in \citet{ormerod_detection_2025}, accounting for the NIRSpec line spread function (LSF). Following \citet{Witstok_2025_primeval_LAE}, we model the Ly$\alpha$ emission line as a skewed Gaussian profile, and determine the EW by dividing the integrated line flux by the median continuum value at $1216.5<\lambda_{\text {emit }}<1225$~{\AA}. In order to classify a galaxy as an LAE, we require a $\geq3\sigma$ detection of Ly$\alpha$ from the NIRSpec PRISM spectrum. Due to the low spectral resolution of the PRISM, we do not aim to characterise the Ly$\alpha$ profile such as its velocity offset ($\Delta v_\mathrm{Ly\alpha}$) and report only the rest-frame EW in the first section of Table \ref{tab:Lya}. 

Using the PRISM spectra, we only identify three galaxies as LAEs. However, for galaxies within the EoR (at $z\sim6-7$), the resolution of the PRISM is near its minimum value. As a result, low-EW Ly$\alpha$ emission will appear similar to a continuum with a blueshifted Ly$\alpha$ break \citep{Jones_2024}. Therefore, we carry out a literature search for galaxies within our sample that may have ground-based or higher resolution \emph{JWST} spectroscopy yielding a Ly$\alpha$ detection. This results in a further four galaxies being identified as LAEs, listed in the second section of Table \ref{tab:Lya}. We additionally determine the Ly$\alpha$ escape fraction ($f_\mathrm{esc}^\mathrm{Ly\alpha}$), which is defined as the ratio between the observed and intrinsic Ly$\alpha$ flux. Following the methodology commonly adopted in the literature \citep[e.g.,][]{Tang_2023, Begley_24_Lya}, we determine the intrinsic Ly$\alpha$ flux from dust-corrected H$\alpha$ fluxes assuming Case B recombination, which gives
\begin{equation}
\label{eqn: fesc lya}
    f_{\mathrm{esc}}^{\mathrm{Ly} \alpha}=\frac{\mathrm{F}_{\mathrm{Ly} \alpha, \mathrm{obs}}}{8.7 \times \mathrm{F}_{\mathrm{H} \alpha, \mathrm{int}}}.
\end{equation}
The derived values range from $f_{\mathrm{esc}}^{\mathrm{Ly} \alpha}= 0.024^{+0.006}_{-0.005}$ to $f_{\mathrm{esc}}^{\mathrm{Ly} \alpha}=0.114^{+0.042}_{-0.041}$, and are presented in Table \ref{tab:Lya}. 
We also include values derived for the high-z and low-z stacks in the third section of Table \ref{tab:Lya}.

\renewcommand{\arraystretch}{1.2}
\begin{table*}
    \centering
    \caption{Ly$\alpha$ properties for the LAEs within our sample. The first section of the table represents LAEs detected from NIRSpec PRISM observations, while the second section shows LAEs identified in the literature, and the third section shows the Ly$\alpha$ properties of the low- and high-z EELG stacks. Columns indicate: the Wide ID of the galaxy, the redshift of the Ly$\alpha$ emission line, the rest-frame Ly$\alpha$ EW measured from NIRSpec (this work) or in the literature, the Ly$\alpha$ velocity offset, the Ly$\alpha$ escape fraction ($f_\mathrm{esc}^\mathrm{Ly\alpha}$), size of the ionized bubble ($R_\mathrm{ion}$), and references. 
    We note that we do not present $z_\mathrm{Ly\alpha}$ or $\Delta v_\mathrm{Ly\alpha}$ where the Ly$\alpha$ line is detected in the NIRSpec PRISM due to the low resolution of the PRISM. Where shown, velocity offsets are quoted from the literature.}
    \begin{ThreePartTable}
    \begin{tabular}{lccccccc} \hline
        Wide ID & $z_\mathrm{Ly\alpha}$ & Ly$\alpha$ EW$_0$ (This work) {\AA} & Ly$\alpha$ EW$_0$ (Literature) {\AA}& $\Delta v_\mathrm{Ly\alpha}~ \rm(km~s^{-1})$ & $f_\mathrm{esc}^\mathrm{Ly\alpha}$& $R_\mathrm{ion}$ (pMpc) & Reference \\ \hline
        Wide-001018 & -- & $17.8^{+4.0}_{-3.2}$ & -- & -- &  $0.024^{+0.006}_{-0.005}$& $<0.301$& -- \\
        Wide-002631 & -- & $46.6^{+13.0}_{-11.8}$ & -- & -- &  $0.114^{+0.042}_{-0.041}$& $<0.519$& -- \\
        Wide-007082 & -- & $39.1^{+9.3}_{-7.9}$ & -- & -- & $0.049^{+0.014}_{-0.014}$& $<0.375$& -- \\ \hline 
        Wide-004267 & $7.485$& -- & $9.5\pm3.1$ & $545\pm184$&  $0.045\pm0.015$& $<0.001$ &S17, T23 \\ 
        Wide-001576 & $7.10813$ & -- & $16\pm8$ & $99\pm32$& $0.030\pm0.016$& $0.211^{+0.058}_{-0.045}$&RB23 \\
        Wide-006308 & $7.5501$ & -- & $11\pm9$ & $\sim154$& <0.055& $<0.232$&C24, T23\\
        Wide-002815 & $7.5078$ & -- & $7.5\pm1.5$ &  -- & -- & -- &F13, S17 \\ \hline
        High-z stack& 7.380${^\ast}$ &$<5.9$  & -- & -- & $<0.003$ & $<0.198$ & -- \\
        Low-z stack & 6.654${^\ast}$ & $16.1^{+4.3}_{-4.3}$ & -- & -- & $0.011^{+0.003}_{-0.003}$ & $<0.248$ & -- \\

        \hline
    \end{tabular}
    \label{tab:Lya}

        \begin{tablenotes}[]
        \footnotesize
        \item References: T23: \citet{Tang_2023}, 
        RB23: \citet{Roberts-Borsani_2023_lya},
        C24: \citet{Cooper_2024_Lya},
        F13: \citet{Finkelstein_2013_Lya},
        S17: \citet{Stark_2017} \\
        ${^\ast}$ Median PRISM redshift ($z_\mathrm{prism}$) of galaxies in the stack.
    \end{tablenotes}
    \end{ThreePartTable}

\end{table*}

\subsubsection{Rest-frame UV Emission Lines}
\label{sec:rest uv}
Through emission line fitting carried out in Section \ref{sec:ppxf}, we identify just four galaxies with C\,\textsc{iii}] emission, with EWs ranging from $\rm EW_0(C\,\textsc{iii}])= 11.7\pm 7.4-37.1 \pm 25.1{\AA}$. In a stack of the EELG sample, we detect C\,\textsc{iii}] at $\sim4\sigma$, with $\rm EW_0(C\,\textsc{iii}]) = 15.5\pm8.2$~{\AA}. We do not detect any C\,\textsc{iv} $\lambda 1550$ emission in individual sources, nor in the EELG stack (see Figure \ref{fig:full stack}).

\begin{figure}
    \centering
    \includegraphics[width=1.0\columnwidth]{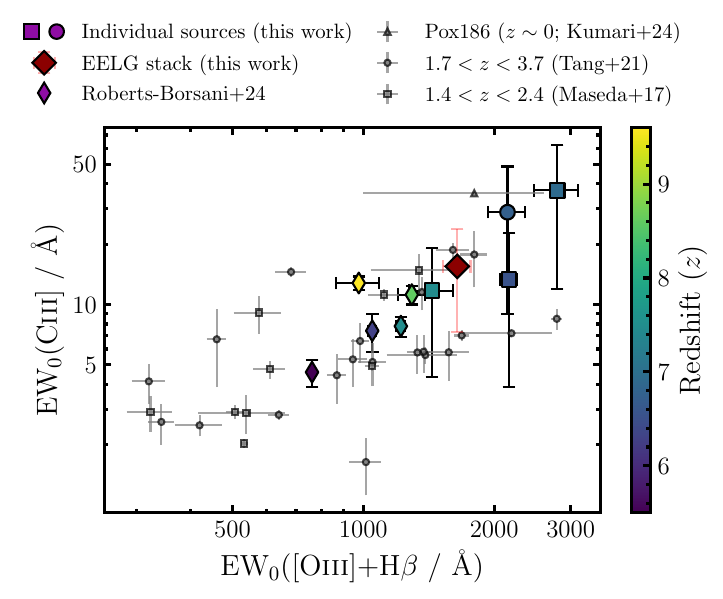}
    \caption{C\,\textsc{iii}] EW as a function of {\oiiihb} EW for the four sources in which C\,\textsc{iii}] is detected in the PRISM spectrum. LAEs are shown as square markers, and all four galaxies are EELGs. We additionally show measurements from stacked spectra in \citet{roberts_borsani_between_the_extremes_2024} as diamond markers. For both individual sources from this work, and high-z stacks from \citet{roberts_borsani_between_the_extremes_2024}, we colour the markers by their redshift. The dark red diamond corresponds to the stack of our EELG sample. For comparison, we show lower redshift sources as grey markers \citep[][]{Kumari_2024_pox186, maseda_2017, tang_2021}.}
    \label{fig:CIII_OIIIHb_EW}
\end{figure}

We show C\,\textsc{iii}] EW as a function of \oiiihb\ EW in Figure \ref{fig:CIII_OIIIHb_EW}, for these individual sources and the EELG stack. The C\,\textsc{iii}] EWs obtained for our sample are large compared to typical values at $z\sim6-7$ \citep[e.g.,][]{roberts_borsani_between_the_extremes_2024}. However, they are consistent with the trend of increasing $\rm EW_0(C\,\textsc{iii}])$ with EW$_0$({\oiiihb}) \citep[e.g.][]{tang_2021, maseda_2017}.

\subsection{AGN Diagnostics}
\label{sec: agn diagnostic}
Emission line diagnostic diagrams, such as the BPT \citep{BPT_1981} and VO87 \citep{VO87} diagrams, are commonly used to distinguish between star forming (SF) galaxies and AGN. In order to exclude potential AGN contamination from our EELG sample, we employ two diagnostic diagrams, the S2-VO87 diagram, and the Mass-Excitation (MEx) diagram.

We first employ the MEx diagram (Figure \ref{fig:MEx}), which utilises the R3 emission line ratio (Equation \ref{eqn: r3}) and galaxy stellar mass to differentiate between star-forming galaxies and AGN \citep[][]{juneau_2011, coil_2015}. Although alternative emission line ratio diagnostic diagrams are often preferred to the MEx diagram due to possible biases in the stellar masses inferred from SED fitting \citep[][]{coil_2015, Cleri_2023a}, we choose to include it as, given the detected emission lines available in our sample, it provides the broadest coverage of our sample. We find that our galaxies lie predominantly within the star-forming region of the MEx diagram, suggesting that the strong EWs are driven by recent star formation rather than AGN. This is consistent with \citet{Llerena_2026}, who find that massive stars are the dominant ionizing source in EELGs. 

We additionally incorporate the S2-VO87 diagram (Appendix \ref{sec:s2-vo87}) which uses the R3 and S2 (Equation \ref{eqn: s2}) emission line ratios. However, the [S\,\textsc{ii}] emission line falls outside the NIRSpec PRISM wavelength coverage for a significant fraction of our sample, and therefore cannot be applied uniformly across our sample in the same way as the MEx diagram. 

In Figure \ref{fig:MEx} one galaxy, Wide-000001, lies within the AGN region. However, this source exhibits no detectable rest-UV emission lines, and [S\,\textsc{ii}] falls outside of NIRSpec wavelength coverage due to its redshift. 
We find five galaxies that show tentative evidence for a blended He$\,\textsc{ii}~\lambda1640$+O\,\textsc{iii}]$\lambda1666$ emission line at $\sim3\sigma$, which may be evidence of a harder UV spectrum, potentially powered by an AGN. However, due to the resolution of the NIRSpec PRISM, it is not possible to conclusively determine whether these emission features are caused by He$\,\textsc{ii}$, or O\,\textsc{iii}]$\lambda1666$. The five galaxies with tentative He$\,\textsc{ii}~\lambda1640$ detections -- Wide-002631, Wide-006482, Wide-007082, Wide-000705, Wide-004267 -- lie within the SF region of the MEx diagram (Figure \ref{fig:MEx}), and [S\,\textsc{ii}] is either not detected or redshifted out of NIRSpec coverage. Therefore, we are unable to conclusively say whether these sources harbour AGN. Furthermore, the stacked spectrum of the full sample (Figure \ref{fig:full stack}) does not show evidence for high ionization rest-UV emission lines, suggesting that on average the ionizing photon production in our sample is dominated by young, massive stars.

In the $R\sim2700$ spectra, the [O\,\textsc{iii}]$\lambda 4363$ and H\,$\gamma$ emission lines are deblended. Through \textsc{ppxf} fitting and visual inspection, we identify one galaxy (Wide-001018) in which both emission lines are robustly detected. Recent work has suggested the use of the O3H$\gamma$ line ratio, where $\mathrm{O3H}\gamma = \log_{10}([\mathrm{O}\,\textsc{iii}]4363/\mathrm{H}\,\gamma)$, in combination with the O32 ratio as a method of identifying galaxies which host an AGN \citep[][]{Mazzolari_AGN_2024}. Although Wide-001018 falls within the AGN only region of the diagram (Figure \ref{fig:o3hg}), its position remains consistent with the SF or AGN region within $1\sigma$. Moreover, its positions on the MEx and S2-VO87 diagrams (Appendix Figures \ref{fig:MEx} and \ref{fig:s2-vo87}) fall within the SF regions, favouring star formation as the origin of extreme line emission. Furthermore, we stack the five galaxies with a robust H$\gamma$ detection in the G395H spectra, and find $\mathrm{O3H}\gamma < -1.55$, consistent with SF as the origin of the extreme equivalent widths. Finally, we do not find any evidence for broad line components in the $R\sim2700$ spectra.

\subsection{Gas-phase Metallicity}
\label{sec:gas phase met}
The preferred method of deriving the gas-phase metallicity is the direct-$T_e$ method through the detection of auroral emission lines such as [O\,\textsc{iii}]$\lambda4363$ \citep[e.g.,][]{Sanders_2024_direct_Te, Laseter_2024}. However, due to the spectral resolution of the NIRSpec PRISM, this is blended with the nearby H$\gamma$ emission line, with alternative auroral lines undetected in our sample. 
[O\,\textsc{iii}]$\lambda4363$ is detected in the high-resolution G395H spectra for just three galaxies within our sample, of which just one is truly robust after visual inspection.
Therefore, following the approach widely used in the literature \citep[e.g.,][]{Rowland_Metallicity_2025}, we utilise strong line calibrations to estimate oxygen abundances ($12+\log(\mathrm{O/H})$) with a consistent methodology across our sample. Many strong line calibrations derived in the local Universe are biased due to the evolution of ionization conditions with redshift, with the ISM conditions moving towards a more extreme state with a harder ionizing spectrum at higher redshift at a fixed metallicity \citep{Sanders_2024_direct_Te, Scholte_2025_EXCELS}. We therefore adopt the strong line calibrations based on auroral line detections at  $1.4 \leq z \leq 8.7$ from \citet{Sanders_2024_direct_Te}. 
 
The R23 ratio is widely used in metallicity determinations, as it is an indication of total excitation due to the inclusion of both singly and doubly ionized oxygen emission lines \citep[][]{cameron_jades_2023, Scholte_2025_EXCELS}. Moreover, the R23 ratio generally shows the least scatter and is sensitive to small changes in metallicity, whereas the O32 ratio is highly dependent on the ionization state \citep[e.g.,][]{nakajima_jwst_2023}.
Despite the O32 ratio primarily tracing the ionization parameter, with only a secondary dependence on metallicity due to the ionization parameter-metallicity relation \citep[e.g.,][]{cameron_jades_2023, Laseter_2024,Scholte_2025_EXCELS}, it can be used to break the degeneracy of double-valued R23 calibrations \citep[e.g.,][]{curti_jades_2024, Rowland_Metallicity_2025}. 

We preferentially employ the R23 ratio where available. In cases where there is no solution for the R23 ratio, or the R23 ratio is an upper limit, we then use the R3 ratio. Similarly, where the R3 ratio yields no solution or is an upper limit, we use the O32 ratio. As the R23 and R2 diagnostics are double-branched, resulting in a low- and high-metallicity solution, we use O32 as a degeneracy breaker to select the appropriate value. We note that there are two galaxies, Wide-000036 and Wide-002815, where the O32 ratio does not conclusively break the degeneracy between the low- and high-metallicity branches. In these cases, we adopt the low-metallicity solution, as this is consistent with the properties of the wider sample and sub-solar oxygen abundances reported for EELGs in the literature \citep[][]{boyett_extreme_2024, Llerena_2026}. For Wide-001576, we adopt the emission line fluxes and ratios obtained in \citet{Ormerod_2026}. The derived oxygen abundances are reported in Table \ref{tab:emission_lines}. 

For our EELG sample, we find an average $12+\log(\rm O/H) = 7.76^{+0.14}_{-0.24}$ ($Z_\mathrm{neb} = 0.12^{+0.04}_{-0.05}Z_\odot$), consistent with the both the high-z EELG stack from \citet{boyett_extreme_2024} and REBELS-IFU EELGs \citep{Rowland_Metallicity_2025}, despite the latter probing more massive, UV-luminous galaxies. Our non-EELG sample exhibits a slightly higher average oxygen abundance ($12+\log(\rm O/H) = 8.01^{+0.21}_{-0.41}$), albeit with increased scatter. Overall, this shows that EELGs are characterised by low gas-phase metallicities, consistent with young stellar populations. We show our sample overlaid on the mass-metallicity relation (MZR) in Figure \ref{fig:mass metallicity}. We additionally show gas-phase metallicity as a function of [O\,\textsc{iii}]$\lambda5007$ in the bottom panel of Figure \ref{fig:o32_ew}. We find a strong correlation between $12+\log \rm (O/H)$ and [O\,\textsc{iii}]$\lambda5007$ EW, such that galaxies with increased EW exhibit lower metallicities, with the best fit relation given by: 

\begin{equation}
\label{eqn:ox_abund_EW}
    \begin{split}
    12+\log(\mathrm{O/H}) = -1.12^{+0.36}_{-0.40} \times \log_{10}[\mathrm{EW_0 ([O\,\textsc{iii}]}\lambda5007/\text{\AA})]  \\
    + 11.23^{+1.25}_{-1.09}.
    \end{split}
\end{equation}
We note that at metallicities less than $12+\log(\mathrm{O/H})\approx7.70$ $(0.1Z_\odot)$, [O\,\textsc{iii}] EWs are greatly reduced \citep[][]{Laseter_2025}, therefore caution should be applied when using Equation \ref{eqn:ox_abund_EW} in the very low metallicity regime.

\subsection{Balmer Break Strength}
\label{sec: balmer break}
An alternative explanation of the observed red IRAC-excess colours in galaxies at $z>7$ is the presence of a Balmer break, indicative of an evolved stellar population \citep[e.g.,][]{eyles_stellar_2007, hashimoto_onset_2018,roberts-borsani_interpreting_2020}. We use NIRSpec PRISM spectroscopy to measure the break strength of our sample with \textsc{ppxf}. We adopt the Balmer break definition utilised in \citet{roberts_borsani_between_the_extremes_2024, witten_rising_2025} of $B = F_{\nu, 4225} / F_{\nu, 3565}$, with windows covering rest-frame 3500-3630{\AA} and 4160-4290{\AA}, avoiding potential emission line contamination. Values of $B>1$ correspond to a Balmer break, while $B<1$ is indicative of an inverse Balmer break, or `Balmer jump', caused by nebular continuum emission from gas ionized by extremely young stars \citep[e.g.,][]{cameron_nebular_2024, katz_21_2025}. The break strengths of our sample are shown in Figure \ref{fig:break strength}. While Balmer jumps are more common than Balmer breaks, one galaxy that shows the weakest EW (Wide-008401) possibly shows a significant Balmer break ($F_\mathrm{\nu,4225}/F_\mathrm{\nu,3565} = 1.66\pm0.59$). However, this galaxy was observed with a significant blue IRAC colour (at $z\sim6.9$), which is difficult to reconcile with a strong Balmer break \citep{roberts-borsani_interpreting_2020}. Furthermore, none of the red IRAC sources at $z>7$ provide evidence for a clear Balmer break, making IRAC colour an inefficient method to select early massive galaxies.

\subsection{SED Fitting}
\label{sec:sed fitting}

We model the spectral energy distribution (SED) of each galaxy using \textsc{bagpipes} v1.3.2 \citep[Bayesian Analysis of Galaxies for Physical Inference and Parameter EStimation;][]{carnall_inferring_2018, carnall_vandels_2019}. We use \citet[][]{bc03} stellar models, with the \citet{Kroupa_2001} initial mass function (IMF). We fit to the PRISM spectra alone, with the exception of Wide-001576 where we adopt the SED fitting results from \citet{Ormerod_2026}.
\textsc{bagpipes} v1.3.2 now enables the fitting of potential damped Lyman-$\alpha$ (DLA) absorption.
This absorption is separate from IGM damping wing absorption, although IGM absorption becomes degenerate with proximate DLA absorption at the low spectral resolution of the PRISM \citep[e.g.,][]{huberty_pitfalls_2025, mason_constraints_2026} and the IGM damping wing is not expected to be very strong at the redshifts considered in this work (see Figure \ref{fig:dla}).
We refer the reader to \citet{Witstok_2025_Lya} for a detailed description, and provide a brief summary below. The absorption cross section of hydrogen is modelled as the Voigt profile approximation from \citet{tasitsiomi_2006}, with a linear correction from \citet{Lee_2013} applied. The redshift of the foreground DLA system is fixed to the systemic redshift throughout fitting. We assume a uniform prior of $19 < \log_{10}(N_{\mathrm{H\,\textsc{i}}}/\mathrm{cm}^{-2}) < 24$  and assume the temperature of the absorbing gas to be $T=1000$ K. 

We adopt a non-parametric star formation history (SFH) from \citet{leja_how_2019}, which fits the star formation rates in fixed time bins, where $\Delta$logSFR between bins is linked by a Student's t-distribution. As in \citet{tacchella_stellar_2022}, we fit a `bursty continuity' model with scale $\sigma = 1$ and $\nu = 2$ degrees of freedom, with the first bins fixed to $0~\rm{Myr} < t < 3~\rm{Myr}$ and $3~\rm{Myr} < t < 10~\rm{Myr}$. This is to allow the SFH to capture extreme line emission, motivated by observational evidence for bursty SFHs seen at high-redshift \citep[e.g.,][]{boyett_extreme_2024, endsley_burstiness_2024, looser_jades_2025}. The remaining 4 bins are equally log-spaced in lookback time until $t(z=20)$, with the inferred SFH largely insensitive to the number of bins used as long as $N_\mathrm{bins}\geq4$ \citep{leja_how_2019}. 
This flexible SFH accommodates both stochastic star formation and underlying older stellar populations, yet still allowing for smooth or declining SFHs if favoured during fitting \citep[see][]{harvey_epochs_2025}. 
We reduce the default star formation rate (SFR) timescale to $10~$Myr to account for the increased specific star formation rate (sSFR), compared to galaxies at lower redshift. We also measure SFRs over timescales of 3 and 30 Myr.

The allowed total stellar mass formed and stellar metallicities are allowed to vary uniformly between $10^5~\rm{M}_\odot < M_\star < 10^{15} ~\rm{M}_\odot$, and $0<Z_*<2.0~ \mathrm{Z}_{\odot}$, respectively. Nebular emission is included using a grid of \textsc{cloudy} \citep{ferland_2017_2017} models, parametrised by the ionization parameter $(-3 < \rm{log}_{10} U < -0.5)$, which \textsc{bagpipes} computes self-consistently. 

We adopt a modified version of the \citet{salim_dust_2018} dust attenuation curve, which parametrizes the dust curve shape with a power-law deviation $\delta$ from the \citet{calzetti_dust_2000} model ($\delta = 0$ for the Calzetti curve, with negative values of $\delta$ producing steeper slopes) and typically includes a Drude profile to model the 2175\AA\ bump. For most sources, we keep the bump strength fixed to $B=0$ and allow only the slope to vary, as the presence of a UV bump has been identified in only two sources \citep{ormerod_detection_2025}. For Wide-006214, which exhibits a tentative UV bump at 3.9$\sigma$ significance, we allow the bump strength to vary uniformly between $0 < B < 10$, assuming a central wavelength of 2175{\AA} rest-frame and bump width of 250{\AA}. 
For Wide-001576, we largely follow the SED fitting procedure in \citet{ormerod_detection_2025}, with minor variations. A full description of the updated SED fitting for Wide-001576 is presented in \citet{Ormerod_2026}, but we summarise the key differences briefly.
Due to the DLA fitting implementation, we no longer mask the Ly-$\alpha$ region. The \textsc{bagpipes} nebular models have been updated to \textsc{cloudy} v25.00, with the removal of ISM dust grains resulting in emission lines up to $\sim1$ dex brighter at ionization parameters $\log_{10}U > -2$ than previously predicted. As a result, we no longer mask the [O\,\textsc{iii}] emission line doublet. 

We use a Gaussian prior on the $V$-band dust attenuation with $\mu_{A_V} = 0.15$ mag, $\sigma_{A_V} = 0.15$ mag and attenuation limited to $0 < A_V < 7$ mag, fixing the fraction of attenuation arising from stellar birth clouds to $60\%$, with the remaining $40\%$ coming from the diffuse ISM \citep{chevallard_simulating_2019}. We also assume a uniform prior on the velocity dispersion in the range $1-1000$ km~s$^{-1}$.   Additionally, we assume that the spectrum follows the PRISM resolution curve based on a point-source morphology, using the resolution curve of an idealised point source generated with \texttt{msafit}, as described in \citet[][Appendix A]{de_graaff_ionised_2024}.
We summarise the priors used in Table \ref{tab:priors}.

Finally, we apply an aperture correction to the stellar masses and SFRs using the ratio $L_\mathrm{UV}/L_\mathrm{UV,prism}$ and quote the corrected values throughout. 
The properties of individual galaxies are listed in Table \ref{tab:sed_properties}. We present median values for the EELG and non-EELG samples in Table \ref{tab:med SED fitting}, showing that EELGs show slightly lower median ages and stellar masses, while $\rm SFR_{10}$ and $A_V$ are very similar for both EELGs and non-EELGs.

{\renewcommand{\arraystretch}{1.5} 
\begin{table}
    \centering
    \caption{Median properties of the EELG and non-EELG sample obtained from \textsc{bagpipes} SED fitting. Errors represent $1\sigma$ uncertainties. Properties are: stellar mass, SFR averaged over the last 10 Myr, $V$-band dust attenuation, and stellar mass weighted age.}
    \begin{tabular}{lcc} \hline
         Property & EELG & Non-EELG  \\ \hline
         $\log_{10}(\mathrm{M_\star/\mathrm{M_\odot}})$ & $8.40^{+0.59}_{-0.30}$ & $8.60^{+0.38}_{-0.16}$ \\
         SFR$_{10}$ (M$_\odot$ yr$^{-1}$) & $17.73^{+34.54}_{-8.01}$ &  $15.36^{+12.60}_{-9.44}$  \\
         $A_V$ (mag) & $0.17^{+0.11}_{-0.09}$ & $0.19^{+0.03}_{-0.12}$ \\ 
         $t_\star$ (Myr) & $19.66^{+12.57}_{-9.91}$ & $27.57^{+64.46}_{-5.04}$ \\
        \hline
    \end{tabular}
    \label{tab:med SED fitting}
\end{table}}

{\renewcommand{\arraystretch}{1.5} 
\begin{table*}
\centering
\caption{Properties of our sample obtained from \textsc{bagpipes} SED fitting. Columns are: galaxy ID, stellar mass, SFR averaged over 3, 10 and 30 Myr timescales, $V$-band dust attenuation, deviation from the slope of the Calzetti dust law, stellar metallicity, and mass-weighted stellar age. All error bars represent $1\sigma$ uncertainties.}
\begin{tabular}{lccccccccc}
\hline
ID & $\log_{10}(\mathrm{M_\star/\mathrm{M_\odot}})$ & SFR$_{3}$ (M$_\odot$ yr$^{-1}$) & SFR$_{10}$ (M$_\odot$ yr$^{-1}$) & SFR$_{30}$ (M$_\odot$ yr$^{-1}$) & $A_V$ (mag) & $\delta$  & $Z_\star$ $(Z_\odot)$ & $t_\star$ (Myr) \\ \hline
Wide-000003 & $8.98^{+ 0.28}_{- 0.26}$ & $30.76^{+ 23.69}_{- 17.62}$ & $57.13^{+ 28.02}_{- 29.79}$ & $26.68^{+ 15.1}_{- 11.22}$ & $0.22^{+ 0.08}_{- 0.07}$ & $-0.24^{+ 0.2}_{- 0.16}$ & $0.36^{+ 0.19}_{- 0.12}$ & $27.57^{+ 79.4}_{- 18.98}$ \\
Wide-000001 & $8.4^{+ 0.31}_{- 0.17}$ & $10.32^{+ 7.88}_{- 3.05}$ & $14.1^{+ 9.54}_{- 5.27}$ & $7.08^{+ 5.22}_{- 1.7}$ & $0.17^{+ 0.08}_{- 0.06}$ & $-0.14^{+ 0.18}_{- 0.2}$ & $0.31^{+ 0.12}_{- 0.09}$ & $27.29^{+ 67.23}_{- 18.19}$ \\
Wide-000036 & $8.6^{+ 0.3}_{- 0.14}$ & $6.13^{+ 10.37}_{- 4.04}$ & $26.75^{+ 7.78}_{- 10.2}$ & $11.87^{+ 4.3}_{- 2.36}$ & $0.2^{+ 0.09}_{- 0.06}$ & $-0.11^{+ 0.15}_{- 0.18}$ & $0.34^{+ 0.16}_{- 0.09}$ & $24.44^{+ 80.63}_{- 17.54}$ \\
Wide-001018 & $8.82^{+ 0.05}_{- 0.06}$ & $46.5^{+ 3.72}_{- 4.72}$ & $17.73^{+ 3.57}_{- 2.0}$ & $23.02^{+ 2.76}_{- 3.9}$ & $0.18^{+ 0.04}_{- 0.04}$ & $0.02^{+ 0.09}_{- 0.1}$ & $0.26^{+ 0.03}_{- 0.03}$ & $21.45^{+ 25.48}_{- 6.05}$ \\
Wide-002631 & $7.86^{+ 0.17}_{- 0.11}$ & $4.23^{+ 1.33}_{- 0.77}$ & $5.52^{+ 1.01}_{- 1.07}$ & $2.34^{+ 1.04}_{- 0.4}$ & $0.06^{+ 0.04}_{- 0.03}$ & $0.04^{+ 0.11}_{- 0.23}$ & $0.2^{+ 0.17}_{- 0.03}$ & $15.79^{+ 28.25}_{- 9.21}$ \\
Wide-000440 & $9.06^{+ 0.41}_{- 0.58}$ & $17.62^{+ 22.09}_{- 11.86}$ & $16.54^{+ 24.3}_{- 9.04}$ & $9.9^{+ 12.77}_{- 5.91}$ & $0.33^{+ 0.08}_{- 0.08}$ & $-0.33^{+ 0.12}_{- 0.1}$ & $0.31^{+ 0.13}_{- 0.08}$ & $153.96^{+ 217.22}_{- 134.63}$ \\
Wide-008401 & $8.88^{+ 0.09}_{- 0.1}$ & $0.92^{+ 1.53}_{- 0.82}$ & $4.45^{+ 1.64}_{- 1.77}$ & $8.78^{+ 5.35}_{- 3.66}$ & $0.12^{+ 0.1}_{- 0.07}$ & $-0.02^{+ 0.15}_{- 0.18}$ & $0.38^{+ 0.22}_{- 0.15}$ & $89.45^{+ 80.91}_{- 42.64}$ \\
Wide-007082 & $8.39^{+ 0.14}_{- 0.21}$ & $27.73^{+ 2.74}_{- 4.26}$ & $9.89^{+ 1.59}_{- 1.64}$ & $5.51^{+ 2.1}_{- 1.57}$ & $0.11^{+ 0.05}_{- 0.05}$ & $0.05^{+ 0.11}_{- 0.22}$ & $0.21^{+ 0.01}_{- 0.01}$ & $38.75^{+ 54.94}_{- 26.13}$ \\
Wide-007437 & $8.95^{+ 0.11}_{- 0.04}$ & $51.38^{+ 11.95}_{- 4.49}$ & $85.04^{+ 6.58}_{- 13.3}$ & $31.21^{+ 10.06}_{- 2.32}$ & $0.13^{+ 0.02}_{- 0.01}$ & $-0.42^{+ 0.08}_{- 0.06}$ & $0.41^{+ 0.02}_{- 0.02}$ & $9.85^{+ 12.65}_{- 4.06}$ \\
Wide-006482 & $8.44^{+ 0.27}_{- 0.18}$ & $1.26^{+ 4.99}_{- 1.21}$ & $15.36^{+ 5.21}_{- 4.94}$ & $7.43^{+ 3.16}_{- 1.75}$ & $0.19^{+ 0.09}_{- 0.06}$ & $-0.07^{+ 0.16}_{- 0.2}$ & $0.31^{+ 0.14}_{- 0.08}$ & $33.91^{+ 64.23}_{- 22.64}$ \\
Wide-000733 & $8.19^{+ 0.09}_{- 0.07}$ & $10.61^{+ 1.67}_{- 1.29}$ & $14.63^{+ 2.43}_{- 1.78}$ & $5.3^{+ 0.98}_{- 0.65}$ & $0.11^{+ 0.06}_{- 0.04}$ & $-0.08^{+ 0.18}_{- 0.23}$ & $0.28^{+ 0.08}_{- 0.07}$ & $9.78^{+ 20.82}_{- 4.1}$ \\
Wide-000749 & $8.1^{+ 0.19}_{- 0.1}$ & $5.53^{+ 2.15}_{- 1.21}$ & $10.2^{+ 2.34}_{- 2.08}$ & $4.07^{+ 1.52}_{- 0.67}$ & $0.06^{+ 0.06}_{- 0.04}$ & $0.06^{+ 0.1}_{- 0.21}$ & $0.27^{+ 0.12}_{- 0.06}$ & $16.12^{+ 44.32}_{- 9.7}$ \\
Wide-002815 & $8.83^{+ 0.11}_{- 0.14}$ & $140.77^{+ 23.22}_{- 30.62}$ & $56.97^{+ 14.79}_{- 14.33}$ & $21.3^{+ 5.4}_{- 5.84}$ & $0.28^{+ 0.08}_{- 0.08}$ & $-0.24^{+ 0.14}_{- 0.15}$ & $0.31^{+ 0.13}_{- 0.09}$ & $9.42^{+ 30.08}_{- 5.85}$ \\
Wide-000715 & $8.13^{+ 0.14}_{- 0.12}$ & $22.94^{+ 6.72}_{- 5.05}$ & $11.95^{+ 3.85}_{- 2.81}$ & $4.5^{+ 1.48}_{- 1.08}$ & $0.2^{+ 0.09}_{- 0.05}$ & $-0.23^{+ 0.19}_{- 0.15}$ & $0.22^{+ 0.09}_{- 0.04}$ & $9.52^{+ 27.61}_{- 5.26}$ \\
Wide-000770 & $8.17^{+ 0.32}_{- 0.27}$ & $9.58^{+ 4.08}_{- 4.11}$ & $8.48^{+ 3.29}_{- 3.51}$ & $4.05^{+ 1.9}_{- 1.61}$ & $0.12^{+ 0.08}_{- 0.07}$ & $0.01^{+ 0.12}_{- 0.2}$ & $0.33^{+ 0.26}_{- 0.12}$ & $32.47^{+ 62.18}_{- 24.64}$ \\
Wide-000705 & $9.06^{+ 0.33}_{- 0.22}$ & $109.17^{+ 20.8}_{- 25.94}$ & $51.63^{+ 20.2}_{- 15.05}$ & $24.03^{+ 7.43}_{- 7.22}$ & $0.57^{+ 0.07}_{- 0.07}$ & $0.09^{+ 0.06}_{- 0.07}$ & $0.66^{+ 0.05}_{- 0.51}$ & $47.83^{+ 137.3}_{- 39.44}$ \\
Wide-001408 & $8.47^{+ 0.21}_{- 0.12}$ & $22.76^{+ 9.09}_{- 3.73}$ & $21.61^{+ 8.17}_{- 4.23}$ & $8.9^{+ 3.96}_{- 1.65}$ & $0.3^{+ 0.09}_{- 0.08}$ & $-0.03^{+ 0.12}_{- 0.17}$ & $0.3^{+ 0.09}_{- 0.07}$ & $21.22^{+ 54.32}_{- 14.56}$ \\
Wide-000767 & $8.97^{+ 0.16}_{- 0.16}$ & $94.61^{+ 13.09}_{- 12.24}$ & $47.84^{+ 12.3}_{- 10.35}$ & $24.37^{+ 7.38}_{- 4.56}$ & $0.41^{+ 0.06}_{- 0.07}$ & $-0.13^{+ 0.07}_{- 0.1}$ & $0.27^{+ 0.04}_{- 0.04}$ & $27.56^{+ 44.42}_{- 17.29}$ \\
Wide-001576 & $8.39^{+ 0.13}_{- 0.09}$ & $9.92^{+ 3.27}_{- 1.78}$ & $20.48^{+ 3.81}_{- 5.00}$ & $8.08^{+ 1.67}_{- 1.29}$ & $0.27^{+ 0.10}_{- 0.09}$ & $-0.04^{+ 0.12}_{- 0.18}$ & $0.39^{+ 0.04}_{- 0.05}$ & $14.37^{+ 39.70}_{- 7.98}$ \\
Wide-001559 & $8.36^{+ 0.12}_{- 0.14}$ & $0.59^{+ 1.83}_{- 0.56}$ & $11.17^{+ 3.33}_{- 2.89}$ & $7.47^{+ 2.58}_{- 2.17}$ & $0.07^{+ 0.06}_{- 0.04}$ & $0.01^{+ 0.13}_{- 0.27}$ & $0.51^{+ 0.13}_{- 0.15}$ & $22.59^{+ 35.35}_{- 9.46}$ \\
Wide-001648 & $8.55^{+ 0.08}_{- 0.17}$ & $4.66^{+ 2.01}_{- 4.25}$ & $5.98^{+ 7.39}_{- 2.72}$ & $12.29^{+ 3.27}_{- 4.43}$ & $0.07^{+ 0.06}_{- 0.03}$ & $-0.07^{+ 0.2}_{- 0.21}$ & $0.37^{+ 0.13}_{- 0.14}$ & $21.13^{+ 20.48}_{- 4.86}$ \\
Wide-006214 & $8.1^{+ 0.07}_{- 0.04}$ & $9.33^{+ 1.3}_{- 1.03}$ & $12.32^{+ 1.03}_{- 1.13}$ & $4.33^{+ 0.58}_{- 0.35}$ & $0.09^{+ 0.03}_{- 0.02}$ & $-0.32^{+ 0.17}_{- 0.12}$ & $0.46^{+ 0.05}_{- 0.1}$ & $8.71^{+ 16.52}_{- 3.12}$ \\
Wide-004221 & $8.52^{+ 0.24}_{- 0.17}$ & $36.07^{+ 9.91}_{- 4.99}$ & $19.76^{+ 6.56}_{- 5.29}$ & $8.51^{+ 3.63}_{- 1.74}$ & $0.18^{+ 0.07}_{- 0.07}$ & $0.03^{+ 0.11}_{- 0.18}$ & $0.24^{+ 0.06}_{- 0.04}$ & $32.19^{+ 65.79}_{- 25.42}$ \\
Wide-006308 & $9.13^{+ 0.22}_{- 0.16}$ & $157.39^{+ 42.19}_{- 27.19}$ & $83.28^{+ 27.17}_{- 22.49}$ & $37.7^{+ 15.91}_{- 8.88}$ & $0.21^{+ 0.07}_{- 0.05}$ & $-0.28^{+ 0.16}_{- 0.13}$ & $0.32^{+ 0.07}_{- 0.07}$ & $23.24^{+ 48.37}_{- 16.79}$ \\
Wide-004267 & $9.05^{+ 0.03}_{- 0.05}$ & $49.22^{+ 4.39}_{- 5.53}$ & $18.88^{+ 2.98}_{- 2.39}$ & $40.67^{+ 4.0}_{- 4.49}$ & $0.15^{+ 0.04}_{- 0.04}$ & $0.11^{+ 0.06}_{- 0.13}$ & $0.16^{+ 0.01}_{- 0.01}$ & $18.27^{+ 11.81}_{- 2.23}$ \\
Wide-007695 & $7.87^{+ 0.2}_{- 0.14}$ & $7.04^{+ 1.65}_{- 1.16}$ & $5.12^{+ 1.14}_{- 1.51}$ & $2.33^{+ 1.31}_{- 0.51}$ & $0.03^{+ 0.03}_{- 0.02}$ & $-0.03^{+ 0.16}_{- 0.21}$ & $0.31^{+ 0.07}_{- 0.06}$ & $19.66^{+ 52.05}_{- 13.06}$ \\ 
\hline
\end{tabular}
\label{tab:sed_properties}
\end{table*}}

\subsection{Damped Lyman Alpha Systems}
\label{sec:Dlya}
A roll over, or softening of the Lyman-$\alpha$ break can be interpreted as strong damped Ly$\alpha$ absorption wings from high column densities of neutral hydrogen. Since the launch of \emph{JWST}, numerous galaxies showing evidence for damped Ly$\alpha$ absorption have been detected at high redshift \citep[e.g.][]{deugenio_jades_2024, umeda_jwst_2024, Heintz_JWST_Primal, heintz_strong_2024, chen_jwst_2024, Rowland_2025_DLA}. To further assess whether our galaxies show evidence of damping wings, we use the Ly$\alpha$ damping parameter \citep{Heintz_JWST_Primal} which is given by: 
\begin{equation}
D_{\mathrm{Ly} \alpha} \equiv \int_{\lambda_{\mathrm{L} y \alpha, \text { low }}}^{\lambda_{\mathrm{L} y \alpha, \text { up }}}\left(1-F_\lambda / F_{\text {cont }}\right) \mathrm{d} \lambda /\left(1+z_{\text {spec }}\right).
\label{eqn:damping}
\end{equation}
This is the equivalent width of the transmitted flux density $F_\lambda$ over the wavelength range $\lambda_{\text {Ly } \alpha, \text { low }}=1180$\AA\ to $\lambda_{\text {Ly } \alpha, \text { up }}=1350 $\AA. The continuum flux is estimated from the UV slopes measured in Section \ref{sec:UV mag slopes}.  We adopt the parameter space defined by \citet{Heintz_JWST_Primal} which corresponds to different physical regimes, and are summarised as follows: 

\begin{itemize}
    \item $D_{\mathrm{Ly} \alpha} < 0$ {\AA}: Strong Ly$\alpha$ emitters (LAEs).
    \item $D_{\mathrm{Ly} \alpha} = 0-35$ {\AA}: Sources with excess continuum flux, which suggests contributions from ionized bubbles or weak LAEs.
    \item $D_{\mathrm{Ly} \alpha} =35-55$ {\AA}: Ly$\alpha$ absorption from an increasingly neutral IGM
    \item $D_{\mathrm{Ly} \alpha} > 55$ {\AA}: Excess damped Ly$\alpha$ absorption from local H\textsc{i} with $N_\mathrm{H\textsc{i}} > 10^{21}~\mathrm{cm}^{-2}$.
\end{itemize}
The measured column densities and damping parameters are shown in Figure \ref{fig:dla}. We show a comparison of the column densities and damping parameters in Appendix \ref{sec:DLA_comparison}. The two parameters are strongly correlated, although at higher values, the column densities cover a wider dynamic range for a given damping parameter.

\begin{figure*}
    \centering
    \includegraphics[width=0.95\textwidth]{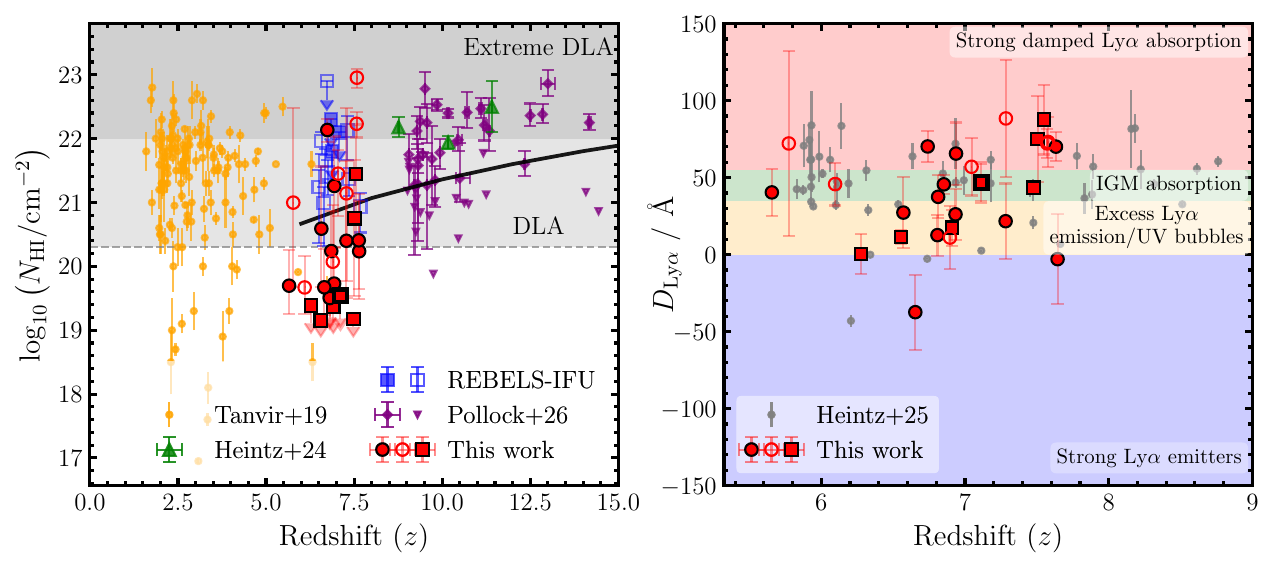}
    \caption{\textit{Left}: H\textsc{i} column densities as a function of redshift. EELGs are shown as solid red markers, while non-EELGs are shown as open red markers. The UV bump galaxy from \citet{ormerod_detection_2025} is shown with a thicker black border. We show LAEs detected in this work and the literature as square markers. For comparison we show gamma-ray burst Ly$\alpha$ absorbers \citep[orange circles;][]{Tanvir_2019}, the REBELS-IFU galaxies \citep[solid blue squares show EELGs while open blue squares show non-EELGs;][]{Rowland_2025_DLA}, and high redshift DLAs \citep[green triangles;][]{heintz_strong_2024}. The light grey shaded region shows the column density $N_{\mathrm{H} 1} \geq 2 \times 10^{20} \mathrm{~cm}^{-2}$, with systems above this value classified as DLAs \citep[][]{Wolfe_2005_DLA}. The darker grey shaded region shows the column density $N_{\mathrm{H} 1} \geq 10^{22} \mathrm{~cm}^{-2}$, corresponding to extreme DLA systems. The solid black line shows the maximum contribution of the IGM towards the column density as a function of redshift at $z\geq6$.
    \textit{Right}: $D_{\mathrm{Ly} \alpha}$ as a function of redshift, compared to galaxies with similar $M_\mathrm{UV}$ from \citet{Heintz_JWST_Primal}. The coloured backgrounds represent the four physical regimes defined in \citet{Heintz_JWST_Primal}.
    \label{fig:dla}
    }
\end{figure*}

\subsection{Ionizing Photon Production Efficiency}
\label{sec:xi ion}
The ionizing photon production efficiency $(\xi_{\rm ion})$ can be measured as the ratio of the hydrogen ionizing photon production rate to the non-ionizing luminosity density. We determine $\xi_\mathrm{ion}$ for our sample following: 
\begin{equation}
    \label{eqn:xi ion}
    \xi_{\rm ion} = \frac{N(\rm H^{0})}{L_{1500}}
\end{equation}
where $N(\mathrm{H}^0)$ is the intrinsic hydrogen ionizing photon production rate in units of s$^{-1}$, and $L_{1500}$ is the intrinsic luminosity density at rest-frame $1500$\AA\ in units of erg s$^{-1}$Hz$^{-1}$, after a dust correction is applied. We correct $L_{1500}$ for dust with the stellar attenuation curves and $A_V$ values obtained from \textsc{bagpipes} SED fitting. 

As in \citet{saxena_jades_2024}, we assume Case B recombination with $T_\mathrm{e} = 10^4~{\rm K}$ and $n_\mathrm{e} = 100~{\rm cm}^{-3}$, giving 
\begin{equation}
    \label{eqn:N(H0)}
    N(\mathrm{H}^0/\mathrm{s}^{-1})\times(1-f_{\rm esc}) = 7.3\times10^{11}L(\mathrm{H}\alpha/\mathrm{erg}~\mathrm{s}^{-1})
\end{equation}
where $f_{\rm esc}$ is the escape fraction of LyC photons \citep[e.g.,][]{Maseda_2020, Simmonds_2023_JEMS_MUSE}, which is $f_\mathrm{esc}=0$ for Case B recombination.
We determine N($\mathrm{H}^0$) from the H$\alpha$ luminosity where possible, and use the H$\beta$ luminosity with an intrinsic ratio of H$\alpha$/H$\beta=2.863$ \citep{osterbrock_ferland} at $z\gtrsim7$ where H$\alpha$ is redshifted out of NIRSpec spectral coverage. We present the ionizing photon production efficiencies in Figure \ref{fig:xi_ion} and Table \ref{tab:emission_lines}. 
Although we do not see a correlation with $M_\mathrm{UV}$, we find that our EELGs show a median $\log_{10}(\xi_\mathrm{ion} / \mathrm{erg^{-1}~Hz})=25.62^{+0.37}_{-0.18}$, similar to the high-z stack in \citet{boyett_extreme_2024}, but at brighter UV luminosities.
As seen in earlier work, we find the ionizing photon production efficiency to be strongly correlated with {\oiiihb} EW \citep[e.g.,][]{tang_mmtmmirs_2019, llerena_ionising_2025}, and obtain a best fit relation:
\begin{equation}
    \label{eqn: xi_ion_EW}
    \begin{split}
       \log_{10}(\xi_\mathrm{ion} / \mathrm{erg^{-1}~Hz}) = {} & (0.00041^{+0.00010}_{-0.00010}) \times \mathrm{EW_0([O\,\textsc{iii}]+H\beta}/\text{\AA})  \\
    & +24.95^{+0.17}_{-0.18}. 
    \end{split}
\end{equation}

We also find a clear, positive correlation between the ionizing production photon efficiency and the ratio of $\rm SFR_3 / SFR_{30}$, a measure of galaxy burstiness, with the best-fit relation given by: 
\begin{equation}
    \label{eqn: xi_ion_sfr_ratio}
    \log_{10}(\xi_\mathrm{ion} / \mathrm{erg^{-1}~Hz}) = (0.28^{+0.08}_{-0.06}) \times \mathrm{(SFR_3 / SFR_{30})}+25.08^{+0.10}_{-0.12}.
\end{equation}

\section{Discussion}
\label{sec:discussion}

\subsection{Excitation Properties}
\label{sec:ism discussions}

In Figure \ref{fig:o32_r23}, we present dust corrected O32 and R23 emission line ratios, tracing ionization state and metallicity, respectively \citep[e.g.,][]{cameron_jades_2023, saxena_jades_2024}. We compare to local SDSS galaxies \citep[][]{York_2000, SDSS_DR7}, the JADES EELG sample \citep{boyett_extreme_2024}, REBELS-IFU galaxies \citep{Rowland_Metallicity_2025} and A2744-YD4 (henceforth referred to as YD4) from \citet{witten_rising_2025}. In red shading, we mark the high metallicity and low ionization region on the R23-O32 plane defined in \citet{witten_rising_2025} in which YD4 lies. YD4 is best described by a rejuvenating SFH, resulting in the presence of both an older  and younger stellar population. This region is comparatively poorly populated, with most galaxies in both our sample and that from \citet{boyett_extreme_2024} occupying a region suggesting higher ionization parameters. The observed offset from the SDSS galaxies indicates a clear difference in the ISM conditions of high-redshift galaxies compared to the typical galaxy population at $z\sim0$, in agreement with \citet{cameron_jades_2023}. 
The scatter in the emission line ratios suggests a diversity of metallicities and ionization conditions within our sample. Although there is some overlap between our sample and the JADES galaxies from \citet{boyett_extreme_2024}, the fainter JADES EELGs are offset from our sample, implying a slightly lower metallicity with higher ionization parameters.  

In Figure \ref{fig:o32_ew}, we compare the O32 ratio to the {\oiii}$\lambda5007$ EW. O32 is known to increase with increasing rest-frame optical emission line EWs \citep[e.g.,][]{tang_mmtmmirs_2019, Tang_2023, boyett_extreme_2024}. 
We find a statistically significant correlation between the O32 ratio and \oiii\ EW ($r_\mathrm{s} = 0.508$, $p=0.009$), consistent with previous work. This suggests that galaxies with higher {\oiii}$\lambda5007$ EW have more highly ionized ISMs, likely driven by the young stellar populations responsible for their extreme nebular emission \citep{boyett_extreme_2024}. 
We fit the O32-EW relation with \textsc{linmix}, fitting to all spectra within our sample, and show the resulting relation as a solid black line with the 16th-84th percentiles indicated in grey shading. Our relation is offset towards lower O32 at fixed EW compared to the relation from \citet{boyett_extreme_2024} (green dashed line), suggesting a lower ionization parameter at a fixed EW within the brighter Wide sample (median $M_\mathrm{UV}=-20.7$ for our IRAC-selected EELG sample, compared to median $M_\mathrm{UV}=-18.7$ for high-z EELGs from \citealt{boyett_extreme_2024}). 
This is further reflected by the high-z EELG stack from \citet{boyett_extreme_2024} showing clear C\,\textsc{iv} emission in its rest-UV spectrum, whereas no rest-UV emission lines are strongly detected in our EELG stack (Figure \ref{fig:full stack}) with the exception of C\,\textsc{iii}], suggesting that the IRAC selected galaxies are more evolved than their fainter counterparts. 
For example, we find that the IRAC-selected EELGs are more massive (median $\log_{10}(\rm M_\star / M_\odot) = 8.40^{+0.59}_{-0.30}$) than the fainter EELGs from \citealt{boyett_extreme_2024}  (median $\log_{10}(\rm M_\star / M_\odot) = 7.45^{+0.37}_{-0.40}$).

While we do not detect a clear Balmer break, outshining by the recent starburst driving the extreme line emission could mask any older stellar population. Moreover, the two samples have similar gas-phase metallicities (Figure \ref{fig:o32_ew}), which may be driven by a recent infall of gas, thus masking any earlier star formation. 

The offset in O32 at fixed \oiii\ EW for the Wide EELGs compared to lower luminosity EELGs may furthermore suggest that our sample may exhibit lower electron density ($n_\mathrm{e}$) and/or star formation rate surface density ($\Sigma_\mathrm{SFR}$), key quantities responsible for the ionization parameter of high redshift galaxies \citep[][]{reddy_jwstnirspec_2023}. 
In the regime $\mathrm{O}32 \gtrsim 1.25$, simulations predict that there is a strong correlation between O32 and density, biasing O32 towards higher values due to the collisional de-excitation of [O\,\textsc{ii}] \citep[][]{choustikov_megatron_2026}.
It has been shown at $4 < z < 10$ that $\Sigma_\mathrm{SFR}$ shows a significant correlation with O32, and that galaxies with higher $\Sigma_\mathrm{SFR}$ should have conditions which facilitate the escape of Lyman continuum (LyC) photons, therefore playing an important role in cosmic reionization \citep[][]{calabro_evolution_2024}. Most of our sample appear extended in NIRCam imaging (Figure \ref{fig:individual slit locs}) which could explain the observed offset in O32 ratios seen in our sample. This may be due to the IRAC selection being sensitive to galaxy wide starbursts, compared to the compact clumps that are more likely to be detected with \emph{JWST}. This effect would be most pronounced in fainter sources, where an extended source at the same SFR as a compact source would be less likely to be detected. \citet{reddy_jwstnirspec_2023} suggest that covering fraction and gas density could both be correlated with $\Sigma_\mathrm{SFR}$ and therefore contribute to differences in O32 for the same ionizing spectrum. The overall low neutral gas column densities in the Wide EELGs suggest a relatively ionized state of the ISM/CGM, making high gas covering fractions less likely.

Galaxies within our EELG sample exhibit similar UV luminosities to the unexpectedly UV luminous $z>10$ galaxies $(M_\mathrm{UV}\lesssim-20)$ uncovered with \emph{JWST} \citep[e.g.,][]{arrabal_haro_spectroscopic_2023,bunker_jades_2023,castellano_jwst_2024, carniani_spectroscopic_2024, kokorev_capers_2025, naidu_cosmic_2026}. Some of these $z>10$ galaxies exhibit extreme ionization parameters $(\log_{10}U\gtrsim-2)$ and high-ionization rest-UV emission lines indicating a harder ionizing spectrum than our EELG sample $(\log_{10}U=-2.32^{+0.19}_{-0.15})$, and an evolution in ISM properties at fixed $L_\mathrm{UV}$ \citep[e.g.,][]{bunker_jades_2023,naidu_cosmic_2026, castellano_investigating_2026}. 

Comparing to EELGs at lower redshift, our sample exhibits systematically higher O32 and {\oiii}$\lambda5007$ EW than the $z\sim2$ galaxies from \citet{tang_mmtmmirs_2019}. This is consistent with the established redshift evolution of \oiiihb\ EWs, which increase from $\simeq 250${\AA} at $z\sim2$ \citep[e.g.,][]{boyett_o_2022}, to $\simeq780${\AA} at $z\sim7$ \citep[e.g.,][]{endsley_o_2021}. This trend has been seen across a number of recent studies \citep[e.g.,][]{Begley_2025, boyett_extreme_2024} and can be attributed to the rise in sSFR but also variation in metallicity and ionization parameter. 
We note that the $z\sim2$ sources from \citet{tang_mmtmmirs_2019} are broadly consistent with our derived relation between O32 and {\oiii}$\lambda5007$ EW, residing below the relation derived in \citet{boyett_extreme_2024}. The sample from \citet{tang_mmtmmirs_2019} has a median stellar mass of $4.9 \times 10^8 \mathrm{M}_{\odot}$, similar to the median stellar mass of our EELG sample, and $\sim1\,\mathrm{dex}$ higher than the median stellar mass of EELGs from \citet{boyett_extreme_2024}. In the local Universe, possible high-redshift analogues have been identified, such as `green peas' \citep[e.g.,][]{cardamone_galaxy_2009,Yang_2017_greenpea} and `blueberries' \citep[][]{Yang_2017_Blueberry}. Green peas are a sample of $z\sim0$ galaxies with strong {\oiii}$\lambda 5007$ emission lines and stellar masses $\log_{10}(\rm M_\star/M_\odot)\approx8.5-9.5$, whereas blueberries are small counterparts to green peas, and have stellar masses $\log_{10}(\rm M_\star/M_\odot)\approx6.5-7.5$ with more extreme O32 ratios. The variation of O32 with stellar mass suggests that, at all redshifts, more evolved galaxies have lower O32 ratios.

\subsection{Cosmic Reionization}
\label{sec:reionization discussion}
It remains an open question whether reionization is driven primarily by faint \citep[e.g.][]{atek_most_2024}, or bright \citep[e.g.][]{naidu_rapid_2020} galaxies. Recent work has begun to shed light on this, finding that brighter galaxies contribute increasingly more at lower redshifts \citep[][]{giovinazzo_reionization_2026, summerfield_thesan-zoom_2026} with luminous galaxies driving reionization by its completion. This picture is supported by model predictions from \citet{bera_toward_2026} which suggest that a substantial contribution from bright galaxies is required to reproduce \emph{JWST} observations of quantities such as the UV luminosity density, ionizing emissivity, and neutral hydrogen fraction, while avoiding a reionization history that ends too early. 

In this context, our uniquely luminous EELG sample allows us to assess the contribution of bright systems during this epoch. Below, we discuss our results in the context of cosmic reionization. 

\subsubsection{Constraints on Ionizing Photon Production}
\label{sec:xi_ion discussion}

\begin{figure*}
    \centering
    \includegraphics[width=0.7\linewidth]{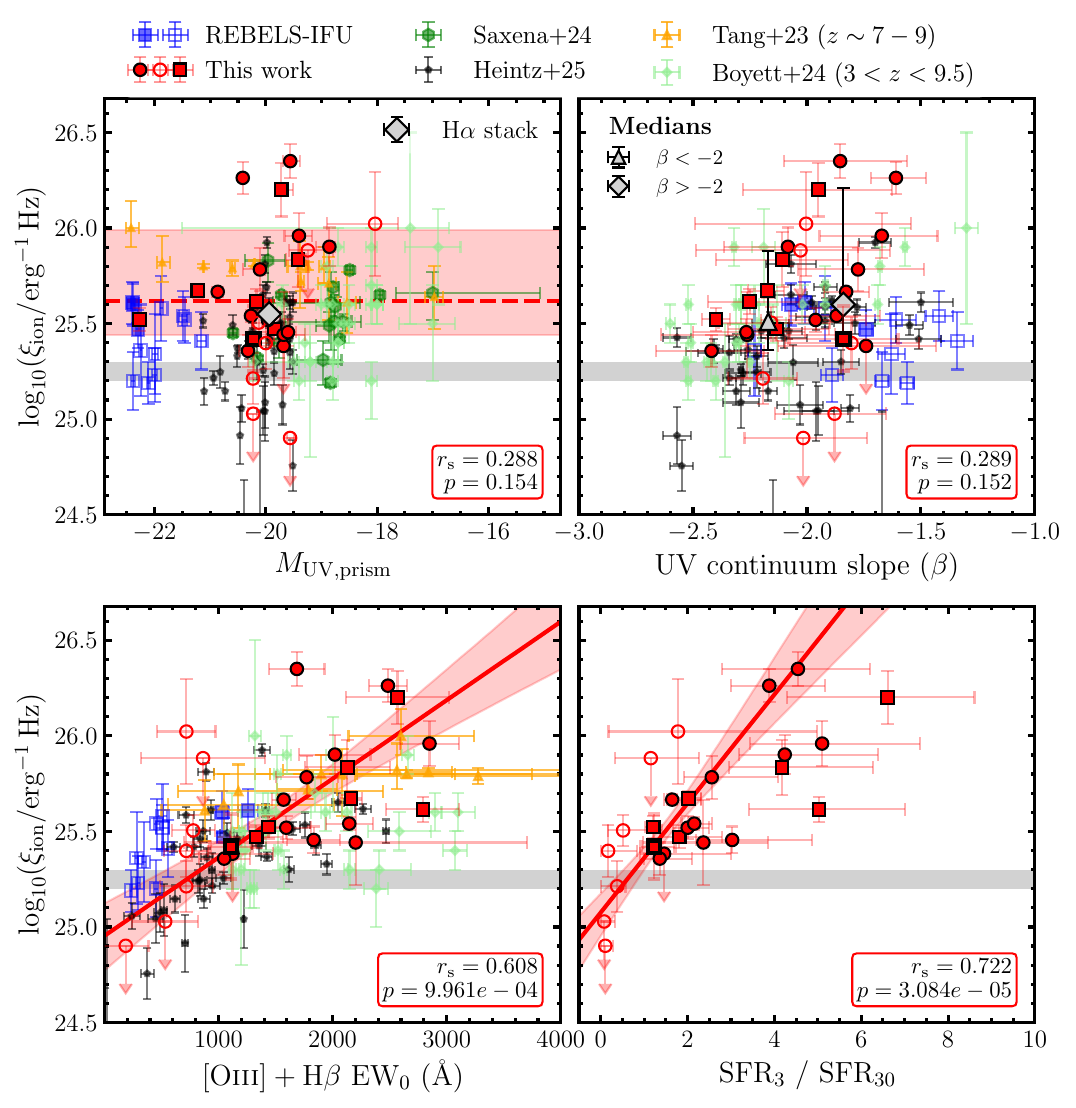}
    \caption{Ionizing photon production efficiency, $\log_{10}(\xi_\mathrm{ion} / \mathrm{erg}^{-1} \mathrm{Hz})$, as a function of $M_\mathrm{UV,prism}$ (top left), UV continuum slope ($\beta$; top right), {\oiiihb} EW (bottom left), and SFR$_3$/SFR$_{30}$ (bottom right). We choose to compare to the $M_\mathrm{UV}$ obtained from the PRISM spectra as this is the value used within the $\xi_\mathrm{ion}$ calculation. The markers used for this work are the same as in previous figures. 
    We compare to $z < 8$ galaxies from \citet{Heintz_JWST_Primal} (black pentagons) at similar $M_\mathrm{UV}$ to our sample, faint Ly$\alpha$ emitters (LAEs) from \citet{saxena_jades_2024} (dark green hexagons), $z\simeq7-9$ galaxies from \citet{Tang_2023} (orange triangles), the REBELS-IFU sample \citep[blue squares;][]{fisher_rebels-ifu_2025, Komarova_REBELS_xi_ion, Rowland_2025_DLA, Rowland_Metallicity_2025}, and JADES EELGs at $3 < z < 9.5$ \citep[light green diamonds;][]{boyett_extreme_2024}.
    In each panel, the grey shaded region shows the canonical values of $\log_{10}(\xi_\mathrm{ion} / \mathrm{erg}^{-1} \mathrm{Hz})=25.2-25.3$ from \citet{Robertson_2013, Robertson_2015}. 
    In the top left panel $(M_\mathrm{UV})$, we show the median $\xi_\mathrm{ion}$ for the EELG sample with the red dashed line, and the red shading indicates the 16-84th percentiles, and the grey diamond shows $\xi_\mathrm{ion}$ obtained from the stack of all galaxies with a H\,$\alpha$ detection.
    In the top right panel, we show the median $\xi_\mathrm{ion}$ for Wide galaxies with $\beta < -2$ (grey triangle) and $\beta > -2$ (grey diamond), with error bars showing the 16-84th percentiles. 
    In the bottom panels, where we find a significant relation between $\xi_\mathrm{ion}$ and \oiiihb\ (bottom left) and SFR$_3$/SFR$_{30}$ (bottom right), the solid red line shows the best fit determined using \textsc{linmix}, with red shading indicating the 16-84th percentiles. The Spearman's correlation coefficient ($r_s$) and $p$-value is shown in the bottom right of each panel.
    }
    \label{fig:xi_ion}
\end{figure*}

The ionizing photon production efficiency is a crucial parameter in estimating the contribution of galaxies to cosmic reionization \citep[e.g.,][]{llerena_physical_2024}. We use the values calculated in Section \ref{sec:xi ion} to determine the contribution of our sample.

The inferred $\log_{10}(\xi_\mathrm{ion}/\mathrm{erg^{-1}~Hz})$ values range from $<24.90$ to $26.35\pm0.09$, with a median value of $\log_{10}(\xi_\mathrm{ion}/\mathrm{erg^{-1}~Hz}) = 25.53_{-0.15}^{+0.43}$ for the full sample, consistent with the REBELS-IFU average of  $\log_{10}(\xi_\mathrm{ion}/\mathrm{erg^{-1}~Hz}) = 25.44\pm0.15$ \citep{Komarova_REBELS_xi_ion}. 
While our median value is consistent with the canonical value of $25.3$, we find that only 5/22 galaxies (excluding those where $\xi_\mathrm{ion}$ is an upper limit) are within $2\sigma$ of this canonical value. 

For EELGs alone, we find a median value of  $\log_{10}(\xi_\mathrm{ion}/\mathrm{erg^{-1}~Hz}) = 25.62_{-0.15}^{+0.43}$, offset from the canonical value of $\log_{10}(\xi_\mathrm{ion}/\mathrm{erg^{-1}~Hz})=25.3$ \citep[][]{Robertson_2013, Robertson_2015}, indicating enhanced ionizing photon production rates compared to typical star forming galaxies. In general, an enhanced $\xi_\mathrm{ion}$ is not unexpected for galaxies exhibiting strong optical emission lines, as previous studies have shown that $\xi_\mathrm{ion}$ increases with [O\,\textsc{iii}]$\lambda5007$ EW \citep[e.g.,][]{tang_mmtmmirs_2019}. Additionally, \citet{boyett_extreme_2024} find an average ionizing photon production efficiency of  $\log_{10}(\xi_\mathrm{ion}/\mathrm{erg^{-1}~Hz}) = 25.5\pm0.2$ for their EELG sample, suggesting that high redshift EELGs represent a population of efficient ionizing photon producers. 
However, values as high as $\log_{10}(\xi_\mathrm{ion}/\mathrm{erg^{-1}~Hz}) \gtrsim 26$ are unexpected \citep[][]{Maseda_2020}. Additionally, \citet{katz_21_2025} predict that the observed ionizing photon production efficiency should not exceed $\log_{10}(\xi_\mathrm{ion}/\mathrm{erg^{-1}~Hz}) \gtrsim 25.8$ if gas is being irradiated by a normal stellar population. Measurements above this value require gas densities high enough to suppress the two-photon emission, an exotic stellar population with an intrinsic $\xi_\mathrm{ion}$ much higher than predictions from normal SPS models, or extreme nebular conditions such as very high temperatures.  

To investigate the physical origin driving the elevated $\xi_\mathrm{ion}$ values seen in our sample, we compare $\xi_\mathrm{ion}$ with a range of galaxy properties in Figure \ref{fig:xi_ion}. Investigating $\xi_\mathrm{ion}$ as a function of the absolute UV magnitude measured from the PRISM spectra ($M_\mathrm{UV,prism}$) tentatively suggests that fainter galaxies show increasing $\xi_\mathrm{ion}$, though we do not find evidence of a significant correlation ($r_s = 0.288, p=0.154$). The observed trend is broadly consistent with results from \citet{llerena_ionising_2025}. Additionally, \citet{Begley_2026_EXCELS} find a shallow relation suggesting that the more $M_\mathrm{UV}$-faint galaxies show increased $\xi_\mathrm{ion}$. Conversely, recent works have also found flat \citep[][]{Austin_2025} and decreasing \citep[such that fainter galaxies show lower $\xi_\mathrm{ion}$;][]{Pahl_2025, Pahl_2026} $\xi_\mathrm{ion}-M_\mathrm{UV}$ relations, highlighting the broad range of results in the literature. Such variation in results between studies may be a result of selection biases, and therefore require large samples with a large range in observed properties to disentangle. 

We next examine the relation between $\xi_\mathrm{ion}$ and the UV continuum slope ($\beta$), which traces the dust content, stellar age, and metallicity of galaxies. 
Bluer galaxies are expected to be younger, have lower metallicities, and have harder ionizing spectra, which may indicate an elevated $\xi_\mathrm{ion}$ compared to redder galaxies \citep[e.g.,][]{bouwens_2014_uv_slopes, Begley_2026_EXCELS}. Therefore, one may expect to see a significant relationship between $\beta$ and $\xi_\mathrm{ion}$, such as that seen in \citet{Begley_2026_EXCELS}.   However, we do not see a significant correlation between $\xi_\mathrm{ion}$ and $\beta$ $(r_s = 0.289, p=0.152)$, and we find a higher median $\xi_\mathrm{ion}$ in redder sources with $\beta>-2$ than bluer sources with $\beta<-2$. In particular, we find the highest $\xi_\mathrm{ion}>26.0$ galaxies have moderately red UV-continuum slopes ($\beta\gtrsim2$). The lack of dependence on $\beta$ is consistent with \citet{Komarova_REBELS_xi_ion} and \citet{Pahl_2025}. This lack of correlation may be a result of nebular continuum effects, whereby strong nebular continuum emission can redden UV slopes \citep[][]{katz_21_2025, narayanan_ultraviolet_2025}. 
Alternatively, a non-zero $f_\mathrm{esc}^{\mathrm{Ly}\alpha}$ will weaken the nebular continuum resulting in a bluer UV slope \citep[][]{Chisholm_2022, Topping_2024_slopes, Saxena_2024_slopes, Jecmen_2026}. Finally, very young star-clusters with high $\xi_\mathrm{ion}$ from their massive stars could still be embedded in dusty molecular clouds \citep[e.g.,][]{Inami2022}.

{\oiiihb} (or [O\,\textsc{iii}]) EW has been suggested to be a powerful diagnostic of $\xi_\mathrm{ion}$ at $z\sim2$ due to the production of the [O\,\textsc{iii}] emission line by young, massive stars  \citep[][]{Chevallard_2018, Reddy_2018b, tang_mmtmmirs_2019}.
Furthermore, it has been shown that EELGs at $z>3.5$ may show $\xi_\mathrm{ion}$ values greater than the canonical values \citep[][]{Nakajima_2016, Onodera_2020, llerena_ionising_2025}.  
We find a strong correlation between $\xi_\mathrm{ion}$ and {\oiiihb} EW $(r_s = 0.608, p=9.961\times10^{-4})$, such that galaxies with higher EW have higher $\xi_\mathrm{ion}$, in agreement with previous work \citep[e.g.,][]{Begley_2026_EXCELS, Pahl_2026, Llerena_2026}. 
The best fit relation is given by Equation \ref{eqn: xi_ion_EW}. This relation is unsurprising given that large EWs are linked with young, metal-poor stellar populations that are expected to produce large amounts of ionizing photons \citep[][]{Chevallard_2018}. However, recent work has shown that the use of [O\,\textsc{iii}] EW alone as a tracer of ionizing photon production efficiency may be limited by its secondary dependence on metallicity, where low gas-phase metallicities may suppress [O\,\textsc{iii}] emission below $12+\log(\rm O/H)\approx 7.70$, despite these galaxies being efficient ionizers \citep[][]{Laseter_2025}.

Finally, we present the relation between $\xi_\mathrm{ion}$ and $\rm SFR_3/SFR_{30}$. $\rm SFR_3/SFR_{30}$ is a measure of the burstiness of a galaxy's SFH, with a larger $\rm SFR_3/SFR_{30}$ indicating a steeper rise in recent star formation. We find a significant relation between $\xi_\mathrm{ion}$ and burstiness ($r_s = 0.722, p=3.084\times10^{-5}$), suggesting that galaxies with a recent upturn in star formation, and therefore a population of O and B stars, are efficient engines of ionizing photons. The best fit relation is given by Equation \ref{eqn: xi_ion_sfr_ratio}.
This is also consistent with a scenario in which EELGs with the most extreme recent bursts of star formation may be in a nebular dominated phase, with extreme {\oiiihb} EWs, high ionizing photon production efficiencies, and UV slopes reddened by nebular continuum emission.

In summary, based on their $\xi_\mathrm{ion}$ alone, bright EELGs have the potential to contribute significantly to cosmic reionization, consistent with \citet{llerena_physical_2024} who find that the majority of EELGs have $\xi_\mathrm{ion}$ greater than the canonical value.

\subsubsection{Neutral Hydrogen Column Densities and DLAs}
\label{sec:nhi discusison}
DLA systems are defined as having large neutral hydrogen (H\textsc{i}) column densities of $N_{\mathrm{H} \textsc{i}} \geq 2 \times 10^{20} \mathrm{~cm}^{-2}$ \citep{Wolfe_2005_DLA}, which are generally attributed to large columns of neutral gas within, or near the galaxy. A subset of these systems reaching column densities of $N_{\mathrm{H} 1} \geq 10^{22} \mathrm{~cm}^{-2}$ are often referred to as `extreme'-DLA systems, with growing numbers being detected by \emph{JWST} \citep[e.g.,][]{heintz_strong_2024, Rowland_2025_DLA, Witstok_2025_Oxygen, Pollock_2026_DLA}. 

We use two methods of assessing whether our sample shows evidence of DLAs: incorporating DLA fitting into our SED fitting analysis (Section \ref{sec:sed fitting}), and measuring the $D_{\mathrm{Ly}\alpha}$ parameter (Section \ref{sec:Dlya}).  We show $D_{\mathrm{Ly}\alpha}$ as a function of redshift in the right panel of Figure \ref{fig:dla}, with colour-coding indicating the four different physical regions defined in \citet{Heintz_JWST_Primal}. With this diagnostic, we find that $\sim38\%$ of our sample lies within the strong DLA region ($D_{\mathrm{Ly}\alpha}>55${\AA}), lower than the DLA fractions presented in \citet{Heintz_JWST_Primal}. 
We find that five out of the seven non-EELG galaxies have $D_{\mathrm{Ly}\alpha}>55${\AA}, indicating strong damped Ly$\alpha$ absorption. This is consistent with a scenario in which the non-EELG galaxies have been unable to ionize their surrounding CGM. In the left panel of Figure \ref{fig:dla}, we show the H\textsc{i} column densities as a function of redshift. We find that $3/26$ spectra are classified as extreme-DLAs: galaxies with $N_\mathrm{H\textsc{i}} \geq 10^{22}~\mathrm{cm}^{-2}$, with only one of these being in the EELG subsample. We classify a further nine spectra as DLAs when defining a DLA as a system with $N_{\mathrm{H} \textsc{i}} \geq 2 \times 10^{20} \mathrm{~cm}^{-2}$. We also overlay the maximum contribution of the IGM to the neutral hydrogen column density as a function of redshift at $z>6$. For galaxies that lie below this line, it becomes hard to disentangle the presence of H$\textsc{i}$ gas near
the galaxy from the contribution of the IGM with $x_\mathrm{HI}\leq1$ \citep[e.g.,][]{Pollock_2026_DLA}. 
However, for galaxies above this line, there must be some contribution from the presence of H$\textsc{i}$ gas within or close to the galaxy.

\begin{figure*}
    \centering
    \includegraphics[width=0.99\linewidth]{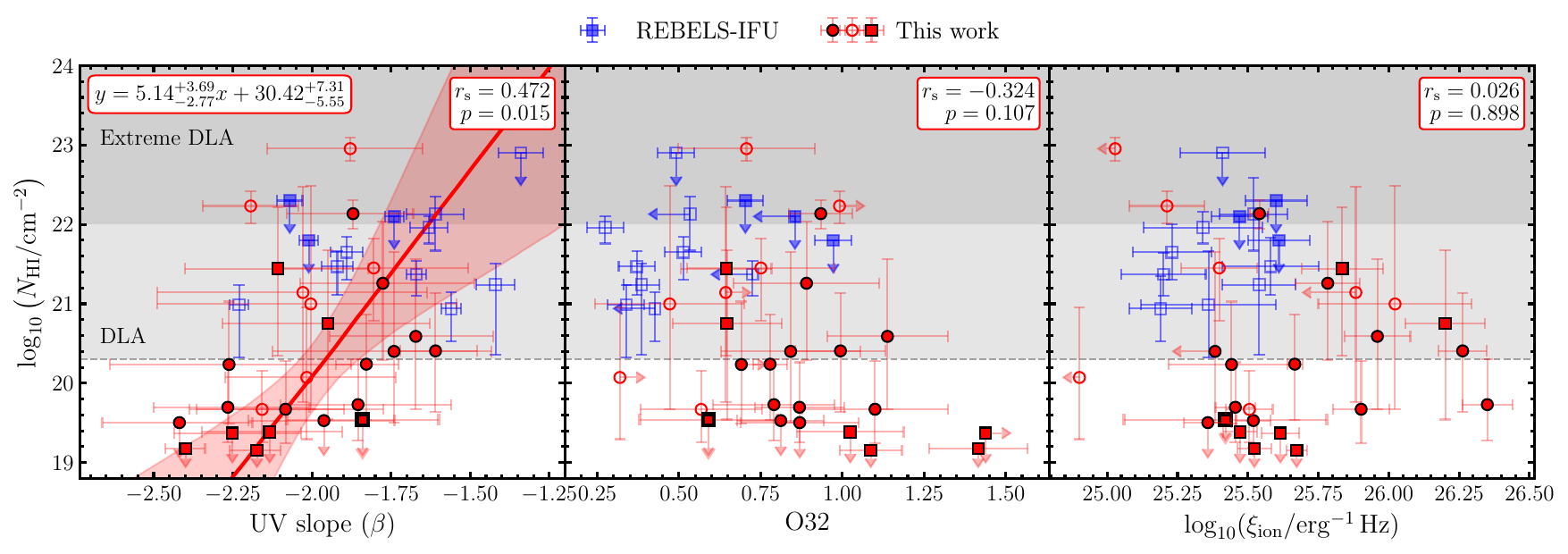}
    \caption{Neutral hydrogen column density as a function of UV continuum slope ($\beta$; left), O32 (centre), and $\xi_\mathrm{ion}$ (right). The markers for this work are the same as shown in previous figures. 
    For comparison, we show the REBELS-IFU galaxies \citep[][]{fisher_rebels-ifu_2025, Rowland_2025_DLA, Rowland_Metallicity_2025, algera_oiii_cii_2025} as blue squares. The Spearman correlation coefficient ($r_\mathrm{s}$) and the $p$-value shown in the top right of each panel. For the UV continuum slope, we show the linear relation for the Wide galaxies with the solid red line, and show the 16-84th percentiles as red shading. 
    The light grey shaded region shows the column density $N_{\mathrm{H}\textsc{i}} \geq 2 \times 10^{20} \mathrm{~cm}^{-2}$, with systems above this value classified as DLAs \citep[][]{Wolfe_2005_DLA}. The darker grey shaded region shows the column density $N_{\mathrm{H}\textsc{i}} \geq 10^{22} \mathrm{~cm}^{-2}$, corresponding to extreme DLA systems.}
    \label{fig:log_nhi_relations}
\end{figure*}

In order to probe the driving cause in the variation of $N_\mathrm{H\textsc{i}}$ across our sample, we examine its relation with three parameters: the UV-continuum slope ($\beta$), the O32 ratio (a proxy for ionization parameter), and $\xi_\mathrm{ion}$, shown in Figure \ref{fig:log_nhi_relations}. 
We find a significant relation only between  $N_\mathrm{H\textsc{i}}$ and $\beta$ ($r_s = 0.472, p = 0.015$), with the most gas-rich galaxies showing more reddening.
The best fit relation is shown in the left panel of Figure \ref{fig:log_nhi_relations} as a solid red line, with the 16th-84th percentiles indicated by red shading. The best-fit relation is given by: 
\begin{equation}
    \label{eqn:nhi_beta}
    \log _{10}\left(N_{\mathrm{HI}} / \mathrm{cm}^{-2}\right) = 5.14_{-2.77}^{+3.69} \times \beta+30.42_{-5.55}^{+7.31}.
\end{equation}

Reddening of the UV-continuum slope is primarily caused by the presence of dust within galaxies, while also being influenced by the age and metallicity of stellar populations, such that younger and more metal poor galaxies exhibit bluer UV slopes \citep[e.g.,][]{wilkins_ultraviolet_2011, Bouwens_2012_UV_slopes, tacchella_stellar_2022}. 
In low redshift galaxy samples, the UV slope has also been linked to the fraction of LyC photons escaping from a galaxy ($f_\mathrm{esc}$) \citep[][]{Chisholm_2022}, with bluer slopes indicating a higher $f_\mathrm{esc}$. In this context, our observed decrease in H\textsc{i} column density towards bluer UV slopes is consistent with a scenario in which these galaxies have ionized their surrounding CGM, allowing LyC photons to escape into the IGM. However, \citet{Heintz_JWST_Primal} compared the $D_{\mathrm{Ly}\alpha}$ parameter to the UV slope at $z>5$, finding no apparent correlation between $\beta$ and DLA absorption, suggesting that larger H\textsc{i} abundances do not translate directly into lower $f_\mathrm{esc}$. Moreover, \citet{Pollock_2026_DLA} do not find a correlation between $\beta$ and $\log_{10}(N_{\mathrm{H} \textsc{i}} /\mathrm{cm}^{-2})$ at $z>9$ and attribute this to possible sightline-to-sightline variations in $N_{\mathrm{H} \textsc{i}}$. A large sightline-to-sightline variation in $\log_{10}(N_{\mathrm{H} \textsc{i}} /\mathrm{cm}^{-2})$ of $0.5-1.5$~dex has also been seen in simulations \citep{Gelli_2025}, which can be attributed to complex ISM morphology arising from bursty SFHs. 

We next investigate the relation between $\log_{10}(N_{\mathrm{H} \textsc{i}} /\mathrm{cm}^{-2})$ and the O32 emission line ratio, which is a commonly used probe of the ionization state of nebular gas within a galaxy, therefore often used as a proxy for the ionization parameter \citep[e.g.,][]{tang_mmtmmirs_2019,Kewley_2019, papovich_2022, cameron_jades_2023}. However, the Spearman correlation coefficient and $p$-value ($r_\mathrm{s}=-0.324, p=0.107$) does not indicate a strong correlation between the two parameters, and may require a larger sample to establish whether the observed trends are statistically significant. 

Finally, we compare  $\log_{10}(N_{\mathrm{H} \textsc{i}} /\mathrm{cm}^{-2})$ and $\xi_\mathrm{ion}$, finding no evidence for any correlation ($r_s=0.026, p=0.898$). Within the Wide sample, including both EELGs and non-EELGs, we see values of $\xi_\mathrm{ion}$ ranging from $\log_{10}(\xi_\mathrm{ion}/\mathrm{erg^{-1}~Hz}) < 25$ to $\log_{10}(\xi_\mathrm{ion}/\mathrm{erg^{-1}~Hz})=26.3\pm0.1$ for galaxies within the DLA or extreme DLA regions, with a median value of  $\log_{10}(\xi_\mathrm{ion}/\mathrm{erg^{-1}~Hz})=25.8$. This broad range of $\xi_\mathrm{ion}$ values for DLA systems is in agreement with  \citet{Heintz_JWST_Primal} which finds that galaxies with strong DLAs occupy most of the $\xi_\mathrm{ion}$ parameter space. Similarly, \citealt{Heintz_JWST_Primal} finds that galaxies with DLAs $(\log_{10}(\xi_\mathrm{ion}/\mathrm{erg^{-1}~Hz})=25.55)$ are offset above the canonical values. The increased offset observed in the Wide sample is likely due to the high fraction of EELGs within the sample. 

\subsubsection{Ionized Bubbles}
\label{sec:Lya discussion}
We use the \textsc{lymana\_absorption} code\footnote{Code available at \href{https://github.com/joriswitstok/lymana_absorption}{https://github.com/joriswitstok/lymana\_absorption}.} to estimate the sizes of potential ionized bubbles surrounding the LAEs within our sample \citep[][]{Witstok_bubble_2024, Witstok_2025_Lya}. For LAEs with a velocity offset measurement (Wide-004267, Wide-001576, Wide-006308), we find ionized bubble sizes of $R_\mathrm{ion} < 0.23$\,physical Mpc (pMpc). The three LAEs identified with PRISM observations do not have a measurement of $\Delta v_\mathrm{Ly\alpha}$, therefore we estimate $R_\mathrm{ion}$ upper limits for $\Delta v_\mathrm{Ly\alpha}=0\,\mathrm{km\,s}^{-1}$, resulting in ionized bubble sizes of $R_\mathrm{ion}<0.52\,\mathrm{pMpc}$.

\begin{figure}
    \centering
    \includegraphics[width=1.0\columnwidth]{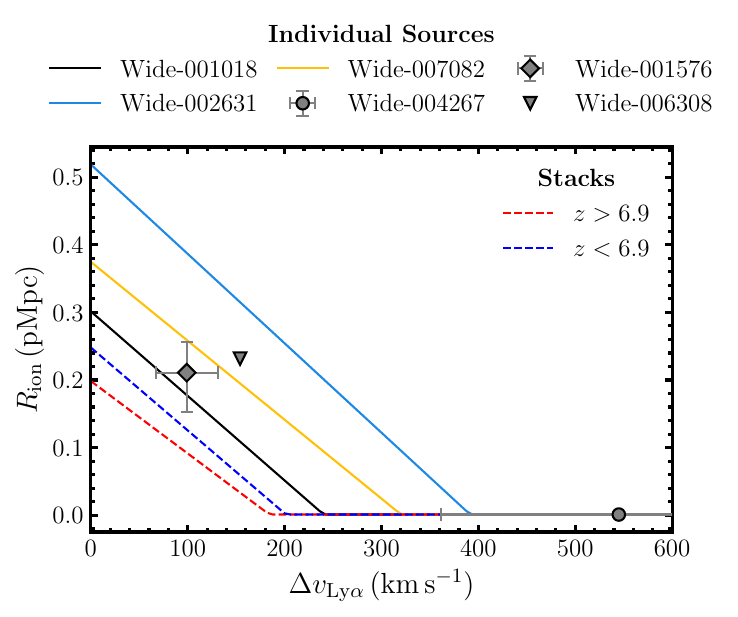}
    \caption{Ionized bubble sizes $(R_\mathrm{ion})$ as a function of Ly$\alpha$ velocity offset ($\Delta v_\mathrm{Ly\alpha}$) for LAEs within our sample. LAEs identified with PRISM observations are shown as solid lines. We show LAEs with velocity offsets as grey markers. We note that Wide-006308, represented by the triangle marker, is an upper limit. Finally, we show the high-z and low-z stacks with the red and blue dashed lines, respectively.}
    \label{fig:R_ion}
\end{figure}

We additionally investigate the ionized bubble sizes of our high-z and low-z stacks, to determine how bubble sizes change towards the later stages of reionization. For the high-z stack, with a median redshift of $z=7.380$, we find $R_\mathrm{ion} < 0.20~\mathrm{pMpc}$. For the low-z stack (median $z=6.654$) we find $R_\mathrm{ion}<0.25~\mathrm{pMpc}$. We show the values $R_\mathrm{ion}$ in Figure \ref{fig:R_ion}, and in Table \ref{tab:Lya}.

Additionally, we use the stacks from Section \ref{sec: stacks} to measure $D_\mathrm{Ly\alpha}$ for the different redshift bins, finding $D_\mathrm{Ly\alpha}=40.5\pm17.5$~{\AA} for the high-z stack, and $D_\mathrm{Ly\alpha}=14.8\pm18.3$~{\AA} for the low-z stack, consistent with an increasingly neutral IGM at higher redshift. The high-z stack does not show any evidence for Ly$\alpha$ emission, indicating that EELGs at $z>6.9$ do not yet reside in large ionized bubbles. While the low-z stack shows evidence for Ly$\alpha$ emission, the relatively modest Lya equivalent width in the stack of $\rm EW_0(Ly\alpha) = 16.1^{+4.3}_{-4.3}\,\AA$ combined with $\log_{10}(\xi_\mathrm{ion} / \mathrm{erg}^{-1}\,\mathrm{Hz)=25.52\pm0.02}$ suggests that most Ly$\alpha$ photons are still scattered out of our line of sight. Given that that we find overall low $\log _{10}(N_{\mathrm{HI}} / \mathrm{cm}^{-2})$, suggesting a possibly ionized CGM, this suggests that the IGM on larger scales must still be largely neutral, as evidenced by the small ionized bubble sizes we find above.  
This might be in tension with models in which brighter galaxies become more important to reionization in the latter stages of the EoR \citep[e.g.][]{summerfield_thesan-zoom_2026, giovinazzo_reionization_2026}.
Additionally, these EELGs may appear very UV-luminous due to their recent strong burst in star formation, as shown in Figure \ref{fig:sfh_figure}, despite being low-mass systems residing in low-mass halos. As low-mass halos are less biased towards overdense regions, their local environments might explain why they are found in such small ionized bubbles. 

Taken together, despite having a high production of ionizing photons and a relatively low neutral hydrogen column density near the galaxy, bright EELGs are unlikely to contribute significantly to cosmic reionization.

\subsection{Diverse Star Formation Histories}
\label{sec:sfh discussion}

Prior to the launch of \emph{JWST}, insights into the stellar ages and mass assembly of galaxies in the EoR relied upon IRAC observations, with IRAC excesses being interpreted as either very strong Balmer breaks \citep[e.g.,][]{Eyles_2005}, or extreme nebular emission lines. The latter interpretation suggested extreme \oiiihb\ EWs, implying the existence of high sSFRs $\gtrsim 2 \times$ greater than those seen at $z\sim2$ \citep[][]{smit_evidence_2014}. 

Recent \emph{JWST} observations have revealed a range of star formation histories within the early Universe, ranging from `mini-quenched' galaxies \citep[e.g.,][]{Looser_2024,looser_jades_2025} with a downturn in star formation in the last 30 Myr, to rapidly rising SFHs in galaxies undergoing a recent burst of star formation \citep[e.g.,][]{Fisher_REBELS_SFH}. Moreover, it has been suggested that the UV-brightest galaxies are often undergoing an upturn in their SFH  \citep[][]{endsley_star-forming_2024, Fisher_REBELS_SFH}. 

\begin{figure}
    \centering
    \includegraphics[width=0.8\columnwidth]{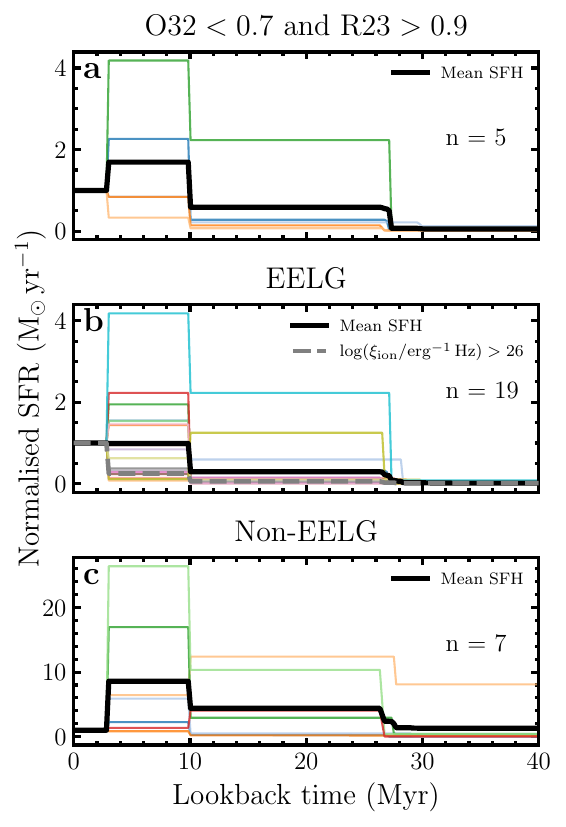}
    \caption{Normalised SFR as a function of lookback time for galaxies with $\rm O32 < 0.7$ and $\rm R23 > 0.9$ (panel \textit{a}), EELGs (panel \textit{b}) and non-EELGs (panel \textit{c}). SFHs are normalised by the most recent time bin of star formation. Individual SFHs are shown by the coloured lines, while the mean SFH is shown by the thick black line. The dashed grey line in panel \textit{b} indicates the mean SFH of the four EELGs with $\log_{10}(\xi_\mathrm{ion}/\mathrm{erg}^{-1}\,\mathrm{Hz})>26$.
    The total number of galaxies in each panel is indicated on the right hand side.}
    \label{fig:sfh_figure}
\end{figure}

\citet{witten_rising_2025} defined a region on the O32 and R23 emission line diagram indicating the high metallicity and low ionization region on the O32-R23 plane, which may contain galaxies hosting an older stellar population or show a rejuvenation in their SFH. We find five spectra within our sample lie within this region, with the remaining 21 spectra lying outside of this of region. We show the SFHs of these galaxies in Figure \ref{fig:sfh_figure}. On average, the galaxies with $\mathrm{O32} < 0.7$ and $\mathrm{R23} > 0.9$ show a downturn in the most recent SFH bin and more stellar mass build up in earlier time bins. We note that due to the small number of spectra within the $\mathrm{O32} < 0.7$ and $\mathrm{R23} > 0.9$ bin ($n=5$) these results should be interpreted with some caution and a larger sample size is required to further understand the properties of galaxies within this relatively poorly populated region of the $\mathrm{O}32-\mathrm{R}23$ parameter space. 

In Figure \ref{fig:sfh_figure}, we also explore the variation in the SFHs of EELGs compared to non-EELGs. We find that on average, EELGs exhibit an increasing star formation history, with a significant upturn in the last 10~Myr, while half the EELG sample has an upturn in the most recent 3~Myr bin.
This is consistent with a recent burst in star formation powered by massive O type stars, providing the ionizing photons driving the strong nebular emission \citep[][]{boyett_extreme_2024}. Conversely, a drop in the SFR over the last 3~Myr results in weaker nebular emission \citep[][]{endsley_star-forming_2024}, consistent with the declining SFHs shown by the non-EELG galaxies. Notably, only 3 of the non-EELGs have declining SFH over the last 10~Myr, while 4 of the non-EELGs have a decline in the last 3~Myr only. This young SFH reflects the low Balmer break strength (Figure \ref{fig:break strength}) of these non-EELGs (hinting at a Balmer jump), combined with a relatively high H$\beta$ EW in two of these sources.

\begin{figure*}
    \centering
    \includegraphics[width=0.95\textwidth]{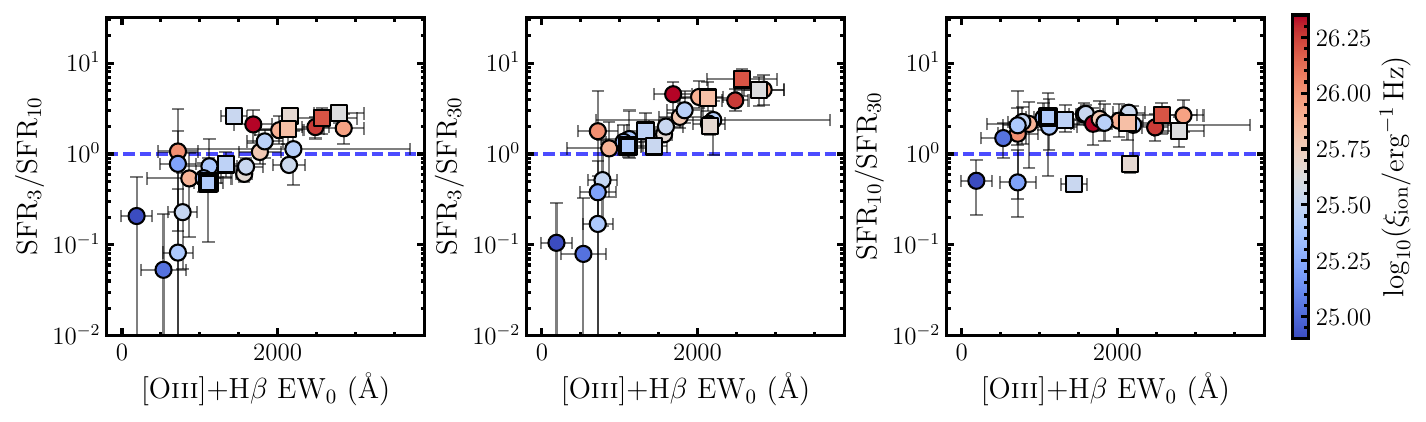}
    \caption{$\rm SFR_3/SFR_{10}$ (left), $\rm SFR_3/SFR_{30}$ (centre), and $\rm SFR_{10}/SFR_{30}$ (right) as a function of \oiiihb\ rest-frame EW, with the points coloured by $\log_{10}(\xi_\mathrm{ion})$. The blue dashed horizontal line indicates where the SFR ratios are equal: galaxies above this line show a rising star formation history on the given timescale, whereas galaxies below this line show a declining star formation history. We show LAEs as square markers, and the UV bump galaxy from \citet{ormerod_detection_2025} is shown with a thicker black border.}
    \label{fig:sfr_ratios}
\end{figure*}

In Figure \ref{fig:sfr_ratios}, we show the ratios of short term to longer term SFRs (tracing the burstiness of a galaxy's SFH) as a function of {\oiiihb} EW. We find that galaxies with more extreme emission lines show increasingly bursty SFHs with a recent upturn \citep[][]{endsley_star-forming_2024, boyett_extreme_2024}. This suggests that EELGs may be observed during a peak in their SFH, whereas non-EELGs are observed during a lull, or decline, in their SFH. 
Figure \ref{fig:sfr_ratios} also suggests that galaxies with elevated $\xi_\mathrm{ion}$ are undergoing an upturn in their SFH. An upturn in the most recent 3~Myr bin corresponds to short lived, massive, O type stars dominating the production of ionizing photons. However, the short lived nature of these stars may suggest that the contribution of EELGs to reionization may vary in line with their SFH. 

\begin{figure}
    \centering
    \includegraphics[width=0.95\columnwidth]{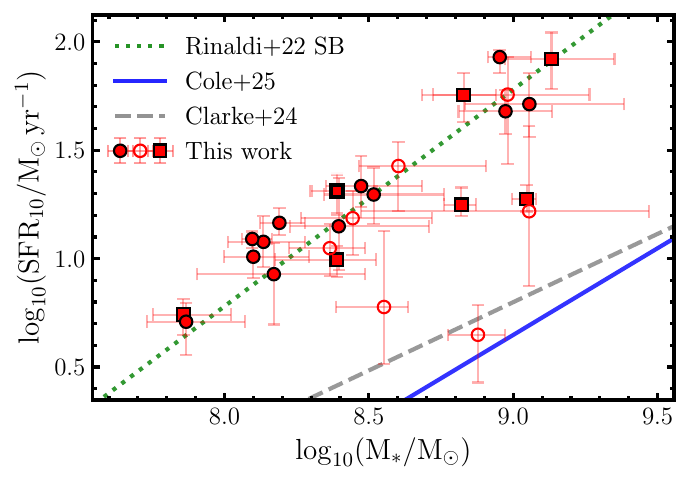}
    \caption{SFR$_{10}$ as a function of stellar mass. Markers are the same as previous figures. We show the SFMS from \citet{Clarke_2024} and \citet{Cole_2025} with grey dashed and blue solid lines, respectively. We also show the starburst relation from \citet{Rinaldi_2022} with the green dotted line.}
    \label{fig:sfr_mass}
\end{figure}
In Figure \ref{fig:sfr_mass}, we show the location of our galaxies in the SFR$_{10}$-stellar mass plane. We find that the majority of galaxies within our sample are offset above the star forming main sequence (SFMS), with most being consistent with the starburst (SB) relation from \citet{Rinaldi_2022} and consistent with the offset towards high sSFRs found by \citet{llerena_physical_2024}. 
We find that four galaxies within the non-EELG sample are consistent with the starburst relation within $1\sigma$. All four of these galaxies show a similar star formation history characterised by a peak of star formation in the $3-10$~Myr bin, followed by a decline in the most recent $0-3$~Myr bin. 

\section{Summary and Conclusions}
\label{sec: conclusion}

We have presented an analysis of 23 \emph{Spitzer}/IRAC excess galaxies at $5.66 < z < 7.64$, observed as part of the NIRSpec Wide GTO survey. Out of our sample, we identify $17/23$ $(\sim74\%)$ galaxies as EELGs, based on a criterion of {\oiiihb}$\geq1000$~{\AA}. This high EELG fraction reflects the IRAC-excess selection, supporting the interpretation that the IRAC excess predominantly arises from strong rest-frame optical nebular emission lines. Our main findings are summarised as follows: 
\begin{itemize}
    \item We measure a median rest-frame {\oiiihb} EW $=1583$~{\AA}, with individual galaxies up to {\oiiihb} EW $=2850$~{\AA}. Through a lack of high-ionization emission lines, an absence of broad emission line components, and through line ratio diagnostic diagrams, we find that the extreme equivalent widths observed in our sample are driven by star formation. However, we cannot rule out an AGN contribution in some individual sources.
    \item We find that EELGs are characterised by moderately high ionisation parameter and sub-solar gas phase metallicities. The O32 ratio correlates significantly with {\oiii} EW, consistent with galaxies with more extreme nebular emission having a more highly ionized ISM. 
    \item We find that our EELGs exhibit more extreme ISM conditions than EELGs at $z\sim2$, but are offset to lower O32 at a fixed {\oiii} EW when compared to faint EELGs from JADES. Combined with the absence of high-ionization rest-UV emission lines in the EELG stacked spectrum, this suggests that these luminous EELGs contain a less extreme radiation field than their fainter counterparts, but lower gas densities could also contribute.
    \item SED fitting suggests that EELGs are undergoing an active burst of star formation, with EELGs characterised by rising SFHs and young stellar mass-weighted ages ($t_\star \sim 20$ Myr). Conversely, non-EELGs are characterised by a downturn in their SFH in the last 3-10~Myr. However, our non-EELGs are still relatively young ($t_\star \sim 28$ Myr), due low Balmer break strengths. Our EELG sample shows a median stellar mass of $\log _{10}\left(\mathrm{M}_{\star} / \mathrm{M}_{\odot}\right) = 8.40_{-0.30}^{+0.59}$ and lies above the star forming main sequence, consistent with starburst relations.
    \item We find a median $\log_{10}(\xi_\mathrm{ion}/\mathrm{erg^{-1}~Hz})=25.62^{+0.37}_{-0.18}$ for our EELG sample, offset above the canonical value of $\log_{10}(\xi_\mathrm{ion}/\mathrm{erg^{-1}~Hz})=25.3$. We find that {\oiiihb} EW and $\rm SFR_3/SFR_{30}$ correlate strongly with $\xi_\mathrm{ion}$, suggesting that bursty EELGs are efficient ionizing photon producers.   
    \item Bright EELGs have low neutral hydrogen column densities, with few galaxies showing evidence for a DLA at the Lyman break. This is consistent with the scenario where they have ionized their CGM. However, stacked spectra do not reveal evidence for Ly$\alpha$ emission at $z>6.9$, and only low-EW Ly$\alpha$ emission ($16.1_{-4.3}^{+4.3}${\AA}) becomes visible in the low-z stack. Given the strong optical line emission, this results in a low Ly$\alpha$ escape fraction ($<2$\%), resulting in  relatively small ionised bubble sizes ($R_\mathrm{ion} < 0.25~\mathrm{pMpc}$). Given the low neutral hydrogen column densities and implications for an ionised CGM, the absence of strong Ly$\alpha$ suggests the IGM surrounding EELGs on larger scales remains largely neutral.
\end{itemize}

Overall, bright EELGs ($\sim L_\mathrm{UV, z=7}^\ast$) are an ideal sample to differentiate between models where bright and faint galaxies contribute most to reionization \citep[e.g.,][]{naidu_rapid_2020, atek_most_2024}. Due to their elevated ionizing photon production efficiency these galaxies have the potential to contribute significantly to the reionization process, however, our analysis of the Ly$\alpha$ break and line emission suggests they are not yet capable of impacting the IGM around them. Given their bursty nature and modest stellar masses, these luminous systems could live in low mass haloes less biased to overdense regions. This could point towards the environment of galaxies having a greater impact on cosmic reionization than the physical properties of individual galaxies, with galaxies at the bright end of the UV luminosity function playing a minor role in reionization.

\section*{Acknowledgements}
KO thanks Callum Witten, Mengtao Tang, and Ryan Endsley for sharing their data, and thanks Francesco D'Eugenio for useful discussions.
KO would like to thank the Science and Technology Facilities Council (STFC) and Faculty of Health, Innovation, Technology and Science (HITS) at Liverpool John Moores University (LJMU) for their studentship. 
RS acknowledges support from an STFC Ernest Rutherford Fellowship (ST/S004831/1). 
JW gratefully acknowledges support from the Cosmic Dawn Center through the DAWN Fellowship. The Cosmic Dawn Center (DAWN) is funded by the Danish National Research Foundation under grant No. 140.
MVM is supported by the National Science Foundation via AAG grant 2205519.
AdG acknowledges support from a Clay Fellowship awarded by the Smithsonian Astrophysical Observatory. AdG acknowledges funding by the European Union (ERC, FIRST-GIANTS, 101221926). AdG is supported by the Lise Meitner Excellence program of the Max Planck Society.
AJB acknowledges funding from the “FirstGalaxies” Advanced Grant from the European Research Council (ERC) under the European Union’s Horizon 2020 research and innovation program (Grant agreement No. 789056). This study made use of Prospero high-performance computing facility at Liverpool John Moores University.

This work is based on observations made with the NASA/ESA/CSA James Webb Space Telescope (JWST). The data were obtained from the Mikulski Archive for Space Telescopes at the Space Telescope Science Institute, which is operated by the Association of Universities for Research in Astronomy, Inc., under NASA contract NAS 5-03127 for JWST. These observations are associated with programmes 1211, 1212, 1213, 1214, and 1215. 
Some of the data products presented herein were retrieved from the Dawn JWST Archive (DJA). DJA is an initiative of the Cosmic Dawn Center (DAWN), which is funded by the Danish National Research Foundation under grant DNRF140.

\section*{Data Availability}

The data underlying this article will be shared on reasonable request to the corresponding author.



\bibliographystyle{mnras}
\bibliography{paper} 




\appendix

\section{Duplicate observations}
\label{sec:duplicate obs}

\begin{figure}
    \centering
    \includegraphics[width=0.99\columnwidth]{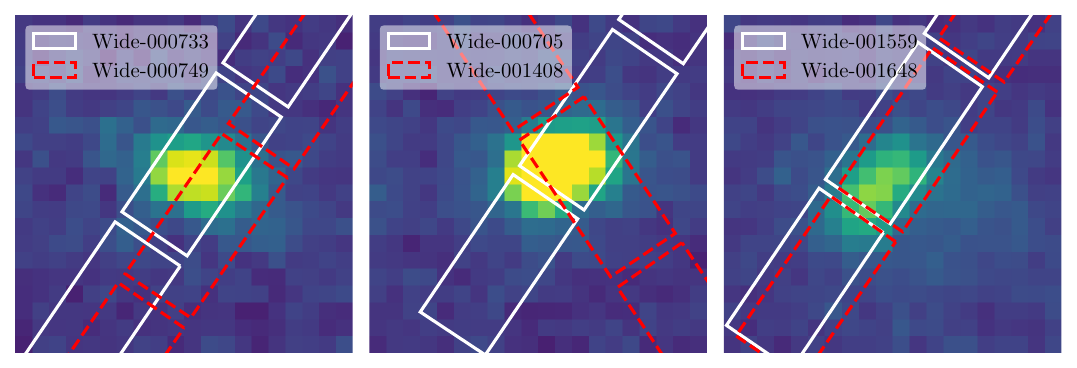}
    \caption{The location of the NIRSpec slits for each of the duplicate observations, overlaid on the NIRCam F444W imaging. All cutouts are $1\arcsec \times 1\arcsec$.}
    \label{fig:slit locs}
\end{figure}

Three galaxies within our sample were observed twice with NIRSpec. Although this results in duplicate observations of three galaxies, we retain both observations separately within our analysis due to the NIRSpec slits sampling different regions of the galaxy. We show the NIRSpec slit locations of the duplicate observations in Figure \ref{fig:slit locs}, overlaid on NIRCam F444W imaging obtained from the DAWN \emph{JWST} Archive \citep[DJA;][]{Valentino_23_DJA_imaging, Brammer_2023_Grizli}.

\begin{figure*}
    \centering
    \includegraphics[width=0.9\linewidth]{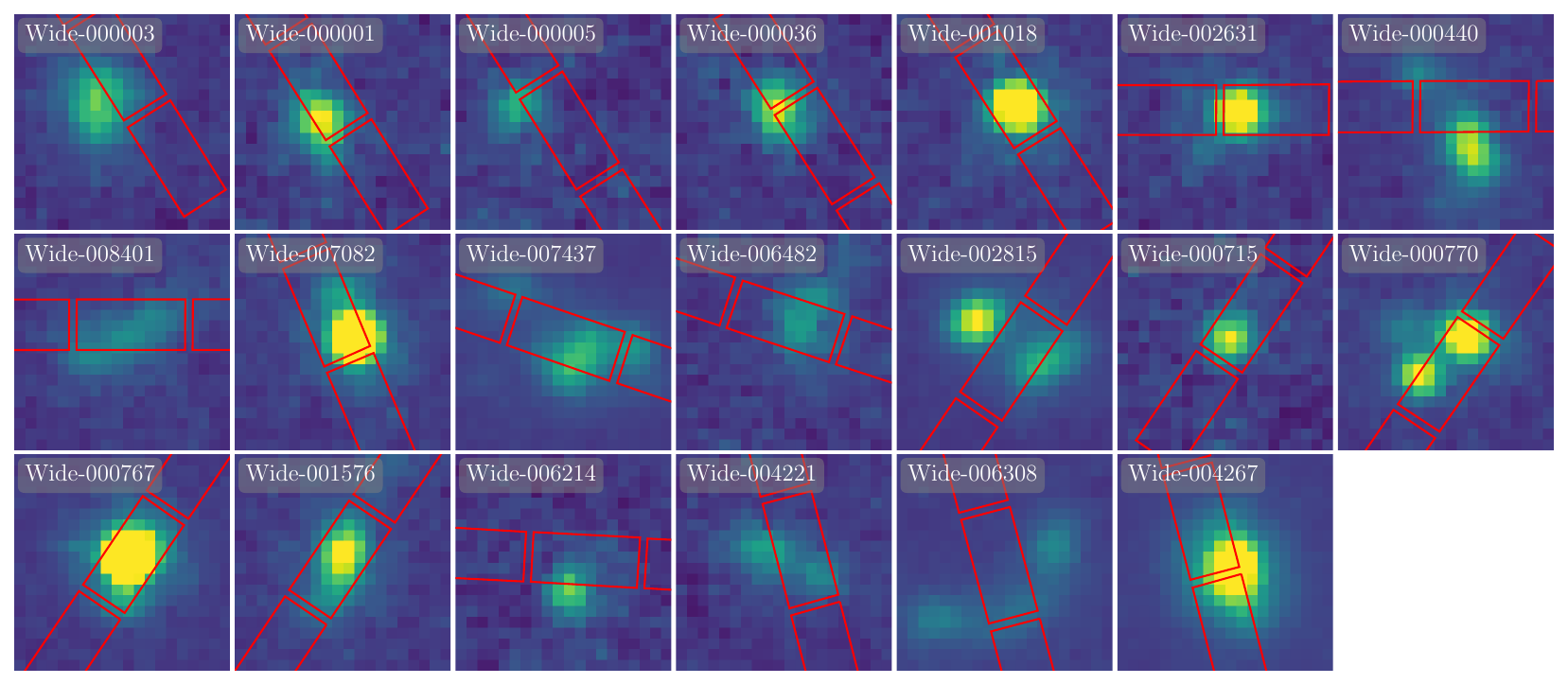}
    \caption{The location of the NIRSpec slits for each galaxy overlaid on the NIRCam F444W imaging. All cutouts are $1\arcsec \times 1\arcsec$. We note that we do not include Wide-007695 as it is not covered by NIRCam imaging available on the DJA.}
    \label{fig:individual slit locs}
\end{figure*}

We show the slit locations for the remaining objects in Figure \ref{fig:individual slit locs}. We note that we do not include Wide-007695 as it is not covered by NIRCam imaging available on the DJA.  

\section{High-resolution spectroscopic redshifts}
\label{sec:high res}
In addition to the PRISM redshifts presented in Table \ref{tab:general}, we present the redshifts obtained from \textsc{ppxf} spectral fitting of the G395H $R\sim2700$ grating ($z_\mathrm{G395H}$) in Table \ref{tab:g395h_z}.  

\begin{table}
    \centering
    \caption{Redshifts and $1\sigma$ errors obtained from \textsc{ppxf} spectral fitting to the G395H high resolution spectra. Galaxies without a redshift do not have any emission lines detected in their G395H spectrum, or \textsc{ppxf} fitting flagged the redshift as a bad fit. }
    \begin{tabular}{l|c} \hline
    Wide ID & $z_\mathrm{G395H}$ \\ \hline
    Wide-000003 & $7.29356 \pm 0.00166$ \\
    Wide-000001 & $7.28295 \pm 0.00021$ \\
    Wide-000036 & $6.09615 \pm 0.00021$ \\
    Wide-001018 & $6.55219 \pm 0.00006$ \\
    Wide-002631 & - \\
    Wide-000440 & - \\
    Wide-008401 & - \\ 
    Wide-007082 & - \\
    Wide-007437 & - \\
    Wide-006482 & - \\
    Wide-000733 & $6.80986 \pm 0.00093$ \\
    Wide-000749 & $6.80801 \pm 0.00009$ \\
    Wide-002815 & - \\
    Wide-000715 & - \\
    Wide-000770 & - \\
    Wide-000705 & $6.93332 \pm 0.00006$ \\
    Wide-001408 & - \\
    Wide-000767 & $7.63059 \pm 0.00007$ \\
    Wide-001576 & $7.10829 \pm 0.00014$ \\
    Wide-001559 & $7.56940 \pm 0.00017$ \\
    Wide-001648 & $7.56896 \pm 0.00016$ \\
    Wide-006214 & $6.73341 \pm 0.00005$ \\
    Wide-004221 & $6.65164 \pm 0.00007$ \\
    Wide-006308 & $7.54347 \pm 0.00009$ \\
    Wide-004267 & $7.47092 \pm 0.00132$ \\
    Wide-007695 & $5.66029 \pm 0.00010$ \\ \hline
    \end{tabular}
    \label{tab:g395h_z}
\end{table}

\section{Balmer Break Strength}
\label{sec: balmer break appendix }
\begin{figure}
    \centering
    \includegraphics[width=0.95\columnwidth]{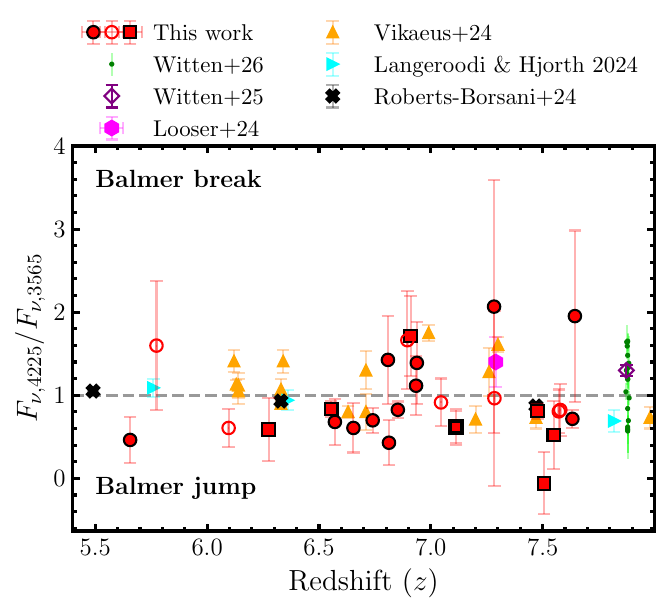}
    \caption{Balmer break strength ($B = F_{\nu, 4225} / F_{\nu, 3565}$) as a function of redshift. Markers are the same as in previous figures. The grey dashed line shows $B=1$ -- the transition from a Balmer break to a Balmer jump. For comparison, we show the mini-quenched galaxy from \citet{Looser_2024} with a pink hexagon, and the protocluster members from \citet{Witten_2026} as green dots, YD4 as an open purple triangle \citep{witten_rising_2025}, galaxies from \citet{vikaeus_be_2024} as orange triangles, binned Balmer break strengths from \citet{langeroodi_nirspec_2024} as blue arrows, and Balmer break strengths measured from stacked spectra from \citet{roberts_borsani_between_the_extremes_2024} are shown as black crosses.}
    \label{fig:break strength}
\end{figure}

We show the Balmer Break strengths obtained in Section \ref{sec: balmer break} in Figure \ref{fig:break strength}.

\section{SED Fitting}

The priors used for our \textsc{bagpipes} SED fitting in Section \ref{sec:sed fitting} are summarised in Table \ref{tab:priors}.

\begin{table}
    \centering
    \caption{Summary of SED fitting parameters and priors used in our \textsc{bagpipes} fitting in Section \ref{sec:sed fitting}. The minimum and maximum values allowed are given in brackets.}
    \begin{tabular}{l|l} \hline
         Parameter & Prior  \\ \hline
         $z$ & Fixed to $z_\mathrm{prism}$\\ 
         $\log_{10}(M_\star/M_\odot)$ & Uniform: (5, 15)\\
         $Z_\star ~(Z_\odot)$ & Uniform: (0, 2.0)\\
         $\log_{10}(U)$ & Uniform: (-3, -0.5)  \\ 
         $A_V$ (mag) & Gaussian: (0, 7), $\mu = 0.15, \sigma=0.15$ \\
         $\sigma_v$ (km s$^{-1}$) & Uniform: (1, 1000)\\ 
         $\Delta\log$(SFR)$_i$ & Student's-t: (-50, 50) \\ 
         \multirow{2}{*}{$B$}
        & Fixed to 0 with the exception of Wide-006214 \\ 
         &  where the prior is uniform: (0, 10) \\ 
         $\delta$ & Uniform: (-0.5, 0.2) \\ 
         $\log _{10}\left(N_{\mathrm{H\textsc{i}}}\right)$& Uniform: (19, 24) \\
         
         \hline
    \end{tabular}
    \label{tab:priors}
    
\end{table}

\section{DLA Comparison}
\label{sec:DLA_comparison}
\begin{figure}
    \centering
    \includegraphics[width=0.8\columnwidth]{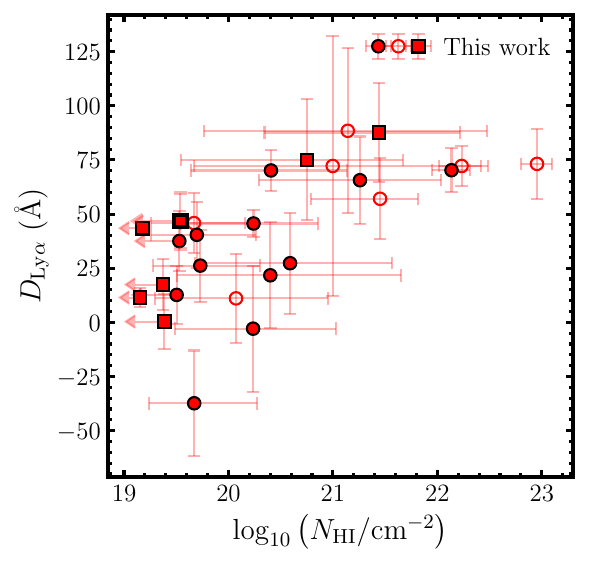}
    \caption{Comparison of $D_\mathrm{Ly\alpha}$ and $\log _{10}\left(N_{\mathrm{H\textsc{i}}}\right)$. The markers are the same as previous figures.}
    \label{fig:Dlya_DLA_comparison}
\end{figure}

We present a comparison between $D_\mathrm{Ly\alpha}$ and $\log _{10}\left(N_{\mathrm{H\textsc{i}}}\right)$ measurements in Figure \ref{fig:Dlya_DLA_comparison}.

\section{AGN Diagnostic Diagrams}
\label{sec:s2-vo87}

At low redshift, demarcation lines from \citet{Kewley_2001} and \citet{Kauffmann_2003} are often used to distinguish between AGN and SF galaxies. However, the increasing ionization parameter at high redshift increases the R3 ratios seen in SF galaxies, moving sources towards these classical demarcation lines \citep[][]{scholtz_2025}. Additionally, the line ratios of narrow-line AGN move towards the SF region at high redshift due to the low metallicity of narrow-line regions, making the use of line ratio diagnostic diagrams more challenging in the early Universe \citep[e.g.,][]{Nakajima_Maiolino_2022, Harikane_2023_AGN, Maiolino_2024_JADES_AGN}. Due to this, we adopt the more conservative demarcation line for the S2-VO87 diagram from \citep[][]{scholtz_2025}. Although the [S\,\textsc{ii}] emission line is not available for the entirety of our sample due to the wavelength coverage of the PRISM spectra, we do not see strong evidence for AGN activity within our sample. One galaxy (Wide-006214) lies within the AGN region, but remains consistent with the star-forming region within $1\sigma$. 

\begin{figure}
    \centering
    \includegraphics[width=0.99\columnwidth]{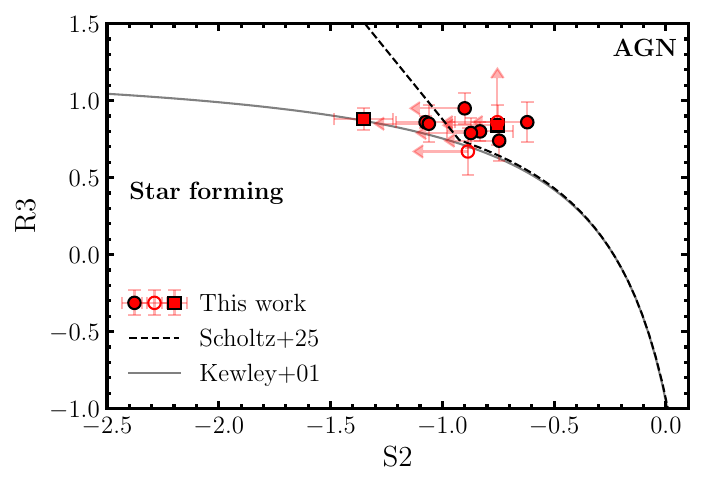}
    \caption{The S2-VO87 diagram (R3 vs S2) for our sample. Markers are the same as previous figures. The classical demarcation line from \citet{Kewley_2001} is shown by the grey solid line, with the more conservative demarcation line utilised in this work, from \citet{scholtz_2025}, shown by the black dashed line. }
    \label{fig:s2-vo87}
\end{figure}

\begin{figure}
    \centering
    \includegraphics[width=0.99\columnwidth]{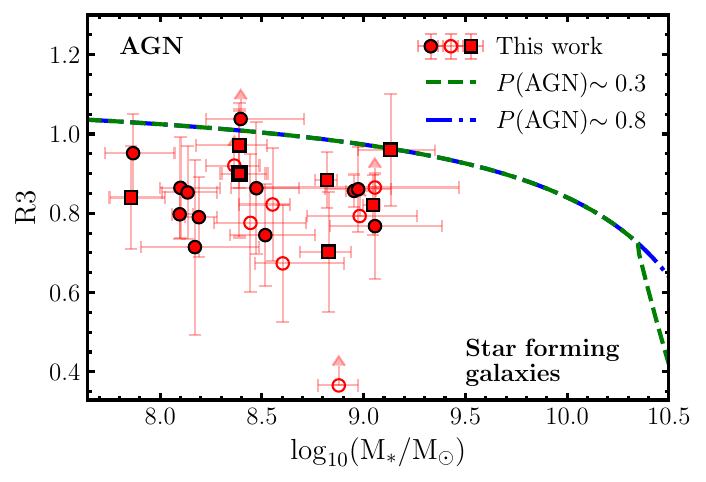}
    \caption{Mass-Excitation (MEx) diagram for our sample, showing the R3 emission line ratio against stellar mass. EELGs are shown as solid red markers, while open circles represent the non-EELG sample. The UV bump galaxy from \citet{ormerod_detection_2025} is shown with a thicker black border.
    The dashed green line and dot-dashed blue line show the divisions between star-forming galaxies and AGN from \citet{juneau_2011}, shifted by $\Delta \log \left(M_* / M_{\odot}\right)=0.75$ as in \citet{coil_2015}. 
    }
    \label{fig:MEx}
\end{figure}

For the one galaxy in which both H\,$\gamma$ and [O\,\textsc{iii}]$\lambda 4364$ are detected robustly in the G395H spectrum (Wide-001018), we utilise the O3H$\gamma$-O32 diagram from \citet{Mazzolari_AGN_2024}. Although Wide-001018 falls within the AGN-only region, its position remains consistent with the SF+AGN region within $1\sigma$.

\begin{figure}
    \centering
    \includegraphics[width=0.8\columnwidth]{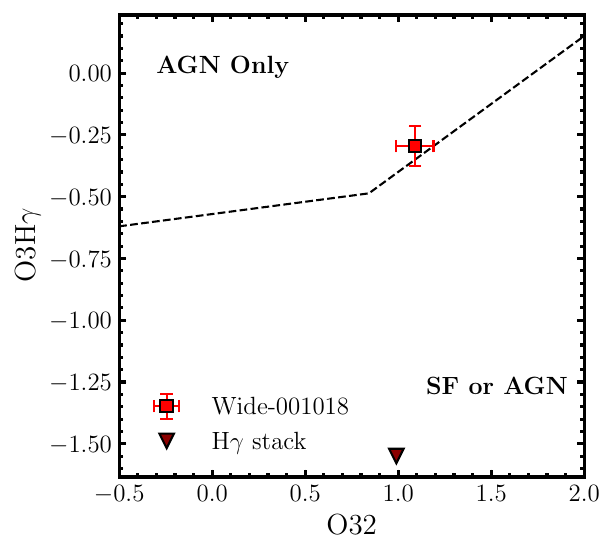}
    \caption{AGN diagnostic diagram showing the O3H$\gamma$ and O32 line ratios for Wide-001018. The dark red triangle indicates the upper limit from the stack of galaxies with an $R\sim2700$ H$\gamma$ detection. The black dashed line indicates the demarcation line from \citet{Mazzolari_AGN_2024}.}
    \label{fig:o3hg}
\end{figure}

\section{Mass-Metallicity Relation}
\label{sec:mass_metallicity}

\begin{figure}
    \centering
    \includegraphics[width=1.0\linewidth]{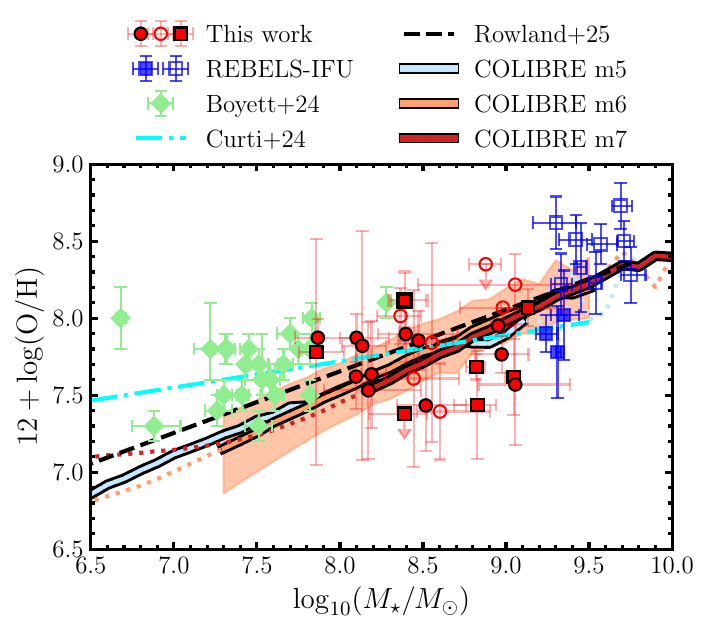}
    \caption{Gas phase metallicity as a function of stellar mass for our sample. Markers are the same as previous figures. We show EELGs from \citet{boyett_extreme_2024} as green diamonds, and the REBELS-IFU sample \citep[Stefanon et al. in prep;][]{Rowland_Metallicity_2025} as blue squares, where open markers represent non-EELGs, and solid markers represent EELGs. We additionally show the mass metallicity relations from \citet{curti_jades_2024} as the cyan dot-dashed line, and \citet{Rowland_Metallicity_2025} as the dashed black line. We additionally show the $z=8$ MZR from star forming galaxies within the COLIBRE simulations at the m5, m6 and m7 mass resolutions \citep[][]{Sharda_2026}. Curves transition from solid to dashed when the number of galaxies drops below 20 in a given mass bin or if the simulation resolution limit is reached, and the orange shading shows the 16th-84th percentiles for the m6 simulation.}
    \label{fig:mass metallicity}
\end{figure}

We show our sample, alongside galaxies from \citet{boyett_extreme_2024} and the REBELS-IFU sample overlaid on best-fit MZR relations from \citet{curti_jades_2024} and \citet{Rowland_Metallicity_2025} in Figure \ref{fig:mass metallicity}. When compared to the MZR from \citet{Rowland_Metallicity_2025}, the majority of our sources fall slightly below the MZR, and show greater consistency with the flatter relation from \citet{curti_jades_2024}.

\section{Full Sample Stack}
\label{sec: stack appendix}

\begin{figure*}
    \centering
    \includegraphics[width=1\linewidth]{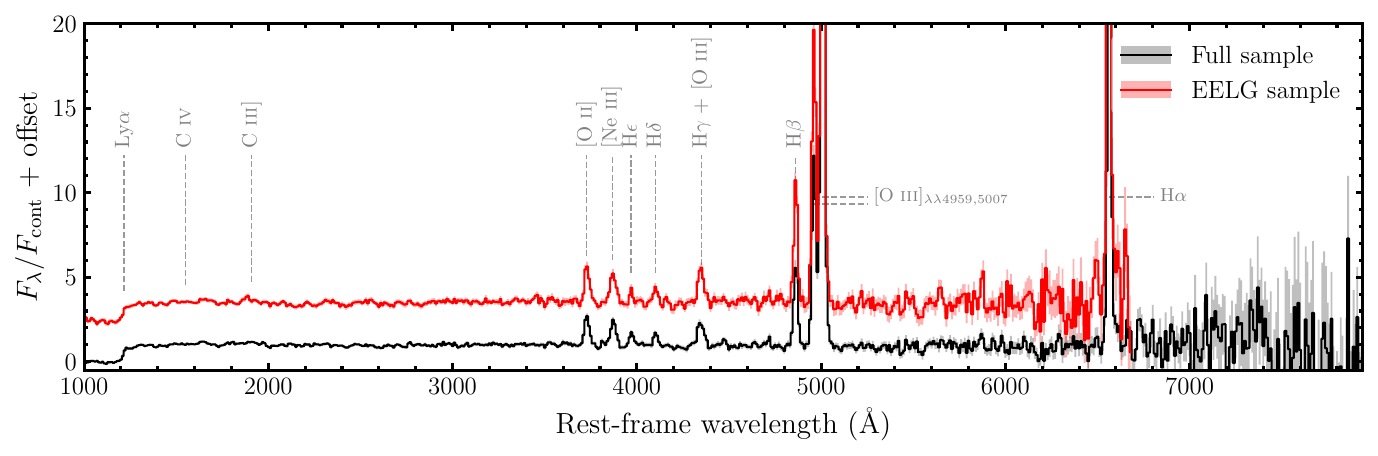}
    \caption{Stack of all 26 galaxies within our sample (black line) and the EELG sample (red line). Prior to stacking, galaxies are normalised according to their UV continuum slope. Shading indicates $1\sigma$ errors. The locations of key emission lines are indicated with grey dashed lines. }
    \label{fig:full stack}
\end{figure*}

We create a stack of the full IRAC-excess sample, including both EELGs and non-EELGs, follow the method described in Section \ref{sec: stacks}. The stack of the full sample is shown in Figure \ref{fig:full stack}, alongside the stack of the EELG sample for comparison. The resulting spectrum is characterised by a lack of rest-UV emission lines, and a lack of high-ionization or broad lines that may indicate the presence of AGN, suggesting that the extreme {\oiiihb} EWs are driven by ionizing photons produced by young, massive stars. 

\section{\emph{Spitzer}/IRAC Excess Colours}
\label{sec:irac appendix}

\begin{figure*}
    \centering
    \includegraphics[width=0.6\linewidth]{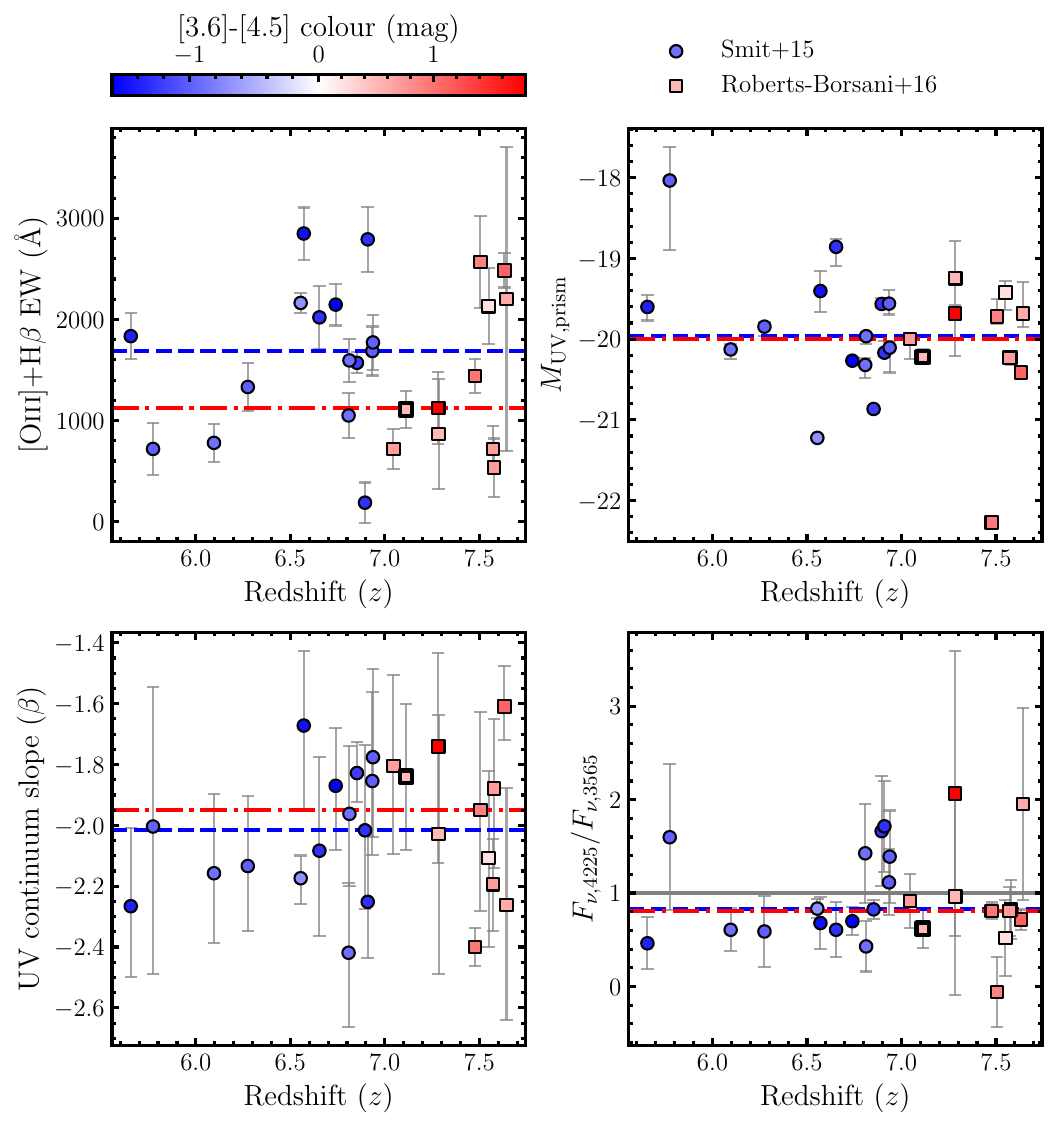}
    \caption{{\oiiihb} EW (top left), $M_\mathrm{UV,prism}$ (top right), UV continuum slope ($\beta$; bottom left), and Balmer break strength (bottom right) as a function of redshift. Galaxies selected from \citetalias{smit_high-precision_2015} are shown as circular markers, while those selected from \citetalias{roberts-borsani_z_2016} are shown as square markers. The UV bump galaxy from \citet{ormerod_detection_2025} is shown with a thicker black border. Markers are coloured according to their $[3.6]-[4.5]$ colour. The blue dashed and red dot-dashed lines indicate the median values for galaxies selected from \citetalias{smit_high-precision_2015} and \citetalias{roberts-borsani_z_2016}, respectively. In the bottom right panel, the solid grey line indicates $F_{\nu,4225}/F_{\nu,3565}=1$, representing the transition from a Balmer break to a Balmer jump.}
    \label{fig:irac_colours}
\end{figure*}

In Figure \ref{fig:irac_colours}, we show various physical properties ({\oiiihb} EW, $M_\mathrm{UV,prism}$, UV continuum slope, and Balmer break strength) as a function of redshift with galaxies coloured by their $[3.6]-[4.5]$ colour.

\bsp	
\label{lastpage}
\end{document}